\documentclass[12pt, oneside]{mnthesis_new}
\usepackage{hyperref}
\usepackage{url}
\usepackage{longtable}
\usepackage{mathrsfs}
\usepackage{multirow}
\usepackage{bigstrut}
\usepackage{amssymb}
\usepackage{graphicx}
\usepackage{amsmath}
\usepackage{amsfonts}

\begin{document}
\bibliographystyle{h-physrev5} 

\phd 
\title{\bf The Dynamics and Thermodynamics of Soft-Wall AdS/QCD}
\author{Thomas Matthew Kelley}
\campus{University of Minnesota} 
\program{Physics} 
\director{Joseph Kapusta} 

\submissionmonth{July} 
\submissionyear{2011} 

\abstract{
Gauge/Gravity dualities open the non-perturbative realms of strongly-coupled gauge theories to analytic treatment. Anti-de Sitter Space/Conformal Field Theory, one way of connecting gravity dual models to gauge theories, is a correspondence between a ten-dimensional Type IIB superstring theory in $AdS_{5}\times S^{5}$ and a $\mathcal{N}=4$ super Yang Mills theory. To describe systems that are experimentally accessible, however, the formal correspondence is modified into a phenomenological duality between a five-dimensional gravity model and a strongly coupled QCD-like gauge theory. This duality is referred to as AdS/QCD.

This work explores aspects of the soft-wall AdS/QCD model. The phrase `soft wall' refers to the means of breaking the conformal symmetry and introducing a mass scale to the gauge side of the duality. We add higher-order terms to the soft-wall Lagrangian and calculate the effect on physical observables. Meson mass spectra gain a more complex structure, exhibiting a better match with the experimental values than previous models. The Gell-Mann--Oakes--Renner relation naturally emerges from the model. We calculate the form factor $F_{\pi}$ and the coupling $g_{\rho\pi\pi}$ as a non-trivial test on the limits of our soft-wall model.

 Introducing a black brane into the gravity dual metric allows us to derive thermodynamic quantities in the gauge theory. As expected at high temperatures, the entropy scales as the cube of the temperature, and the speed of sound reaches its conformal limit of one-third. Thermal condensates contribute leading-order terms, modifying the temperature, entropy, and free energy behavior. We find that the system undergoes a phase transition from hadronic matter to a strongly coupled quark-gluon plasma at a critical temperature.
 
}
\copyrightpage 
\acknowledgements{

Many people have contributed to my success at the University of Minnesota. First, I would like to thank several fellow graduate students. I thank Brian Batell for taking time from his own research and helping me start mine. Daniel Sword worked through the more confusing aspects and mathematical intricacies of gauge/gravity dualities with me. And most of all, I must thank Todd Springer for the invaluable input and help throughout my tenure.

I thank Dan Cronin-Hennessy for taking a chance on a young graduate student. I also thank Misha Stephanov and Francesco Nitti who provided guidance and insights into my thermodynamic work. 

I also thank my two advisors:  Tony Gherghetta, who despite going halfway around the world still gave me his time and effort, and Joe Kapusta, who was generous enough to take me on as a student halfway through my graduate career.

Of course, none of this work would have been accomplished if it were not for my patient spouse, Laura. She endured through the frustrations and the late nights of work. Her unconditional love and support proved invaluable during my great physics quest.  

This work has been supported by the School of Physics and Astronomy at the University of Minnesota, the US Department of Energy (DOE) under Grant No. 
DE-FG02-87ER40328, and the Graduate School at the University of Minnesota under the Doctoral Dissertation Fellowship.

}
\dedication{To my parents, whose support made my education possible.
}


\beforepreface 

\figurespage
\tablespage

\afterpreface            



\chapter{Introduction}
\label{intro_chapter}

\begin{flushright}
``You have to have an idea of what you are going to do, but it should be a vague idea.''\\
-Pablo Picasso

\end{flushright}

All the matter and energy in the 100 billion stars of the Milky Way galaxy and the 100 billion galaxies in the universe came from one infinitely dense point. Some 13.1 to 13.7 billion years ago, the point began expanding after the event now known as the Big Bang \cite{Hinshaw:2008kr}. According to modern thinking, one unified force governed the universe for the first 10$^{-43}$ seconds of its existence \cite{deBoer:1994dg, pdg}. Over the next picosecond (10$^{-12}$ s), the unified force fractured into the four known forces that rule the cosmos today: strong, electromagnetic, weak, and gravitational.

 Today, we have two prevailing theories describing all of nature. The Standard Model encapsulates the strong, electromagnetic, and the weak forces, describing them in terms of so-called gauge theories \cite{Lykken:2010mc}. General Relativity describes the spacetime curvature responsible for gravitational interactions, but is inherently incompatible with the Standard Model. Many physicists, past and present, seek theoretical constructs that unify the Standard Model with general relativity into a grand unified theory that depicts the beginning of the universe. One such grand unified scheme that has gained notoriety for its mathematical complexity and exclusion from experimental verification is string theory. However, ideas from string theory allow for a correspondence between extra-dimensional gravity models and strongly coupled gauge theories. Such a correspondence allows calculations done in gravity models to be interpreted as gauge theory quantities. In our research, we concentrate on how gauge/gravity dualities describe the strongly coupled regions of the strong force. 

A gauge theory called quantum chromodynamics (QCD) describes the strong force. QCD characterizes the interactions among quarks, the constituents of nucleons, and gluons, the color-charge force carriers. Using QCD, one can develop effective theories describing hadronic matter. QCD distinguishes itself from other gauge theories with two characteristics: confinement and asymptotic freedom. Confinement ensures that only color-charge singlets exist, meaning only colorless objects are physically allowed. Asymptotic freedom refers to the high-energy behavior of quarks. At high energies, the quarks within the hadrons are weakly-interacting particles, allowing perturbative calculations.

Traditionally, calculations within gauge theories have been done in series expansions. For instance, if we wanted to calculate the cross section of an interaction $\sigma_{I}$, then we could express it as a perturbative expansion in powers of the coupling constant $g$, where $q$ is the energy-momentum vector of the collision,
\begin{equation}\label{equpert}
\sigma_{I} = F_{1}(q^2)g + F_{2}(q^2) g^{2}+ F_{3}(q^2) g^{3} + \ldots.
\end{equation}
As long as $g\ll 1$, higher-order terms in $g$ may be ignored, and calculations consider only the first few terms. A problem arises when $g\geq 1$; subsequent terms add ever-increasing values to the expansion (\ref{equpert}). Thus for strongly coupled systems ($g\geq 1$), perturbative techniques break down. 

In QCD, as in all gauge theories in the Standard Model, the coupling constant changes with energy, $g\rightarrow g(q^2)$; this phenomenon is termed the running of the coupling. As the energies involved increase, the coupling constant of QCD decreases. At energies less than $\mathcal{O}(1$ GeV$)$, perturbative calculations fail, and new means of calculating physical observables must be found \cite{Peskin:qft}. Gauge/gravity correspondence provides a remarkable mathematical tool to carry out such nonperturbative calculations in strongly coupled theories. 

\begin{figure}[h!]
\begin{center}
\includegraphics[scale=0.60]{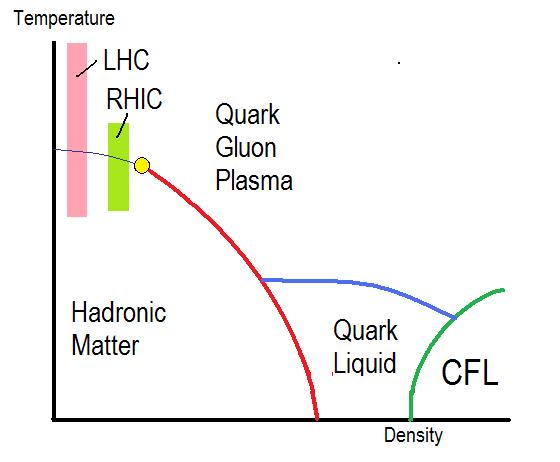}
\caption{The phase diagram of QCD \cite{pdg}. Hadronic matter experiences a cross-over transition to quark-gluon plasma as it reaches a critical temperature. Experiments at the RHIC and LHC probe temperatures around the cross-over.}
\label{figQCDdiagram}
\end{center}
\end{figure}

According to recent experiments, nonperturbative calculations are still needed as the temperature is increased in physical systems. In QCD, the protons and neutrons melt and become quark-gluon plasma (QGP) with increasing temperatures. Just as water turns to steam, as temperatures increase beyond a critical point, the confined hadronic matter undergoes a cross-over phase change to deconfined state of the QGP as seen in the phase diagram in Figure \ref{figQCDdiagram}. In this state, quarks and gluons are no longer bound within hadrons but associate freely \cite{qgp, qgp3}. Such a substance has been created at the Relativistic Heavy-Ion Collider (RHIC) at Brookhaven National Laboratory and most recently at the Large Hadron Collider (LHC) at CERN. Experimental results suggest that the QGP is strongly coupled and mimics the behavior of fluids \cite{Adams:2005dq, Adcox:2004mh, Arsene:2004fa, Back:2004je}. As a result, ideas from fluid dynamics are useful in gauge/gravity dualities at finite temperature. 

In this thesis, we will discuss a particular gauge/gravity duality called Anti-de Sitter/Conformal Field Theory (AdS/CFT) and how it gives rise to an effective correspondence called Anti-de Sitter/Quantum Chromodynamics (AdS/QCD). This work proceeds as follows
\begin{itemize}
\item{In Chapter \ref{chbackground} and \ref{chApplications}, we review the literature concerning the correspondence dictionary of AdS/CFT and explain how it yields the effective correspondence AdS/QCD. We go into detail on how to modify the gravity dual model in AdS space to describe a QCD-like theory with a large number of colors, $N_{c}$. We explore how physicists have applied the gauge/gravity duality to model condensates, chiral symmetry breaking (CSB), and meson mass spectra of QCD. We also detail how the duality characterizes the QGP and how the thermodynamics of this QGP-like fluid are encoded in a black-hole metric of the gravity model.}

\item{In Chapter \ref{chzero}, we modify existing AdS/QCD models at zero temperature. We improve the description of CSB done in previous works. These improvements are expanded to the mass spectra of the scalar, pseudoscalars, vectors, and the axial-vectors. Concluding this chapter, we explore AdS/QCD calculations of $g_{\rho\pi\pi}$, the form factor $F_{\pi}$, and the Gell-Mann--Oakes--Renner relation. }

\item{In Chapter \ref{chthermo}, we expand our AdS/QCD model by introducing a black brane into the AdS space. We then explore the entropy, speed of sound, and free energy of the QGP-like fluid created. Investigating the actions of the systems, we show how the transition from the AdS thermal metric to a black-hole metric corresponds to the transition from the confined phase to the deconfined phase of QCD. }

\item{Finally, in Chapter \ref{conclusion}, we discuss the main conclusions of the analysis, open questions yet to be answered, and future prospects for this area of research.}
\end{itemize}

The majority of the work presented here has been previously published in \cite{Gherghetta:2009ac} and \cite{Kelley:2010mu}. The material in Chapter \ref{chthermo} and Appendix \ref{appPhenom} will be published at a later date. Additional details and expanded discussions are included.



\chapter{AdS/CFT: Overview of the Gauge/Gravity Duality}\label{chbackground}

\begin{flushright}
``This is your last chance. After this, there is no turning back. You take the blue pill - the story ends, you wake up in your bed and believe whatever you want to believe. You take the red pill - you stay in Wonderland and I show you how deep the rabbit-hole goes.''\\
Morpheus, \emph{The Matrix}(1999)
\end{flushright}

In this chapter, we explain how the AdS/CFT correspondence connects to QCD. The focus of this thesis is modeling particular dynamics and thermodynamics of a QCD-like gauge theory using a soft-wall AdS/QCD. Explaining the connection between AdS/CFT to AdS/QCD and then the phenomenology leading up to the original research in Chapter \ref{chzero} and \ref{chthermo} covers a great deal of material. For additional background, several reviews have been written \cite{Aharony:1999ti, Klebanov:2005mh, Myers:2008fv, Edelstein:2009iv, Gubser:2009md, Gubser:2011qv}. We begin by exploring the basics of the AdS/CFT correspondence. From there, we introduce several phenomenological modifications meant to deform the CFT into a QCD-like theory with a large number of colors. 

\section{The Basics of AdS/CFT}
The basis of all dualities is the ability to describe a physical system in two different but equivalent ways. The principles of AdS/CFT correspondence were established late last century \cite{Maldacena:1997re, Gubser:1998bc, Witten:1998qj, Klebanov:1999tb}. The seminal works drew a direct correspondence between Type IIB string theory in $AdS_{5}\times S^{5}$ with a $\mathcal{N}=4$ supersymmetric Yang-Mills (SYM) theory by investigating the behavior of so-called branes in the low-energy and strong coupling limits. Because AdS/CFT uses a lower-dimensional theory to construct a higher-dimensional action, it is often referred to as a holographic model.

\begin{figure}[h!]
\begin{center}
\includegraphics[scale=0.60]{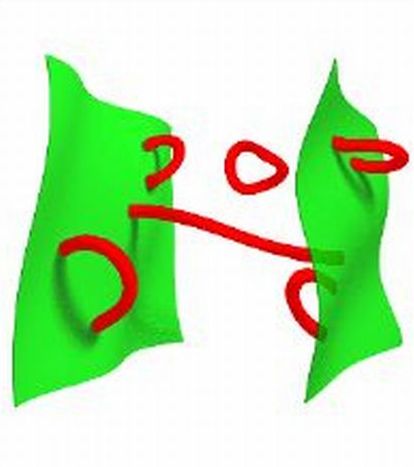}
\caption{A simple illustration of closed strings (loops) and open strings which have endpoints on the D-branes.}
\label{figdbrane}
\end{center}
\end{figure}

To adequately explain the system leading to the correspondence, we must understand a few essential ideas of string theory. The fundamental objects of string theory are called $p$-branes, where $p$ specifies the dimensionality of the brane. Two important excited 1-branes are the closed and open strings. As the name suggests, closed strings form a loop and propagate freely through space, whereas open strings have endpoints that must terminate on a objects called Dirichlet $p$-branes, or just D$p$-branes. A simple illustration of these objects is shown in Figure \ref{figdbrane}. Open and closed strings are characterized by their lengths $l_{s}$ and a coupling constant $g_{s}$ that determines the strength of their interactions. The string length is often expressed in terms of the Regge slope parameter $\alpha'$ \cite{Zwiebach}, 
\begin{equation}\label{equalphals}
\alpha' = l_{s}^{2}.
\end{equation}

 Given a stack of $N$ parallel D3-branes in a type IIB string theory, two low-energy descriptions arise depending on the permutation of the low-energy and strong-coupling limits. Taking the low-energy limit of the open string excitations first, we see that a SU($N$) gauge theory results, specifically the $\mathcal{N}=4$ SYM theory with a gauge coupling $g_{YM}$. The relation between the gauge coupling and the string coupling becomes \cite{Strominger:1995cz, Witten:1995im}
\begin{equation}\label{equCouplingRelate}
g_{YM}^{2} = 4\pi g_{s}.
\end{equation}
Assuming that the string length is small, we neglect the interactions of the closed strings which exist away from the D-branes in the bulk, that is far from $r=\infty$. By taking the strong coupling limit next, we obtain a strongly coupled SYM theory. 

 On the other hand, taking the strong coupling limit first means the back-reaction from the coincident $N$ branes cannot be neglected and the resulting curvature of spacetime must be accounted for using general relativity. One finds black-hole solutions to the supergravity equations where the classical metric takes the form as in \cite{Strominger:1995cz, Witten:1995im, Horowitz:1991cd},
\begin{equation} \label{equSGmetric}
ds^{2} = \frac{1}{\sqrt{1+\frac{R^{4}}{r^{4}}}}\left( -dt^{2} + d\vec{x}^{2}\right) + \sqrt{1+\frac{R^{4}}{r^{4}}}(dr^{2}+r^{2}d\Omega_{5}),
\end{equation}
where the curvature radius $R$ is given by \cite{Witten:1995im, Horowitz:1991cd},
\begin{equation}\label{equBaseCorr}
R^{4} = 4 \pi g_{s}N \alpha'^{2}.
\end{equation}
Because of gravitational red-shifting, the low-energy limit corresponds to moving away from the D3-branes at the ultraviolet (UV) boundary located at $r=\infty$. Taking the horizon limit when $r\ll R$, the supergravity metric (\ref{equSGmetric}) takes the form,
\begin{equation} \label{equ10drmetric}
ds^{2} = \frac{r^{2}}{R^{2}} (-dt^{2}+d\vec{x}^{2}) + \frac{R^{2}}{r^{2}}(dr^{2} + r^{2} d\Omega_{5}),
\end{equation}
where $r$ is strictly positive.
We can separate the metric (\ref{equ10drmetric}) into $S^{5}$,
\begin{equation}
R^{2} d\Omega_{5},\nonumber
\end{equation} 
and  $AdS_{5}$,
\begin{equation}\label{equ5drmetric}
\frac{r^{2}}{R^{2}}(-dt^{2}+d\vec{x}^{2})+ \frac{R^{2}}{r^{2}}dr^{2},
\end{equation}
giving an $AdS_{5}\times S^{5}$ string theory. We use an alternate form of the $AdS_{5}$ metric where
\begin{equation}
z = \frac{R^{2}}{r}.
\end{equation}
This transformation gives a metric with a single warp factor,
\begin{equation}\label{equ5dzmetric}
ds^{2} = a(z)^{2}\left(-dt^{2} + d\vec{x}^{2} + dz^{2}\right),
\end{equation}
where $a(z)=R/z$, $z>0$, with a UV cut-off residing at an infinitesimal $z$-value, $z=R_{0}$. 

\begin{figure}[h!]
\begin{center}
\includegraphics[scale=0.43]{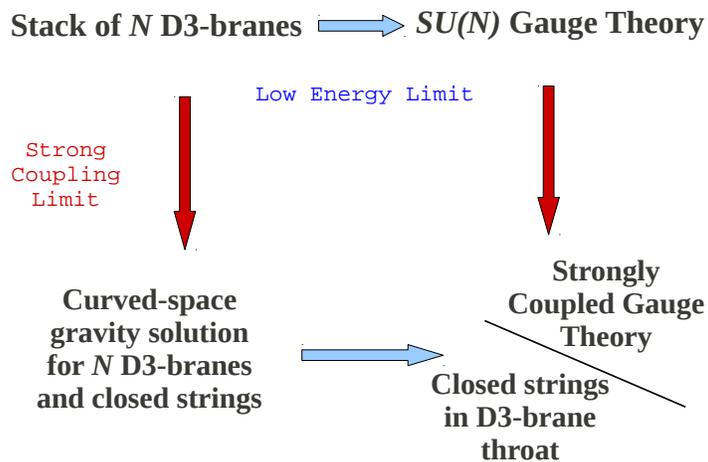}
\caption{The order in which the limits are taken cause the stack of D3-branes to either look like a strongly coupled gauge theory or a low-energy supergravity theory \cite{Myers:2008fv}.}
\label{figchart}
\end{center}
\end{figure}

 As shown in Figure \ref{figchart}, we began with a stack of D3-branes and end with two low-energy, strongly coupled systems. In AdS/CFT, classical gravity using the metric (\ref{equ10drmetric}) adequately describes the full ten-dimensional string theory as long as the radius of curvature is much larger than the string length,
\begin{equation}\label{equrls}
R\gg l_{s}.
\end{equation}
Using (\ref{equalphals}), (\ref{equCouplingRelate}), and (\ref{equBaseCorr}), we see that
\begin{equation}
\frac{R^{4}}{l_{s}^{4}} = g_{YM}^{2}N.
\end{equation}
Thus, (\ref{equrls}) requires that the quantity $g_{YM}N$, often referred to as the 't Hooft coupling $\lambda$, is large, 
\begin{equation}
g_{YM}^{2}N\equiv \lambda\gg 1.
\end{equation}
In other words, classical gravity is valid whenever the corresponding gauge theory is strongly coupled. The duality between string theory and the SYM theory should not need any corrections from stringy effects. We note here that $\lambda$ plays the role of the effective gauge coupling. The gauge coupling is usually tied to a scalar field in the string theory called the dilaton, $\Phi$, where
\begin{equation}\label{equdillambda}
\Phi = \log{\lambda}.
\end{equation} 

In order to work with this correspondence, we reduce the ten-dimensional metric (\ref{equ10drmetric}) to a five-dimensional AdS metric. Eliminating the $S^{5}$ manifold on the gravity side removes some symmetry on the gauge theory side. The R-charges associated with supersymmetry happen to have a $SO(6)$ symmetry, the same symmetry as the five-sphere. Removing this symmetry, we effectively reduce the SYM theory to a non-supersymmetric CFT. 
By disregarding the 5-sphere $S^{5}$, we reduce the ten-dimensional string theory to a five-dimensional classical gravity solution and the $\mathcal{N}=4$ SYM theory to a simpler four-dimensional, conformal Yang-Mills (YM) field theory, our AdS/CFT. 

\section{The Correspondence Dictionary}

While no rigorous mathematical proof establishing the AdS/CFT correspondence exists, it passes many nontrivial tests that have lead to the creation of a correspondence dictionary \cite{Gherghetta:2006ha}. In general, the dictionary allows one to draw connections between a strongly coupled gauge theory in $d$ dimensions and a gravity dual formulated in $AdS_{d+1}$. There are two methods for constructing the duality: top-down and bottom-up. Top-down models start with a critical or non-critical string theory and reduce to a lower-dimensional supergravity theory. A correspondence is then drawn to a strongly coupled gauge theory. Bottom-up construction begins with the precepts of a strongly coupled gauge theory and builds an effective gravity dual in AdS space. All the work presented in this thesis uses bottom-up construction.

Essentially, the correspondence equates the generating functional of gauge-invariant operators of the gauge theory with the minimum of the supergravity action, subject to boundary conditions of the supergravity fields coinciding with sources \cite{Gubser:1998bc,Erdmenger:2007cm}. In other words, 
\begin{equation}
\langle {\rm e}^{\int_{\partial AdS} d^{d}x\,\phi^{0}(\vec{x})\mathcal{O}(\vec{x)}}\rangle_{{\rm CFT}}  =  {\rm e}^{i S_{{\rm SUGRA}}(\phi)}|_{\phi=\phi_{0}}~,
\end{equation}
where $\phi^{0}$ is a source term residing on the boundary of the SUGRA model. The gauge theory resides on the boundary, $\partial AdS$, of the gravity dual. Thus, for every operator of conformal dimension $\Delta$ in the gauge theory, we have a coupling term, $\phi^{0}\mathcal{O}$. As a result, we insert a scalar field with mass $m$ into the $AdS_{d+1}$ gravity dual. Considering the Lagrangian of a free scalar field,
\begin{equation}
\mathcal{L} = g^{MN}\partial_{M}\phi\partial_{N}\phi - m^{2}\phi,
\end{equation}
the momentum space field equation becomes
\begin{equation}
z^{1+d}\partial_{z}\left(z^{1-d}\partial_{z}\phi\right) - \left(m^{2} + k^{2} z^{2}\right)\phi,
\end{equation}
where $\partial_{i}^{2}\phi = k^{2}\phi$. Near the $z=0$ boundary, the $k^{2}$ term is neglected, and the field behaves asymptotically as \cite{Maldacena:1997re, Witten:1998qj}
\begin{equation}\label{equstandardform}
\phi \approx \phi^{0}z^{\Delta_{-}} + \langle\mathcal{O}\rangle z^{\Delta_{+}},
\end{equation}
where the two roots are
\begin{equation}\label{equRoots}
\Delta_{\pm} = \frac{d}{2}\pm \sqrt{\frac{d^{2}}{4}+m^{2}R^{2}}.
\end{equation}
In most cases, we specify the operator dimension $\Delta$ and $d$ such that the mass is determined by
\begin{equation}\label{equSmass}
m^{2} = \Delta(\Delta-d),
\end{equation}
giving the two possible solutions of (\ref{equRoots}),
\begin{equation}\label{equDeltas}
\Delta_{+} = d,\quad\quad\quad\quad \Delta_{-} = d-\Delta.
\end{equation}
It is also possible, however, to couple an operator with a $p$-form $\mathcal{B}$,
\begin{equation}
\int_{M_d}{\mathcal{B}\wedge \mathcal{O}}.
\end{equation}
In this case, the scalar field in the gravity dual must take a mass of \cite{Witten:1998qj}
\begin{equation}\label{equSmasswpform}
m^{2} = (\Delta-p)(\Delta-p-d).
\end{equation}
At first glance, it seems that we are able to model a limited subset of gauge field operators, namely when $\Delta>p+d$, to avoid instabilities from tachyons. However, it has been shown that tachyons above a certain mass,
\begin{equation}
m^{2}>-\frac{d^{2}}{4},
\end{equation}
are allowed \cite{Breitenlohner:1982bm}.

In addition, the correspondence dictionary allows for symmetries of the gauge field to be mapped to the gravity dual. If the gauge theory possesses a global symmetry represented by a transformation $U = {\rm e}^{i \eta}$, then $\eta$ is a constant on the boundary of the gravity dual. In the bulk ($z\ne 0$), nothing prevents $\eta$ from becoming a function of spacetime coordinates, $x$ and $z$.  Therefore, the global symmetry in the gauge theory becomes a local symmetry in the gravity dual. In that case, one inserts a gauge field into the gravity dual for every relevant global symmetry.

\section{The AdS/QCD Correspondence}

Now that we have addressed the basis of the AdS/CFT correspondence, we will discuss how it relates to QCD. One immediately sees problems in connecting the CFT to QCD. By definition, the CFT has no intrinsic energy scale and no massive particles, whereas QCD has a scale set by $\Lambda_{{\rm QCD}}$ and a number of massive resonances. Similarly, a conformal theory has no running coupling, whereas QCD has a coupling which increases its strength as energy decreases, leading to confinement. Before we explain the means of breaking the conformal symmetry and incorporating confinement, we summarize some of the characteristics of QCD with a large number of colors $N_{c}$, the gauge theory that will correspond to the AdS gravity dual in Chapters \ref{chzero} and \ref{chthermo}.    

\subsection{Large-$N_{c}$ QCD}

The precise motivation for connecting QCD (or any gauge theory) comes from considering the 't Hooft large-$N_{c}$ limit \cite{tHooft:1973jz}. QCD possesses a SU(3) gauge symmetry, but no expansion parameter. By replacing the gauge group with SU($N_{c}$) and taking the $N_{c}\rightarrow\infty$ limit, 't Hooft generalized QCD at infinite colors and performed expansions out in powers of $1/N_{c}$. The large-$N_{c}$ limit is a natural extension of a low-energy string theory construction, as we will show \cite{Mateos:2007ay}.

\begin{figure}[h!]
\begin{center}
\includegraphics[scale=0.3]{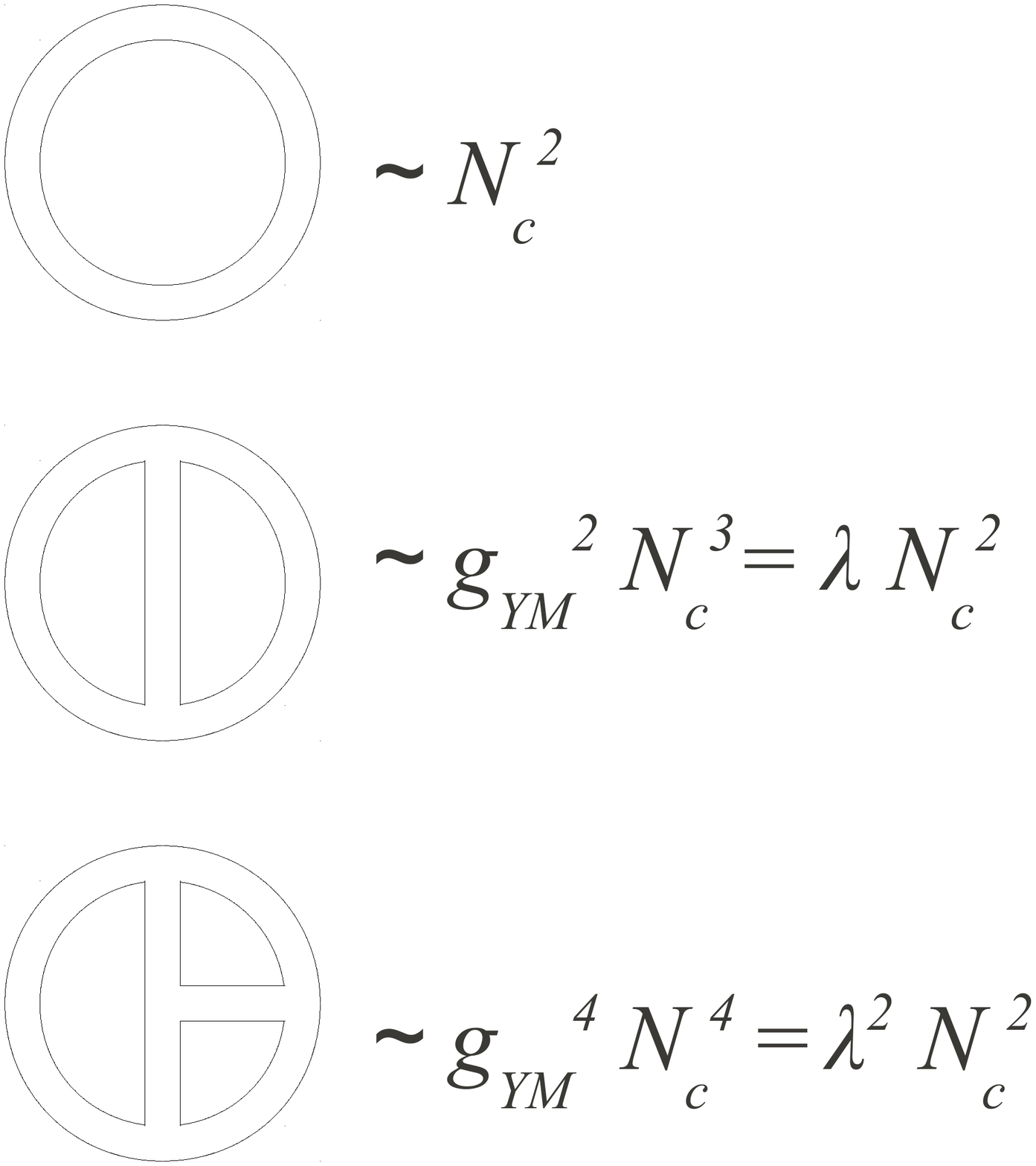}
\caption{Three examples of the most simple planar diagrams in double-line notation \cite{Mateos:2007ay}. Each free color index contributes a factor of $N_{c}$ and each vertex contributes a factor of $g_{YM}$.}
\label{figplanar}
\end{center}
\end{figure}

\begin{figure}[h!]
\begin{center}
\includegraphics[scale=0.2]{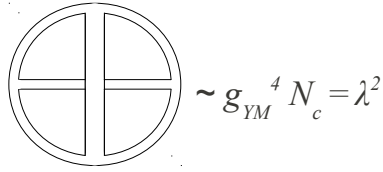}
\caption{The most simple non-planar diagram in double-line notation is shown \cite{Mateos:2007ay}. The amplitude of this diagram is suppressed by a factor of $N_{c}^{2}$ relative to the planar diagrams.}
\label{fignonplanar}
\end{center}
\end{figure}

In large-$N_{c}$ QCD, in which we assume $N_{c}\gg N_{f}$, gluons dominate the gauge theory. When drawing the relevant Feynman diagrams, it is best to consider them in double-line notation where one line represents one color charge so that gluons are represented by a double line. Each free color index then contributes a factor of $N_{c}$ and each vertex a factor of $g_{YM}$ to the amplitude. The diagrams naturally fall into two categories: planar and non-planar. Planar diagrams are those drawn without crossing lines \cite{Jenkins:2009wm}, as shown in Figure \ref{figplanar}. Each planar diagram scales as $N_{c}^{2}$ in a power series of $\lambda^{l-1}$, where $l$ is the number of loops in the diagram.
Non-planar diagrams, such as the one shown in Figure \ref{fignonplanar}, also expand in powers of $N_{c}$ and $\lambda$. However, the non-planar diagrams are suppressed by a factor of $N_{c}^{2}$ relative to the planar diagrams in Figure \ref{figplanar}.

The connection with string theory occurs when associating each Feynman diagram to a Riemann surface. In double-line notation, each line is a closed loop defining a boundary of a two-dimensional plane. We construct the Riemann surface by gluing these planes along their boundaries according to its Feynman diagram \cite{Mateos:2007ay}. We find that the amplitude, $\mathcal{A}$, of any given diagram is an expansion in terms of $N_{c}$ and $\lambda$ whose powers are fixed by the genus, $g$, of the Riemann surface \cite{Aharony:1999ti},
\begin{equation}\label{equlargeNA}
\mathcal{A} = \sum_{g=0}^{\infty}N_{c}^{2-2g}\,f_{g}(\lambda).
\end{equation}
All the planar diagrams correspond to Riemann surfaces with g=0, whereas non-planar diagrams must map to surfaces with $g>0$. According to (\ref{equlargeNA}), diagrams associated with surfaces of $g>0$ are suppressed by powers of $N_{c}^{2}$. 

To summarize, we see that each diagram in the gauge theory is associated with a Riemann surface. All non-planar diagrams are suppressed as $N_{c}\rightarrow\infty$. Thus, the planar limit of the gauge theory corresponds to the classical limit of the string theory where only Riemann surfaces of $g=0$ are considered. This means only a subset of Feynman diagrams are relevant for the large-$N_{c}$ limit of QCD \cite{Mateos:2007ay,Aharony:1999ti,Jenkins:2009wm}.

When considering the large-$N_{c}$ limit, some characteristics of QCD mesons are modified. The decay width of mesons scales as 1/$N_{c}$ so that as $N_{c}\rightarrow\infty$ they become infinitely narrow states. It is important to note, however, that for highly excited states where $n\sim N_{c}$, widths can become large. Also in large-$N_{c}$ QCD, we can express the two-point correlators as
\begin{equation}\label{equLargeNoperator}
\langle\mathcal{O}(k)\mathcal{O}(-k)\rangle = \sum_{n=1}^{\infty}\frac{F_{n}^{2}}{k^{2}+m_{n}^{2}},
\end{equation}
where the matrix element for $\mathcal{O}$ to create the $n$th meson of mass $m_{n}$ is $F_{n}=\langle 0|\mathcal{O}|n\rangle$. The sum in (\ref{equLargeNoperator}) must be infinite since the two-point function itself diverges logarithmically for large $k^{2}$ \cite{tHooft:1973jz, Witten:1979kh}. The infinite tower of masses created in the large-$N_{c}$ limit possesses similar qualitative features of the so-called Kaluza-Klein tower. As we show in Chapter \ref{chApplications}, Kaluza-Klein decomposition is an important method in analyzing the meson mass spectra.

\subsection{Conformal Symmetry Breaking and Confinement}

AdS space is conformally invariant; therefore, the string theory corresponds to a conformally invariant gauge theory with characteristics of large-$N_{c}$ QCD. For the gravity dual to describe a gauge theory with desired properties, we must break the conformal symmetry on the gauge side. Since pure AdS space on the string side ensures conformal symmetry on the gauge side, the simplest solution is to modify the AdS space in some fashion. It should be noted, however, that QCD becomes roughly conformal at high energies, so we should recover AdS space in the limit $z\rightarrow 0$. Imposing a hard wall or a soft wall are two methods of modifying AdS space on which we will concentrate. Breaking conformal symmetry in this manner also mimics confinement by restricting the effective range of the gauge theory.

\subsubsection{The Hard Wall}

The hard wall refers to the simplest means of breaking the conformal symmetry of the gauge theory. First investigated in \cite{Polchinski:2001tt, Polchinski:2002jw}, further work was performed in \cite{DaRold:2005vr, DaRold:2005zs, Erlich:2005qh, Csaki:2008dt, Csaki:2006ji}. In hard-wall AdS/QCD models, one imposes two branes at $z=R_{0}$ and $z=R_{1}$ called the UV and IR brane, respectively. The cutoff at $R_{1}$ serves two purposes: (i) introducing a scale,
\begin{equation}
R_{1} \approx \frac{1}{\Lambda_{{\rm QCD}}}, 
\end{equation}
and (ii) mimicking confinement.

As a phenomenological model, the hard wall was successful in describing a number of QCD properties, such as form factors, effective coupling constants, and correlation functions. However, it failed to accurately model meson mass spectra. Regge trajectories, the behavior of mass spectra scaling $m_{n}^{2}\sim n$, is well-established experimentally, but AdS/QCD models using the hard-wall geometry produced mass spectra exhibiting the scaling $m_n\sim n$.   

\subsubsection{The Soft Wall}  

Introduced as a refinement to the hard wall \cite{Karch:2006pv,Batell:2008me, Colangelo:2008us, Evans:2006ea, Evans:2006dj, Zuo:2009dz}, the soft-wall geometry imposes a $z$-dependent dilaton field to break conformal symmetry and mimic confinement. In string theory, standard decomposition of the closed string sector yields a one-particle state called the dilaton. When constructing a effective gravity dual in $AdS_{d}$ space, the dilaton enters in the action as
\begin{equation}
S = \int d^{d}x\,{\rm e}^{-2\Phi(z)}\sqrt{-g}\mathcal{L}_{{\rm grav}} + \int d^{d}x\,{\rm e}^{-\Phi(z)} \sqrt{-g}\mathcal{L}_{{\rm matter}} +\ldots.
\end{equation}
The dilaton modifies the equations of motion; therefore, it modifies meson mass spectra, form factors, correlation functions, and any other property derived from the action. 

In most works, the dilaton usually has a power solution,
\begin{equation}\label{equgeneraldilaton}
\Phi(z) = (\mu z)^{\nu},
\end{equation}
where $\mu$ introduces a scale for the gauge theory. For simplicity, $\mu$ is often set as 
\begin{equation}
\mu = \Lambda_{{\rm QCD}}.
\end{equation}
Given (\ref{equgeneraldilaton}), the meson mass spectra behaves as \cite{Batell:2008me}
\begin{equation}
m_{n}^{2} \sim n^{2-\frac{2}{\nu}}~.
\end{equation}
If $\nu=2$, we recover Regge trajectories in the mass spectra. We comment on this case in Chapter \ref{chApplications}.

\subsection{Running of the Coupling}

At high energies, QCD is weakly-coupled and has a coupling with a logarithmic running,
\begin{equation}\label{equalpha}
\alpha_{s}(Q) =  \frac{\alpha_{s}(q)}{1+\frac{\beta}{4\pi}\alpha_{s}(q)\log{\frac{Q^{2}}{q^{2}}}},
\end{equation}
where $\alpha_{s}$ is proportional to the square of the coupling constant g, and \cite{Zee:qft} 
\begin{equation}
\beta = -\frac{11}{3}N_{c}+\frac{2}{3}N_{f}.
\end{equation}
The $\beta$ function becomes much more complicated at low energies and cannot be written in closed form. Experimental evidence, however, has shown that the QCD coupling increases at low energies.

While conformal symmetry is broken in either the soft- or hard-wall model, only the soft-wall model allows for the coupling constant to run. In string theory, the string coupling often is not constant \cite{Zwiebach}. The coupling is determined by the dilaton field $\Phi$,
\begin{equation}
g_{s}N_{c} \sim {\rm e}^{\Phi(z)},
\end{equation}
making the 't Hooft coupling $\lambda$,
\begin{equation}\label{equrundefine}
\lambda = {\rm e}^{\Phi(z)}.
\end{equation}  
By defining the coupling in this way, we incorporate a running coupling, an advantage of the soft-wall model. In general, we could specify (\ref{equrundefine}) to mimic the running seen in QCD. 

\section{Conclusion}

In this chapter, we outlined the precepts of the AdS/CFT correspondence dictionary, and tied the gauge side of the duality to a large-$N_{c}$ QCD theory. We covered two key means of building a gravity dual model from a gauge theory: (i) gauge theory operators map to scalar fields in the gravity action and (ii) global symmetries in the gauge theory map to local symmetries in the gravity action.  In order to deform (break) the AdS space (conformal symmetry) and introduce a mass scale, we introduced methods of sharply cutting off the extra dimension with a IR brane at a point $z=R_{1}$ (the hard-wall model) and effectively cutting off $z$ using a dilaton background field (the soft-wall model). Running of the coupling only occurs in the soft-wall model.



\chapter{Applications of AdS/QCD}\label{chApplications}
\begin{flushright}
``History proves abundantly that pure science, undertaken without regard to applications to human needs, is usually ultimately of direct benefit to mankind.''
-Irving Langmuir
\end{flushright}

Now that the basics of AdS/CFT and its phenomenological cousin AdS/QCD have been reviewed, we detail some of the models that have been formulated in the last decade using the gauge/gravity duality. We comment on models addressing hadronic properties in Section \ref{secHadrons}. We then outline how to incorporate thermodynamics into the AdS/QCD picture in Section \ref{secThermodynamics}.   

\section{Hadronic Properties}\label{secHadrons}

Work addressing phenomenological models\footnote{In this context, phenomenological models refers to those that do not dynamically generate a background geometry, but assume one.}of hadronic properties are prolific \cite{Gherghetta:2009ac, Kelley:2010mu, Sui:2009xe, Kwee:2007nq, DaRold:2005vr, DaRold:2005zs, Erlich:2005qh, Evans:2006ea,  Csaki:2006ji, Karch:2006pv, Batell:2008me, Colangelo:2008us, Cherman:2008eh,  Evans:2006dj, dePaula:2008fp, Zuo:2009dz,  Jugeau:2009qa, Jugeau:2009aa, Huang:2007fv, DeFazio:2008mb, Grigoryan:2007my, Grigoryan:2008cc, Galow:2009kw, Babington:2003vm, Hong:2005np, McNees:2008km, BallonBayona:2007vp, Soda:2010si, Yakhshiev:2010bd}. The models that concern our work usually begin with a relatively simple action incorporating a limited number of QCD properties: (i) chiral symmetry, (ii) quark and gluon condensates, and (iii) a background dilaton field to model a running coupling.

The string-frame action for these models follows the standard AdS/QCD dictionary. Gauge fields $L_{M}$ and $R_{M}$ model the SU($N_{f}$)$_L\times$ SU($N_{f}$)$_R$ global chiral symmetry of QCD with $N_{f}$ flavors of quarks. We also include a bifundamental field $X^{ab}$ that is dual to the operator $q^{a}\bar{q^{b}}$ and takes on a $z$-dependent vacuum expectation value (VEV) $v(z)$ to break the chiral symmetry. Since $X^{ab}$ is dual to a dimension-3 operator, its mass becomes
\begin{equation}
m_{X}^{2}R^{2} \equiv \Delta(\Delta-d) = 3(3-4)= -3.
\end{equation}
The field $X$ actually becomes a complex field to incorporate the scalar $S$ and the pseudoscalar $P$ fields,
\begin{equation}\label{equX}
X^{ab} = \left(\frac{v(z)}{2}\emph{1}^{ab}+S^{a}t^{b}\right){\rm e}^{iP^{a}t^{b}}. 
\end{equation}
We only will write the indices $a,b$ of the field $X$ when needed. The five-dimensional action takes the form,
\begin{equation} \label{action0}
S_{5}=-\int d^{5}x \sqrt{-g}\,{\rm e}^{-\phi(z)}{\rm Tr}\left[|D X|^{2}+ m_{X}^{2} |X|^{2}+\frac{1}{4 g_{5}^{2}}(F_{L}^{2}+F_{R}^{2})\right],
\end{equation}
where $g_5^2$ is the five-dimensional coupling constant rescaling the field strength tensors $F_{L}$ and $F_{R}$, which are defined as
\begin{eqnarray}
F_{L}^{MN}&=&\partial^{M}{L^{N}}-\partial^{N}{L^{M}}-i[L^{M},L^{N}],\nonumber\\
F_{R}^{MN}&=&\partial^{M}{R^{N}}-\partial^{N}{R^{M}}-i[R^{M},R^{N}],\label{equstrtensors}
\end{eqnarray} 
where $L^{M}= L^{Ma} t^a$ and Tr$[t^{a}t^{b}]=\delta^{ab}/2$. The covariant derivative becomes 
\begin{equation}
D^M X=\partial^M X-i L^M X+iX R^M.
\end{equation}
To describe the vector and axial-vector fields, we simply transform the $L$ and $R$ gauge fields into the vector ($V$) and axial-vector ($A$) fields where 
\begin{eqnarray}
L^{M}&=&V^{M}+A^{M},\label{equLplusRV}\\
R^{M}&=&V^{M}-A^{M}.\label{equLminusRA}
\end{eqnarray}
Substituting equations (\ref{equLplusRV}) and (\ref{equLminusRA}) into the last term of the action (\ref{action0}),
\begin{eqnarray}
F_{L}^{2}+F_{R}^{2} &=& 2\left(\partial^{M}L^{N}\partial_{M}L_{N} - \partial^{M}L^{N}\partial_{N}L_{M}+\frac{1}{2}[L^{M},L^{N}][L_{M},L_{N}]\right)\nonumber\\
&+& 2\left(\partial^{M}R^{N}\partial_{M}R_{N} - \partial^{M}R^{N}\partial_{N}R_{M}+\frac{1}{2}[R^{M},R^{N}][R_{M},R_{N}]\right),\nonumber\\
&=& 4\left(\partial^{M}V^{N}\partial_{M}V_{N} - \partial^{M}V^{N}\partial_{N}V_{M}+\frac{1}{2}[V^{M},V^{N}][V_{M},V_{N}]\right)\nonumber\\
&+& 4\left(\partial^{M}A^{N}\partial_{M}V_{N} - \partial^{M}A^{N}\partial_{N}A_{M}+\frac{1}{2}[A^{M},A^{N}][A_{M},A_{N}]\right),\nonumber\\
&=& 2 \left(F_{V}^{2}+  F_{A}^{2}\right),\label{equchiraltovector}
\end{eqnarray}
where now
\begin{eqnarray}
F_{V}^{MN}&=&\partial^{M}{V^{N}}-\partial^{N}{V^{M}}-\frac{i}{\sqrt{2}}[V^{M},V^{N}],\\
F_{A}^{MN}&=&\partial^{M}{A^{N}}-\partial^{N}{A^{M}}-\frac{i}{\sqrt{2}}[A^{M},A^{N}].
\end{eqnarray}
Equation (\ref{equchiraltovector}) allows us to rewrite the action (\ref{action0}) in terms of the vector and axial-vector fields,
\begin{equation}\label{action0simple}
S_{5}=-\int d^{5}x \sqrt{-g}\,{\rm e}^{-\phi(z)}{\rm Tr}\left[|D X|^{2}+ m_{X}^{2} |X|^{2}+\frac{1}{2g_{5}^{2}}(F_{V}^{2}+F_{A}^{2})\right],
\end{equation}
where the covariant derivative now becomes 
\begin{equation}
D^M X=\partial^M X-i[V^{M},X]-i\{A^{M},X\}.
\end{equation}

The 5D coupling constant is inserted in order to match the vector two-point function $\Pi_{V}(q^{2})$ from the operator product expansion of QCD \cite{Cherman:2008eh},
\begin{equation}
\int{d^{4}x \langle J_{\mu}^{a}(x)J_{\nu}^{b}(0)\rangle} = \delta^{ab}(k_{\mu}k_{\nu} - q^{2}g_{\mu\nu})\Pi_{V}(k^{2}),
\end{equation}
 where $k$ is the 4D momentum and $J^{a}$ is the vector current. As shown in \cite{Erlich:2005qh, Kim:2008ff}, the two-point function near the boundary, which corresponds to large scales in the field theory, becomes
\begin{equation}
\Pi_{V}(k^{2}) = -\frac{1}{2 g_{5}^{2}}\log{k^{2}}.
\end{equation}
Matching to the large-$N_{c}$ QCD perturbative result,
\begin{equation}
\Pi_{V}(k^{2}) = -\frac{N_{c}}{24\pi^{2}}\log{k^{2}},
\end{equation} 
we find that 
\begin{equation}\label{equg5}
g_{5}^{2} = \frac{12 \pi^{2}}{N_{c}}.
\end{equation}

To obtain linear mass trajectories from this action, we must introduce a background dilaton field $\phi$ with the asymptotic behavior
\begin{equation} \label{dilatonlz}
\phi(z\rightarrow \infty) \simeq \mu^{2} z^2,
\end{equation}
where $\mu$ sets the mass scale for the meson spectrum. The $z$-dependent dilaton field also ensures broken conformal symmetry and a running coupling. In Section \ref{secVEV}, we show that if chiral symmetry remains broken as excitation number increases, indicated by the behavior of $v(z)$, then the quadratic behavior of the dilaton arises naturally.

\subsection{Equations of Motion}

By varying the action (\ref{action0simple}), we find the equations of motion for the scalar VEV, scalar field, pseudoscalar field, vector field, and the axial-vector field. We only consider the tree-level terms since we are using a weakly coupled gravity dual. In every sector, we derive the mass spectrum in the hard-wall model and the soft-wall model. In the hard-wall model, it is the customary to find mass eigenvalues from the poles of the two-point correlator \cite{DaRold:2005vr, DaRold:2005zs}. In AdS/QCD, the two-point correlator is found by taking two functional derivatives of the action with respect to a source field, $\psi^{0}$, at the boundary,
\begin{equation}
\frac{\delta S}{\delta \psi^{0}\delta \psi^{0} }\Big|_{R_{0}} = \langle \mathcal{O}\mathcal{O}\rangle.
\end{equation}
In the soft wall, Kaluza-Klein (KK) decomposition is the preferred method for finding mass eigenvalues. In KK decomposition, we break the field of interest, $\psi$ for instance, into an infinite tower of 4D components and purely $z$-dependent parts,
\begin{equation}
\psi(x,z) = \sum_{n=0}^{\infty}\Psi_{n}(t,\vec{x})\psi_{n}(z).
\end{equation}
Then using Proca's equation \cite{Ryder:qft},
\begin{equation}\label{equProca}
\partial_{i}\partial^{i}\Psi_{n} -m_{n}^{2}\Psi_{n} = 0,
\end{equation}
 we formulate a mass eigenvalue equation from the equation of motion.

\subsubsection{Scalar VEV}
Since the scalar VEV only depends on the bulk coordinate $z$, it does not require KK decomposition. We merely use (\ref{equX}) and vary the Lagrangian with respect to $v(z)$,
\begin{eqnarray}
\delta \mathcal{L}_{VEV} &=& -\delta\left({\rm e}^{-\phi}\sqrt{-g}\,{\rm Tr}\left[\frac{g^{zz}}{4}\partial_{z}v\partial_{z}v + m_{X}^{2}\frac{v^{2}}{4}\right]\right)\nonumber \\
&=& -\delta\left(\frac{1}{2}{\rm e}^{-\phi}a(z)^{5}\left(a(z)^{-2}\partial_{z}v\partial_{z}v +m_{X}^{2} v^{2} \right)\right)\nonumber\\
&=&  -{\rm e}^{-\phi}a^{3}\partial_{z}v\delta\partial_{z} v - {\rm e}^{-\phi} a^{5} m_{X}^{2} v \delta v \nonumber\\
&=& \left(\partial_{z}\left({\rm e}^{-\phi}a^{3}\partial_{z}v\right) - {\rm e}^{-\phi}a^{5}m_{X}^{2} v\right) \delta v,
\end{eqnarray}
finding the equation of motion to be
\begin{equation}
\partial_{z}^{2}v -\partial_{z}\phi\partial_{z}v + \frac{\partial_{z}a}{a}\partial_{z}v - a^{2}m_{X}^{2}v = 0.
\end{equation}
For the AdS metric and $m_{X}^{2}R^{2} = -3$, we see that the equation of motion reduces to
\begin{equation}\label{equVEVfullsimple}
v'' - \phi'v'-\frac{3}{z}v'+\frac{3}{z^{2}}v=0,
\end{equation}
where ($'$) denotes a derivative with respect to $z$.

 In the hard wall model where $\phi'=0$, we find that the exact solution to (\ref{equVEVfullsimple}) is
\begin{equation}
v_{hw}(z) = c_{1} z + c_{2} z^{3},
\end{equation}
where $c_1$ and $c_2$ are integration constants. Comparing this solution to the expected behavior (\ref{equstandardform}), $c_{1}$  and $c_{2}$ correspond to the source term and the operator expectation value, respectively,
\begin{eqnarray}
c_{1} &\sim& m_{q}, \\
c_{2} &\sim& \langle q\bar{q}\rangle \equiv \sigma.
\end{eqnarray}
However, \cite{DaRold:2005vr, DaRold:2005zs} argue for an alternative method. Placing a UV (IR) brane at $z=R_{0}$ ($z=R_{1}$), they specify the hard wall in the range $R_{0}<z<R_{1}$, where
\begin{equation}
m_{q} = \frac{R}{R_{0}}v_{hw}(R_{0}),\quad\quad\quad \sigma = R \,v_{hw}(R_{1}),
\end{equation}
making
\begin{eqnarray}
c_{1} &=& \frac{m_{q}R_{1}^{3} - \sigma R_{0}^{2}}{R\,R_{1}\left(R_{1}^{2} - R_{0}^{2}\right)},\\
c_{2} &=&  \frac{\sigma - m_{q}R_{1}}{R\,R_{1}\left(R_{1}^{2} - R_{0}^{2}\right)}.
\end{eqnarray}

 In the standard soft-wall model where $\phi=(\mu z)^{2}$, the solution to (\ref{equVEVfullsimple}), is given by~\cite{Karch:2006pv, Colangelo:2008us}
\begin{equation} \label{equvevswsol}
v_{sw}(z)= \frac{m_q}{R} z \,\Gamma\left(\frac{3}{2}\right) U\left(\frac{1}{2},0,\mu^{2} z^{2}\right),
\end{equation}
where $U(a,b,y)$ is the Tricomi confluent hypergeometric function. One possible solution is disregarded since the action must be finite in the IR. In the small-$z$ limit, (\ref{equvevswsol}) expands to \cite{Colangelo:2008us}
\begin{equation}\label{equvevswsolexp}
v_{sw}(z)\rightarrow \frac{m_{q}}{R}z - \frac{\mu^{2}m_{q}}{2R}z^{3}\left(1-2\gamma_{E} - 2 \log(\mu z) - \psi\left(\frac{3}{2}\right)\right),
\end{equation}
 where $\psi$ is the Euler function. However, comparing (\ref{equvevswsolexp}) to the standard form (\ref{equstandardform}), one sees that the chiral condensate is proportional to the quark mass, in contradiction to QCD. This proportionality stems from the fact that only one solution survives the imposed constraints. Furthermore, the solution (\ref{equvevswsol}) has an asymptotic limit $v(z)\rightarrow m_{q}/\mu$ for large values of $z$, which influence the large-$n$ excitations. Since $v$ is responsible for the mass splittings between the $\rho$'s and $a_{1}$'s, the meson masses become degenerate in the chiral limit, $m_{q}\rightarrow 0$. this does not occur in QCD. The problem of coupling between explicit chiral symmetry and spontaneous chiral symmetry breaking is addressed in Chapter \ref{chzero}.

A particular normalization scheme for the scalar VEV is implemented in the small-$z$ limit \cite{Cherman:2008eh}, which we use in Chapter \ref{chzero}. The normalization scheme arises from the fact that when coupling an operator $\mathcal{O}$ to a source $\phi^{0}$ one always has the freedom to redefine $\mathcal{O}\rightarrow \zeta \mathcal{O}$ and $\phi^{0}\rightarrow \phi^{0}/\zeta$, keeping the coupling term $\mathcal{O}\phi_{0}^{\mathcal{O}}$ unchanged \cite{Cherman:2008eh}. In the case of $v(z)$, we define a new parametrization,
\begin{equation}\label{equnormalv}
v(z) = \frac{\zeta m_{q}}{2} z + \frac{\sigma}{2\zeta}z^{3}.
\end{equation}
We fix $\zeta$ by matching the two-point correlation function of the scalar field between the QCD result at large Euclidean momentum, $k^{2}$, and that calculated from the gravity dual. In QCD, the two-point correlation function is
\begin{equation}
\int d^{4}x {\rm e}^{ikx}\langle q\bar{q}(x) q\bar{q}(0)\rangle = \frac{N_{c}}{8\pi^{2}}k^{2} \log{k^{2}} + \ldots,
\end{equation}
whereas the dual model yields
\begin{equation} \label{equqqqqcorr}
\int d^{4}x {\rm e}^{ikx}\langle q\bar{q}(x) q\bar{q}(0)\rangle = \frac{\zeta^{2}}{2}k^{2} \log{k^{2}} + \ldots.
\end{equation}
This suggests that the normalization in (\ref{equnormalv}) is
\begin{equation}\label{equzeta}
\zeta = \frac{\sqrt{N_{c}}}{2\pi}~;
\end{equation}
we assume this relation in Chapter \ref{chzero}.

\subsubsection{Scalar Sector}
Mass eigenvalues arising from the scalar sector represent the $f_{0}$ mesons.
The scalar field $S$ has a similar equation of motion to $v(z)$. The variation of the Lagrangian in terms of $S$ gives
\begin{eqnarray}
\delta \mathcal{L}_{S} &=& \delta\left({\rm e}^{-\phi}\sqrt{-g} \,{\rm Tr}\left[g^{MN}\partial_{M}S(x,z)\partial_{N}S(x,z) + m_{X}^{2}S(x,z)^{2}\right]\right)\nonumber\\
&=& \delta\left({\rm e}^{-\phi}\sqrt{-g}\,{\rm Tr}(t^{a}t^{b})\left(g^{\mu\nu}\partial_{\mu}S\partial_{\nu}S + g^{zz}\partial_{z}S\partial_{z}S + m_{X}^{2} S^{2}\right) \right)\nonumber\\
&=& {\rm e}^{-\phi}a^{3}\eta^{\mu\nu}\partial_{\mu}S\, \delta \partial_{\nu}S + {\rm e}^{-\phi}a^{3}\partial_{z}S \,\delta \partial_{z}S + {\rm e}^{-\phi} a^{5} m_{X}^{2}S\, \delta S\nonumber\\
&=& \left(-{\rm e}^{-\phi} a^{3} \partial_{\mu}\partial^{\mu}S - \partial_{z}\left({\rm e}^{-\phi}a^{3}\partial_{z}S\right) + {\rm e}^{-\phi}a^{5}m_{X}^{2} S\right)\delta S, \label{equlastS}
\end{eqnarray}  
leaving the equation of motion,
\begin{equation}\label{equScalargenEOM}
{\rm e}^{\phi}a^{-3}\partial_{z}\left({\rm e}^{-\phi}a^3\partial_{z}S\right) + \partial_{\mu}\partial^{\mu}S - a^{2}m_{X}^{2}S = 0.
\end{equation}

In the hard-wall model, we need to find the poles of the scalar two-point function $\Pi_{S}$. Taking two functional derivatives of the effective action with respect to the scalar source field $S^{0}$ gives the two-point function $\Pi_{S}(k)$,
\begin{equation}
\Pi_{S} = \frac{\delta\mathcal{S}_{eff}}{\delta S^{0}\delta S^{0}}|_{R_{0}},
\end{equation}
where in this case,
\begin{equation}
\mathcal{S}_{eff} = \int{d^{4}x \sqrt{-g}g^{zz} S\partial_{z}S}.
\end{equation}
After a Fourier transform $S(k,z) = \int d^{4}x {\rm e}^{ikx}\,S(x,z)$, it is convenient to define \cite{Krikun:2008tf}
\begin{equation}
 S(k,R_{0})=S^{0}(k),
\end{equation}
making (\ref{equScalargenEOM}),
\begin{equation}\label{equScalarhwEOM}
a^{-3}\partial_{z}\left(a^3\partial_{z}S(k,z)\right) - k^{2}S(k,z) - a^{2}m_{X}^{2}S(k,z) = 0.
\end{equation}
We solve (\ref{equScalarhwEOM}) directly to find that
\begin{equation} \label{equgenShwsol}
S(k, z) = C_{1}z^{2} J_{1}(ik z) + C_{2} z^{2}Y_{1}(ik z), 
\end{equation}
where $J_{1}$ and $Y_{1}$ are Bessel functions of the first kind. In order to obtain a zero mode from (\ref{equgenShwsol}), we modify the Neumann boundary conditions to obtain a boundary mass term \cite{Gherghetta:2006ha},
\begin{equation}\label{equSmassboundary}
\left( \partial_{z} S - \frac{\xi}{R}S \right)\Big|_{z=R_{0},R_{1}}=0,
\end{equation} 
where $\xi$ is a constant parametrizing the boundary mass.
It is convenient to rearrange the constants in (\ref{equgenShwsol}) such that 
\begin{equation}\label{equcoeffS}
S(z) = C_{S}\,z^{2}\left(J_{1}(ik z) + b_{S}(k) Y_{1}(ik z)\right).
\end{equation}
Applying the boundary conditions to (\ref{equcoeffS}), we find that
\begin{eqnarray}
b_{S}(k)&=& -\frac{ik R_{1} J_{0}(ik R_{1})+ (1-\frac{\xi}{R} R_{1})J_{1}(ik R_{1})}{ik R_{1} Y_{0}(ik R_{1})+ (1-\frac{\xi}{R} R_{1})Y_{1}(ik R_{1})},\\
C_{S}&=& \frac{\xi}{R}\frac{S^{0}}{R_{0}^{2}\left(J_{1}(ik R_{0}) + b_{S}(k) Y_{1}(ik R_{0})\right)}.
\end{eqnarray}
The effective action on the boundary then becomes
\begin{equation}\label{equeffS}
\mathcal{S}_{eff}\Big|_{R_{0}} = \int a^{3} S^{0} S^{0}\frac{\xi}{R R_{0}}\left[1+ ik R_{0} \frac{J_{0}(ik R_{0})+b_{S}(k)Y_{0}(ik R_{0})}{J_{1}(ik R_{0})+ b_{S}(k) Y_{1}(ik R_{0})}\right].
\end{equation}
Taking two functional derivatives of (\ref{equeffS}), we are left with the two-point function,
\begin{equation}
\Pi_{S} = \frac{R^{2}\xi}{R_{0}^{4}}\left(1+ik R_{0}\frac{J_{0}(ik R_{0})+b_{S}(k)Y_{0}(ik R_{0})}{J_{1}(ik R_{0})+ b_{S}(k) Y_{1}(ik R_{0})}\right).
\end{equation}

In the limit that $k R_{1}\gg 1$, the poles of $\Pi_{S}$ occur when
\begin{equation}
\frac{J_{0}(ik R_{1})}{Y_{0}(ik R_{1})}=0.
\end{equation}
The scalar mass   $m_{n}$, where $k^{2}= -m_{n}^{2}$,  are equal to the poles of the two-point correlator. We find an approximate mass tower of 
\begin{equation}
m_{n} \approx  \left(n+\frac{3}{4}\right)\frac{\pi}{R_{1}},\quad\quad n=0,1,2,\ldots,
\end{equation}
where $R_{1}$ is setting the mass scale in the hard-wall model. We see that indeed the hard-wall model fails to produce Regge trajectories.

In the soft-wall model, where again we take $\phi=\mu^{2}z^{2}$, we use a KK decomposition, 
\begin{equation}
S(x,z) = \sum_{n=0}^{\infty}S_{n}(z)\mathcal{S}_{n}(x),
\end{equation}
and Proca's equation (\ref{equProca}) to simplify (\ref{equlastS}),
\begin{eqnarray}
\left(-{\rm e}^{\phi}a^{-3}\partial_{z} \left({\rm e}^{-\phi}a^{3} \partial_{z} S_{n}(z)\right) - m_{n}^{2}S_{n}(z) - a^{2} m_{X}^{2}S_{n}(z)\right)\mathcal{S}_{n}(x)&=& 0, \nonumber\\
-{\rm e}^{\phi}a^{-3}\partial_{z} \left({\rm e}^{-\phi}a^{3} \partial_{z} S_{n}(z)\right) - (m_{n}^{2}+ a^{2} m_{X}^{2})S_{n}(z) &=& 0,\nonumber\\ 
-\partial_{z}^{2}S_{n} + \partial_{z}\phi\partial_{z}S_{n} - 3\frac{\partial_{z}a}{a}\partial_{z}S_{n} - a^{2}m_{X}^{2}S_{n} - m_{n}^{2}S_{n}&=& 0.\label{equlastS2}
\end{eqnarray}
In the AdS metric, the scalar mass eigenvalue equation becomes
\begin{equation}\label{equSfullsimple}
-S_{n}'' + \phi'S_{n}' + \frac{3}{z}S_{n}' - \frac{3}{z^{2}}S_{n} = m_{n}^{2} S_{n},
\end{equation}
where ($'$) denotes a derivative w.r.t to $z$. Following the procedure outlined in Appendix \ref{appSchTransform} and substituting
\begin{equation}
S_{n} = s_{n} {\rm e}^{\mu^{2} z^{2} + 3 \log{z}}
\end{equation}
into (\ref{equSfullsimple}), the equation of motion then simplifies to 
\begin{equation}
s_{n}'' + \left(\mu^{4}z^{2} + \frac{3}{4z^{2}} + 2\right)s_{n} = m_{n}^{2}s_{n}.
\end{equation}
Using the form outlined in Appendix \ref{appdiffeq}, we calculate the scalar eigenfunction and mass eigenvalues,
\begin{eqnarray}
s_{n}(z) &=& {\rm e}^{-\mu^{2}z^{2}}\sqrt{\frac{2z \,\varrho!}{(1+\varrho)!}}L_{\varrho}^{1}(\mu^{2}z^{2}),\quad\quad \varrho = \frac{m_{n}^{2}+2}{4\mu^{2}},\\
m_{n}^{2} &=& (4 n + 6)\mu^{2}, \quad\quad n=0,1,2,\ldots.
\end{eqnarray}
Hence, in the pure soft-wall model, the square of the scalar mass eigenvalues are linear in $n$. We shall compare this result to the results from our modified soft-wall model in Chapter \ref{chzero}.

\subsubsection{Pseudoscalar Sector}

The pseudoscalar mass eigenvalues correspond to the pions, the pseudo--Nambu-Goldstone bosons. Very few papers actually address the mass eigenvalues of the pseudoscalars \cite{DaRold:2005vr, DaRold:2005zs}. The pseudoscalar sector is the most difficult to investigate because of the many ways to define the pseudoscalar field in terms of the physical $\pi$ field and because the pseudoscalars couple to a component of the axial-vector field. The field $P(x,z)$ has been parametrized as $2 \pi(x,z)$ and as $\pi(x,z)/v(z)$. One can even abandon the exponential representation (\ref{equX}) and use the linear representation \cite{Kaplan:2009kr},
\begin{equation}\label{equlinearrep}
X^{ab} = \frac{v(z)}{2}\emph{1}^{ab}+ S^{a}(x,z)t^{b} + i \pi^{a}(x,z)t^{b}.
\end{equation} 
In addition, depending on the gauge, the pseudoscalar field couples to $A_{z}$ or the longitudinal component of $A_{\mu}$,
\begin{equation}
A_{\mu} =  A_{\mu\perp} + \partial_{\mu} \varphi,
\end{equation}
generating a system of equations that can only be solved by numerical techniques. The various issues surrounding the pseudoscalar sector are addressed in Section \ref{secPseudo}.

\subsubsection{Vector Sector}\label{secbkVector}
 
The vector field corresponds to the $\rho$ mesons. The vector gauge field, $V_{M}$ is introduced into the gravity dual action through the field tensors $F_{L,R}$ used to model the global chiral symmetry of QCD. Equivalently, the vector field is dual to the 1-form vector current $\bar{q}\gamma_{\mu}q$. According to (\ref{equSmasswpform}), the mass term $m_{V}^{2}$. In either case, the relevant terms from the action are
\begin{equation}
\mathcal{S}_{V} = -\int{d^{5}x {\rm e}^{-\phi}\sqrt{-g}\left(\partial_{M}V_{N} \partial^{M}V^{N} - \partial_{M}V_{N}\partial^{N}V^{M}\right)}.
\end{equation}
In this extra-dimensional model, we analyze the equations of motion in the so-called axial gauge while imposing a Lorentz gauge on the transverse components,
\begin{equation}
V_{z} =0, \quad\quad \partial_{\mu}V_{\perp}^{\mu}=0.
\end{equation} 
We drop the subscript $_{\perp}$ since we ignore the longitudinal part.
Varying the action, we find that 
\begin{eqnarray}
\delta \mathcal{S}_{V} &=& -\delta\left({\rm e}^{-\phi}\sqrt{-g}g^{\mu\rho}g^{\nu\sigma}\left(\partial_{\mu}V_{\nu}\partial_{\rho}V_{\sigma} - \partial_{\mu}V_{\nu}\partial_{\sigma}V_{\rho}\right)+{\rm e}^{-\phi}\sqrt{-g}g^{zz}g^{\mu\nu}\partial_{z}V_{\mu}\partial_{z}V_{\nu}\right) \nonumber\\
&=& -{\rm e}^{-\phi}\sqrt{-g}g^{\mu\rho}g^{\nu\sigma}\left(\partial_{\mu}V_{\nu}\delta\partial_{\rho}V_{\sigma} - \partial_{\mu}V_{\nu}\delta\partial_{\sigma}V_{\rho}\right) - {\rm e}^{-\phi}\sqrt{-g}g^{zz}g^{\mu\nu}\partial_{z}V_{\mu}\delta\partial_{z}V_{\nu}\nonumber\\
&=& {\rm e}^{-\phi}a(z) (\partial^{2}V_{\mu}\delta V^{\mu} - \partial_{\mu}\partial^{\nu}V_{\nu}\delta V^{\mu}) + \partial_{z}\left({\rm e}^{-\phi} a(z)\partial_{z} V_{\mu}\right)\delta V^{\mu}\nonumber\\
&=& \left({\rm e}^{-\phi} a(z) \partial_{\nu}\partial^{\nu}V_{\mu} + \partial_{z}\left({\rm e}^{-\phi} a(z)\partial_{z} V_{\mu}\right)\right)\delta V^{\mu},\label{equVectorstep}
\end{eqnarray}
leaving 
\begin{equation}\label{equVectorhwEOM}
{\rm e}^{\phi}a^{-1}\partial_{z}\left({\rm e}^{-\phi} a(z)\partial_{z} V_{\mu}\right)+ \partial_{\nu}\partial^{\nu}V_{\mu} =0.
\end{equation}

In the hard wall model, just as in the scalar sector, we find the two-point correlator $\Pi_{V}$, whose poles give the mass eigenvalues $m_{n}$. After a Fourier transform $V_{\mu}(k,z) = \int d^{4}x\,{\rm e}^{ikx}\,V_{\mu}(x,z)$, we again factor the vector field,
\begin{equation}\label{equVectorFourier}
V_{\mu}(k,z) = V_{\mu}^{0}(k)V(k,z), \quad\quad V(k,R_{0})=1. 
\end{equation}
Substitution (\ref{equVectorFourier}) into (\ref{equVectorhwEOM}), we see that
\begin{equation}\label{equvectorhwsimple}
a^{-1}\partial_{z} a(z)\partial_{z} V(k,z) -k^{2}V(k,z) =0,
\end{equation} 
whose general solution is expressed as
\begin{equation}
V(k,z) = C_{V}\, z\left(J_{1}(ik z) + b_{V}(k) Y_{1}(ik z)\right).
\end{equation}
Applying the boundary conditions,
\begin{equation}
\partial_{z}V(k,z)\Big|_{z=R_{0},R_{1}}=0,
\end{equation}
we find that 
\begin{eqnarray}
b_{V}(k) &=&  -\frac{J_{0}(ik R_{1})}{Y_{0}(ik R_{1})},\nonumber\\
C_{V} &=& \frac{V_{\mu}^{0}}{R_{0}\left( J_{1}(ik R_{0}) +b_{V}(k)Y_{1}(ik R_{0})\right)}.
\end{eqnarray}
The effective action in the vector sector is
\begin{eqnarray}
\mathcal{S}_{eff} &=& \int{d^{4}x\, a(z)\, V_{\mu}\partial_{z} V^{\mu}}|_{z=R_{0}} \nonumber\\
&=& \int d^{4}x \frac{ik R}{R_{0}} V_{\mu}^{0}V^{\mu,0}\frac{J_{0}(ik R_{0}) + b_{V}(k)Y_{0}(ik R_{0})}{ J_{1}(ik R_{0}) +b_{V}(k)Y_{1}(ik R_{0})},
\end{eqnarray}
making the two-point vector correlation function,
\begin{equation}\label{equ2pointVcorr}
\Pi_{V}(k) = \frac{ik R}{R_{0}}\frac{J_{0}(ik R_{0}) + b_{V}(k)Y_{0}(ik R_{0})}{J_{1}(ik R_{0}) +b_{V}(k)Y_{1}(ik R_{0})}.
\end{equation}
In the limit as $R_{0}\rightarrow 0$ and $k R_{1}\gg 1$, the poles of (\ref{equ2pointVcorr}) occur when
\begin{equation}
b_{V}(k)Y_{1}(ik R_{0}) = -\frac{J_{0}(ik R_{1})}{Y_{0}(ik R_{1})}Y_{1}(ik R_{0}) = 0.
\end{equation}
The condition looks similar to the one in the scalar sector, and we obtain a similar asymptotic behavior for vector mass excitations,
\begin{equation}
m_{n} \approx \left(n+\frac{3}{4}\right)\frac{\pi}{R_{1}},\quad\quad n=0,1,2,\ldots.
\end{equation}

In the soft-wall model, we again use a KK decomposition,
\begin{equation}
V_{\mu}(x,z) = \sum_{n=0}^{\infty}V_{n}(z)\mathcal{V}_{\mu,n}(x),
\end{equation} 
and Proca's equation (\ref{equProca}) to find that the equation of motion becomes
\begin{equation}
m_{n}^{2}V_{n} \mathcal{V}_{n} + {\rm e}^{\phi} a^{-1}\partial_{z}\left({\rm e}^{-\phi} a(z)\partial_{z} V_{\mu}\right) \mathcal{V}_{n} = 0.
\end{equation}
In the AdS metric, the equation of motion simplifies to 
\begin{equation} \label{equVector1}
V_{n}'' -\phi' V_{n} - \frac{1}{z} V_{n} + m_{n}^{2} V_{n} = 0,
\end{equation}
where ($'$) denotes a derivative with respect to $z$.
We transform (\ref{equVector1}) using the process described in Appendix \ref{appSchTransform}, where
\begin{equation}
V_{n} = v_{n} {\rm e}^{\phi(z)+\log{z}}.
\end{equation}
The equation of motion then becomes
\begin{equation}
-v_{n}'' + \left(\mu^{4} z^{2} + \frac{3}{4 z^{2}} \right)v_{n} = m_{n}^{2} v_{n}.
\end{equation}
Matching the form in Appendix \ref{appdiffeq}, we see that the eigenfunction and mass eigenvalues are 
\begin{eqnarray}
v_{n}(z) &=& {\rm e}^{-\mu^{2}z^{2}}\sqrt{\frac{2z\,\varrho!}{(1+\varrho)!}}L_{\varrho}^{1}(\mu^{2}z^{2}),\quad\quad \varrho = \frac{m_{n}^{2}}{4\mu^{2}},\\
m_{n}^{2} &=& 4(n+1)\mu^{2},\quad\quad n=0,1,2,\ldots,
\end{eqnarray}
where the vector sector shows the same linear trajectory when plotting $m_{n}^{2}$ vs $n$.

\subsubsection{Axial-vector Sector}\label{secbkAxial}

The axial-vector mass eigenvalues should match the mass values of the $a_{1}$ meson. The equation of motion for $A_{M}$ is obtained in the same manner as the vector field $V_{M}$, however, we see the symmetry between the vector and axial-vector is broken by the scalar VEV $v(z)$. Now we see why the function $v(z)$ contains the parameters for explicit and spontaneous chiral symmetry breaking: it couples to the axial-vector field, breaking the symmetry.

We again use the axial gauge, imposing a Lorentz condition on the transverse component,
\begin{equation}
A_{z} = 0, \quad\quad \partial_{\mu}A_{\perp}^{\mu}=0.
\end{equation} 
Following the same steps as the vector field analysis, we obtain the exact form in (\ref{equVectorstep}) with a symmetry breaking term $v^{2}A_{\mu}$,
\begin{equation}
\delta\mathcal{S}_{A} = \left({\rm e}^{-\phi} a(z) \partial^{2}A_{\mu} + \partial_{z}\left({\rm e}^{-\phi} a(z)\partial_{z} A_{\mu}\right)+ {\rm e}^{-\phi}g_{5}^{2}a(z)^{3}v^{2}A_{\mu}\right)\delta A^{\mu}. 
\end{equation} 
Using the vector sector as a guide, this reduces to a familiar equation of motion,
\begin{equation} \label{equAxial1}
{\rm e}^{\phi}a^{-1}\partial_{z}\left({\rm e}^{-\phi} a(z)\partial_{z} A_{\mu}\right)+ \partial_{\nu}\partial^{\nu}A_{\mu} - g_{5}^{2}a^{2}v^{2} A_{\mu} = 0,
\end{equation}
where the derivative ($'$) is with respect to $z$. As $v\rightarrow 0$, (\ref{equVector1}) and (\ref{equAxial1}) are essentially equivalent differential equations; therefore, the vector and axial-vector sector contain a degenerate tower of masses. The scalar VEV controls the mass splittings between the two meson mass eigenvalues.

In the hard-wall model, we again break the axial-vector field up into 
\begin{equation}
A_{\mu} = A_{\mu}^{0}(k) A(k,z), \quad\quad A(k,R_{0}) =1, 
\end{equation}  
making the equation of motion,
\begin{equation}\label{equAxialhw}
-A'' +\frac{1}{z}A' + k^{2}A + g_{5}^{2}a^{2}v^{2} A=0.
\end{equation}
The solution to (\ref{equAxialhw}) depends greatly upon the form of $v(z)$. We are concerned with the interesting cases of a constant or linear scalar VEV  \cite{Karch:2006pv}; therefore, we assume
\begin{eqnarray}
v_{1}(z) &=& \gamma, \label{equAvhw1}\\
v_{2}(z) &=& \Gamma z, \label{equAvhw2}
\end{eqnarray}
where $\gamma$ and $\Gamma$ are positive constants. Of course, (\ref{equAvhw1}) and (\ref{equAvhw2}) may only be the large-$z$ asymptotic behavior of the VEV, since we have stated that at small $z$, $v\sim m_{q} z + \sigma z^{3}$. The large-$z$ solution corresponds to the asymptotic behavior of high excitation modes $n$ since the large-$z$ limit of the effective potential affects the higher modes more than the lower modes. The equation of motion (\ref{equAxialhw}) becomes a  modified form of the vector equation (\ref{equvectorhwsimple}). Using $v_{1}$, the axial-vector solution becomes
\begin{equation}
A(k,z) = C_{A}z\left[J_{\alpha}(ik z) +b_{A}(k)Y_{\alpha}(ik z)\right],\quad\quad \alpha=\sqrt{1+g_{5}^{2}R^{2}\gamma^{2}}.
\end{equation}
This solution leads to an asymptotic mass spectrum,
\begin{equation}
m_{n} = \left(n+\frac{3}{4} + \sqrt{1+g_{5}^{2}R^{2}\gamma^{2}}\right)\frac{\pi^{2}}{R_{1}^{2}}.
\end{equation}
Using $v_{2}$, we find the solution becomes
\begin{equation}
A(k,z) = C_{A}z\left[J_{1}\left(z\,i\sqrt{k^{2} + g_{5}^{2}R^{2}\Gamma^{2}}\right) + b_{A}(k) Y_{1}\left(z\,i\sqrt{k^{2} + g_{5}^{2}R^{2}\Gamma^{2}}\right)\right],
\end{equation}
which gives 
\begin{equation}
m_{n}^{2} = \left(n+\frac{3}{4}\right)^{2}\frac{\pi^{2}}{R_{1}^{2}} + g_{5}^{2}R^{2}\Gamma^{2}.
\end{equation}
As expected, the axial-vector mass spectrum is shifted relative to the vector masses because of the CSB function $v(z)$. Also anticipated, the hard wall does not produce Regge trajectories in the axial-vector sector. A more general treatment of the first- and second-order corrections to $\Pi_{A}$ is done in \cite{Krikun:2008tf}.

In the soft wall, we proceed as in the vector sector. We introduce the KK decomposition,
\begin{equation}
A_{\mu}(x,z) = \sum_{n=0}^{\infty}a_{n}(z)\mathcal{A}_{\mu}(x).
\end{equation}
 Using the transform in Appendix \ref{appSchTransform}, (\ref{equAxial1}) becomes
\begin{equation}
-a_{n}'' +\left(\mu^{4}z^{2} + \frac{3}{4z^{2}} - \frac{v^{2}R^{2}}{z^{2}}\right)a_{n} = m_{n}^{2}a_{n}.
\end{equation}
To obtain a solution to $a_{n}$ and $m_{n}^{2}$, one must again specify the function $v(z)$. We consider two forms, (\ref{equAvhw1}) and (\ref{equAvhw2}). For the constant VEV (\ref{equAvhw1}), the eigenfunction and eigenvalues are
\begin{eqnarray}
a_{n}(z) &=& {\rm e}^{-\mu^{2}z^{2}}\sqrt{\frac{2z \,\varrho!}{(\sqrt{1+g_{5}^{2}\gamma^{2}R^{2}}+\varrho)!}}L_{\varrho}^{\sqrt{1+g_{5}^{2}\gamma^{2}R^{2}}}(\mu^{2}z^{2}), \label{equbkaxial1}\\
m_{n}^{2} &=& \left(4 n + 2\sqrt{1+g_{5}^{2}\gamma^{2} R^{2}} + 2\right)\mu^{2},\label{equmassaxial1}
\end{eqnarray} 
where 
\begin{equation}
\varrho = \frac{m_{n}^{2}}{4\mu^{2}}.
\end{equation}
Using a linear VEV (\ref{equAvhw2}), the eigenfunctions and eigenvalues become
\begin{eqnarray}
a_{n}(z) &=& {\rm e}^{-\mu^{2}z^{2}}\sqrt{\frac{2z\,(\varrho+\varpi)!}{(1+\varrho + \varpi)!}}L_{\varrho + \varpi}^{1}(\mu^{2}z^{2}),\\
m_{n}^{2} &=& \left(4 n + 4 + g_{5}^{2}\Gamma^{2}R^{2}\right)\mu^{2},
\end{eqnarray}
where
\begin{equation}
\varpi = \frac{g_{5}^{2}\Gamma^{2}R^{2}}{4\mu^{2}}.
\end{equation}
One sees that indeed the vector and axial-vector mesons are degenerate if $v\rightarrow 0$. In addition, their Regge trajectories slopes are parallel, precisely what experimental mass values of $\rho$'s and $a_{1}$'s suggest \cite{pdg}.

\section{Thermodynamics}\label{secThermodynamics}

Thermodynamic quantities become relevant in the AdS/QCD framework when addressing different physical phases of QCD. At a critical temperature $T_{c}$ hadronic matter melts into a strongly coupled plasma where quarks and gluons freely associate \cite{qgp, qgp3, phases}. We describe the thermodynamics of this quark-gluon plasma (QGP) by modifying the metric of the gravity dual model. The gravity dual metric (\ref{equ5dzmetric}) corresponds to the gauge theory below $T_{c}$. Above $T_{c}$, we must consider a new metric, 
\begin{equation}\label{equbhmetric}
ds^{2} = a(z)^{2} \left(-f(z) dt^{2} + d\vec{x}^{2} + \frac{dz^{2}}{f(z)}\right),
\end{equation}
with boundary conditions $f(0)=1$ and $f(z_h)=0$. The boundary conditions of $f$ allow for a useful parametrization that simplify expressions,
\begin{equation}\label{equfintheP}
f(z) =  1 - \frac{\mathcal{P}(z)}{\mathcal{P}(z_h)},
\end{equation}
where 
\begin{equation}
\mathcal{P}(z) = \int_{0}^{z}dx\, \frac{1}{a(x)^{3}}.
\end{equation}
The black-hole thermodynamics, outlined in \cite{Bekenstein:1973, Bekenstein:1974, Hawking:1974sw} then play the role of the QGP thermodynamics. The position of the black-hole horizon $z_h$ determines the temperature, and, ultimately, influences all the thermodynamic quantities of the gauge theory. 

We consider energy expressions to derive the black-hole thermodynamics. For a normal gas of temperature $T$, pressure $p$, entropy $S$, and volume $V$, the differential change of the internal energy is simply
\begin{equation}\label{equfirstLaw}
dE = T dS - p dV.
\end{equation}
Analogously, the energy of a black hole depends on the three quantities an observer can measure: charge $q$, horizon area $A$, and angular momentum $J$,
\begin{equation}\label{equBHfirstLaw}
dE =  \frac{\kappa}{8\pi G_{d}}dA + q\,d\varphi + \Omega\,dJ,
\end{equation}
where $G_{d}$ is the gravitational constant in $d$ dimensions, $\kappa$ is the surface gravity, $\varphi$ the electrostatic potential, and $\Omega$ the angular velocity \cite{Bekenstein:1973}. Comparing (\ref{equfirstLaw}) and (\ref{equBHfirstLaw}), we draw a correspondence between the quantities,
\begin{eqnarray}
T dS \rightarrow \frac{\kappa}{8\pi G_{d}}dA\label{equBHcorresp},\\
- p \,dV \rightarrow q \,d\varphi + \Omega \,dJ.\label{equUseless}
\end{eqnarray}
For the purposes of our work, we are only concerned with (\ref{equBHcorresp}), where conventions dictate that
\begin{eqnarray}
T = \frac{\kappa}{2},\\
S = \frac{A}{4 G_{d}}.
\end{eqnarray}
The area of the black-hole horizon is calculated from the induced metric $\gamma$,
\begin{equation}
A = \int{dV \sqrt{\gamma}},
\end{equation}
where $dV$ is the induced volume element. Using the metric (\ref{equbhmetric}), we find that the entropy density is given by
\begin{equation}
s \equiv \frac{S}{V} = \frac{a(z_h)^{p}}{4 G_{d}},
\end{equation}
where $p$ is the number of spatial dimensions.

We stated above that all thermodynamic quantities ultimately come from the metric (\ref{equbhmetric}) evaluated at the position of the horizon $z_h$. To see this, we rewrite the $t$- and $z$-components of the black-hole metric in an expanded form around $z_h$ as done in \cite{Zee:qft, Springer:thesis},
\begin{equation}
ds^{2} = -\gamma_{t}(z-z_{h})dt^{2} + \frac{\gamma_{z}}{z-z_{h}}d\rho^{2} +\ldots,
\end{equation}
and make the substitutions,
\begin{eqnarray}
\rho = 2\sqrt{\gamma_{z}(z-z_{h})},\\
t=i\tau,
\end{eqnarray}
making
\begin{eqnarray}
d\rho^{2} = \frac{\gamma_{z}}{z-z_{h}}dz^{2}.
\end{eqnarray}
In these coordinates, the metric then appears in polar form,
\begin{equation}
ds^{2} = \frac{\gamma_{t}}{4\gamma_{z}}\rho d\tau^{2} + d\rho^{2} + \ldots,
\end{equation}
with the variable $\tau$ and $\rho$ playing the role of the polar and radial coordinate, respectively. Examining an arc length, $\alpha$, in this coordinate system, we see that at constant $\rho=\rho_{0}$ and a period of $\beta$,
\begin{equation}
\alpha^{2} = \frac{\gamma_{t}}{\gamma_{z}}\rho^{2}\beta^{2}.
\end{equation} 
The arc length should be equal to the circumference $C = 2\pi\rho_{0}$ in order to ensure regularity at the origin,
\begin{eqnarray}
(2\pi\rho_{0})^{2} = \frac{\gamma_{t}}{\gamma_{z}}\rho^{2}\beta^{2},\\
\frac{1}{\beta} = \frac{1}{4 \pi}\sqrt{\frac{\gamma_{t}}{\gamma_{z}}}.
\end{eqnarray}
Following the conventions of finite temperature field theory, we equate the period in Euclidean time with the inverse of the temperature,
\begin{equation}
T=\frac{1}{\beta},
\end{equation}
giving us the Hawking temperature expression,
\begin{equation}
T=\frac{1}{4\pi}\sqrt{\frac{\gamma_{t}}{\gamma_{z}}}.
\end{equation}
Using the metric (\ref{equbhmetric}), the temperature can be expressed as 
\begin{equation}
T = -\frac{1}{4\pi}\frac{\partial f}{\partial z}\Big|_{z=z_h} = \frac{1}{4\pi\mathcal{P}(z_h)}\frac{\partial\mathcal{P}(z)}{\partial z}\Big|_{z=z_h},
\end{equation}
giving a direct relationship between $T$ and $z_h$.

Knowing the entropy and the temperature allows us to find other useful quantities that emerge in Chapter \ref{chthermo}, such as the free energy $\mathcal{F}$ and the speed of sound through the QGP-like medium $v_{s}$. From thermodynamics, the free energy is defined as
\begin{equation}\label{equfree}
\mathcal{F} = -\int{S\, dT}.
\end{equation} 
We can express (\ref{equfree}) strictly in terms of the metric,
\begin{equation}
\mathcal{F} = \frac{V}{4 G_{d}}\int{a(z_h)^{p} \frac{\partial}{\partial z_{h}}\left(\frac{1}{4\pi \,a(z_h)\,\mathcal{P}(z_h)}\right)}.
\end{equation} 
The speed of sound through the medium is also dependent upon $S$ and $T$,
\begin{eqnarray}
v_{s}^{2} &=& \frac{\partial P}{\partial \epsilon} = \frac{d P/d T}{d \epsilon/d T} = \frac{s}{T ds/dT}\nonumber\\
&=& \frac{d\log{T}}{d\log{s}},
\end{eqnarray}
where $P$ is the pressure and $\epsilon$ is the energy density. Equipped with these relations, we are able to calculate much about the deconfined phase of the gauge side of the gauge/gravity duality.

\subsection{Finite Temperature Action}\label{secbktempaction}
When considering the thermodynamics of AdS/QCD, we are interested in dynamically generating the geometry and the fields. Dynamic solutions are not easy to find in AdS/QCD models; most work focuses on phenomenological models to describe the thermodynamics of the corresponding gauge theory. The gravitational action needed to generate the metric and fields has an intrinsic coupling with the dilaton,
\begin{equation}
S = \int{d^{d}x\, {\rm e}^{-2\Phi}\sqrt{-g}\,\mathcal{L}_{{\rm grav}}} + \int{d^{d}x\, {\rm e}^{-\Phi} \sqrt{-g}\,\mathcal{L}_{{\rm matter}}}.
\end{equation} 
The action considered in this section is widely used \cite{ BallonBayona:2007vp, Kovtun:2005ev, Herzog:2006ra, Kajantie:2006hv, Son:2007vk, Evans:2008tu, Gursoy:2008za, Gursoy:2008bu, Gursoy:2009kk, Panero:2009tv, Alanen:2009na, Alanen:2010tg, Li:2011hp} and contains simple elements reduced from a critical string theory to five dimensions \cite{Son:2007vk},
\begin{equation}\label{equbkstringL}
\mathcal{S}_{grav} = -\frac{1}{16\pi G_{5}}\int{d^{5} x} {\rm e}^{-2 \Phi}\sqrt{-g} \left(R - \Lambda + 4 g^{MN}\partial_{M}\Phi\partial_{N}\Phi - V_{s}(\Phi)\right),
\end{equation}
where (\ref{equbkstringL}) is defined in the string frame.

We use the gauge/gravity duality to relate the 5D gravitational coupling $G_5$ to the 
number of colors $N_c$ in the 4D gauge theory. First, consider an $AdS_5\times S^5$ background. 
The ten-dimensional gravitational constant $G_{10} =8\pi^6 g_s^2 l_s^8$, where $g_s$ is the string 
coupling and $l_s$ is the string length~\cite{Son:2007vk}. After compacting to five dimensions
we obtain 
\begin{equation}
G_{10} = G_5 V_5 = G_5 \pi^3 R^5.
\label{equG5step}
\end{equation}
By the AdS/CFT correspondence, we also have the relation~\cite{Maldacena:1997re},
\begin{equation} \label{equStringtoField}
4\pi g_{s} N_{c} = \frac{R^{4}}{l_{s}^{4}}.
\end{equation}
Combining (\ref{equG5step}) and (\ref{equStringtoField}), we can then relate $G_{5}$ to the number of colors, 
$N_c$ 
\begin{equation}\label{equG5}
\frac{1}{16\pi G_{5}} =  \frac{N_{c}^{2}}{8\pi^{2} R^{3}}.
\end{equation}
This relation is specific to the $AdS_5\times S^5$ background, but the numerical factors can be different in the case of a pure AdS manifold. Analogous to (\ref{equG5}), we define in our five-dimensional case,
\begin{equation}
\frac{1}{16\pi G_5} =  M_5^3 N_c^2 = \upsilon_0 \frac{N_c^2}{R^3},
\end{equation}
where $M_5$ is the 5D Planck scale and $\upsilon_0$ is a constant. The constant $\upsilon_0$ is fixed
by matching the free energy to that obtained in the high temperature limit of QCD. The result 
is~\cite{Gursoy:2008za}
\begin{equation}\label{equpropconstant}
           \upsilon_0= (M_5 R)^3 = \frac{1}{45\pi^2}.
\end{equation}
We use (\ref{equpropconstant}) for our holographic model in Chapter \ref{chthermo}.

Since we are interested in the gravitational properties of this action, it is most useful to put (\ref{equbkstringL}) in the Einstein frame. Switching from the string to the Einstein frame occurs with a conformal transformation,
\begin{equation}
g_{MN}^{s} = {\rm e}^{\frac{4}{3}\Phi}g_{MN}^{E},
\end{equation}
giving
\begin{equation}\label{equActionE}
S_E =-\frac{1}{16 \pi G_{5}} \int d^5x \sqrt{-g}\left(R_{E}  -\frac{1}{2}g^{MN}_{E}\partial_M\phi\partial_N\phi - V_{E}(\phi)\right),
\end{equation}
where we make the substitutions,
\begin{eqnarray}
\phi = \sqrt{\frac{8}{3}}\Phi,\\
 V_{E} = V_{s}\,{\rm e}^{\frac{4}{3}\Phi}.
\end{eqnarray}
We drop the $E$ subscript from this point.
In general, the action (\ref{equActionE}) contains no mass term for the dilaton; therefore, according to the correspondence dictionary, the dilaton must be dual to a dimension-4 operator. The simplest dimension-4 operator is the gluonic operator, Tr$\left(F_{\mu\nu}F^{\mu\nu}\right)$.

In the Einstein frame, we are able to use the unaltered Einstein equations,
\begin{equation}\label{equEinsteinE}
G_{MN} = 8\pi G_{5}T_{MN}.
\end{equation}
The metric and all the fields of the system are generated from the system of three equations, two from (\ref{equEinsteinE}) and one from the direct variation of the action (\ref{equActionE}) with respect to the dilaton $\phi$. Using the Einstein metric,
\begin{equation}
a(z)^{2}\left(-f(z)dt^{2} + d\vec{x}^{2} + \frac{dz^{2}}{f(z)}\right),
\end{equation}
they become
\begin{eqnarray}
\frac{f''}{f'} + 3\frac{a'}{a} &=& 0, \label{equfindf} \\
12 \frac{a'^{2}}{a^{2}} - 6\frac{a''}{a} &=& \phi'^{2} , \label{equfinitefields} \\
6\frac{a'^{2}}{a^{2}} + 3\frac{a''}{a} + 3\frac{a'}{a}\frac{f'}{f} &=& -\frac{a^{2}}{f} V(z).\label{equfindpot}
\end{eqnarray}
The system becomes extremely hard to solve for all but the simplest cases. We examine two cases that have been investigated in \cite{Gursoy:2008za}: the conformal case and the power solution. 

\subsubsection{Conformal Case}
In the conformal case, we set the dilaton to zero. As a result, (\ref{equfinitefields}) gives pure AdS space,
\begin{equation}
a(z) = \frac{R}{z},
\end{equation} 
(\ref{equfindf}) is satisfied when
\begin{equation}
f(z) = 1- \frac{z^{4}}{z_h^{4}},
\end{equation}
and (\ref{equfindpot}) holds true for 
\begin{equation}
V=-\frac{12}{R^{2}}.
\end{equation} 
This describes a perfectly conformal gas. The temperature is tied to the position of the black-hole horizon,
\begin{equation}\label{equconformaltemp}
T = -\frac{f'(z_h)}{4\pi} = \frac{1}{\pi z_h}.
\end{equation}
Entropy scales with temperature as expected,
\begin{equation}
S = \frac{a(z_h)^{3}}{4 G_{5}} = \frac{4 N_{c}^{2}}{45 \pi z_{h}^{3}} = \frac{4\pi^{2}}{45}N_{c}^{2}T^{3}.\label{equconformalentropy}
\end{equation}
The velocity of sound through this thermal gas is characterized by
\begin{eqnarray}
v_{s}^{2} &=& \frac{d\log{T}}{d\log{s}}\nonumber\\
 &=& \frac{d}{d\log{s}}\left(\frac{1}{3}\log{\frac{45}{4\pi^{2}N_{c}^{2}}} + \frac{1}{3}\log{s}\right) =\frac{1}{3}.
\end{eqnarray}
The free energy is found from the entropy (\ref{equconformalentropy}),
\begin{equation}
\mathcal{F}=-\int{S dT} = -\frac{\pi^{2}}{45}N_{c}^{2}T^{4}.
\end{equation}
Again, we see that a pure AdS gravity dual produces properties of a perfect fluid.

\subsubsection{Dilaton Power Solution}
Because the system of equations (\ref{equfindf}), (\ref{equfinitefields}), and (\ref{equfindpot}) is difficult to handle, we define $\phi$ as a simple power solution,
\begin{equation} \label{equbkgeneralphi}
\phi = \sqrt\frac{8}{3} (\mu z)^{\nu}.
\end{equation}
With this dilaton, however, there is no obvious way to satisfy (\ref{equfindf}),(\ref{equfinitefields}), and (\ref{equfindpot}) with a temperature-independent expression for $V(z)$. A temperature-dependent $V$ invalidates any useful traditional thermodynamic relations. Much has been written on this point; some writings focus on extremely complicated forms for $\phi$ and $a(z)$ \cite{Gursoy:2008za, Gursoy:2008bu, Gubser:2008ny}, but a closed-form, temperature-independent solution has yet to be found. Currently, only phenomenological models work when assuming (\ref{equbkgeneralphi}). In Chapter \ref{chthermo}, we attempt a novel approach to generating a dynamic solution in the soft-wall model.

\begin{figure}[h!]
\begin{center}
\includegraphics[scale=0.40]{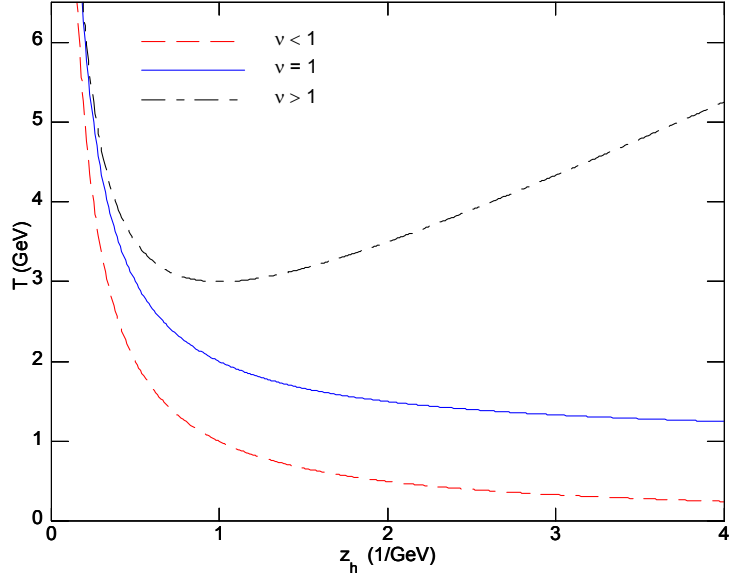}
\caption{Depending on the value $\nu$ takes, the temperature and the horizon position have 3 distinct relationships. The $\nu<1$ case produces conformal results. When $\nu=1$, we find that temperature approaches a constant as $z_h\rightarrow\infty$. When $\nu>1$, we find a stable and unstable black-hole solution.  }
\label{fig3temp}
\end{center}
\end{figure}

 Given (\ref{equbkgeneralphi}), we have
\begin{equation}
f(z) = \frac{\Gamma\left(\frac{4}{\nu},-\frac{3}{\sqrt{6}} (\mu z_{h})^{\nu}\right)-\Gamma\left(\frac{4}{\nu},-\frac{3}{\sqrt{6}} (\mu z)^{\nu}\right)}{\Gamma\left(\frac{4}{\nu},-\frac{3}{\sqrt{6}} (\mu z_{h})^{\nu}\right)-\Gamma\left(\frac{4}{\nu}\right)},
\end{equation}  
which gives a temperature of
\begin{equation}\label{equTgeneral}
T(z_{h}) = \left(\frac{9}{4}\right)^{\frac{1}{\nu}}\frac{\nu \mu^{4} z_{h}^{4} {\rm e}^{\frac{3}{\sqrt{6}}\mu^{\nu}z_{h}^{\nu}}}{4\pi z_{h}\left(1-\Gamma\left(\frac{4}{\nu},-\frac{3}{\sqrt{6}}\mu^{\nu}z_{h}^{\nu}\right)\right)}.
\end{equation}
We see that (\ref{equTgeneral}) has a minimum and has two distinct asymptotic limits,
\begin{eqnarray}
T(z_{h}\rightarrow 0) &=& \frac{1}{\pi z_{h}} \label{equT0},\\
T(z_{h}\rightarrow\infty) &=& \frac{\nu\mu^{\nu}}{4 \pi}z_{h}^{\nu-1}\label{equTinfty}.
\end{eqnarray}
Three distinct temperature-horizon relationships arise depending upon on the value of $\nu$ as shown in Figure \ref{fig3temp}. The $\nu<1$ case is effectively equivalent to the conformal case, which was already discussed. The $\nu=1$ case has the $T(z_h)$ approaching a constant as $z_h\rightarrow\infty$. Confining solutions exist for $\nu>1$ \cite{Gursoy:2008za}. They contain a minimum temperature below which no black hole exists. Above the minimum temperature a stable black hole and an unstable black hole can form depending upon the slope of the $T$ vs $z_h$ plot,
\begin{eqnarray}
\frac{dT}{d z_{h}}&<& 0\quad\quad \text{stable},\\
\frac{dT}{d z_{h}}&>& 0 \quad\quad \text{unstable}.
\end{eqnarray}

Stability issues become obvious as we examine the entropy, speed of sound, and the free energy for confining theories describing the QGP-like fluid. The Bekenstein entropy is
\begin{eqnarray} \label{equSzh}
S = \frac{4 N_{c}^{2}}{45\pi}\frac{{\rm e}^{-\frac{3}{\sqrt{6}}\mu^{\nu}z_{h}^{nu}}}{z_{h}^{3}}.
\end{eqnarray}
The entropy behaves quite differently depending on the region of $z_{h}$,
\begin{eqnarray}
S(z_{h}\rightarrow 0,T) &\approx&  \frac{4 \pi^{2} N_{c}^{2}}{45}\left(T^{3} - \frac{3\mu^{\nu}}{\sqrt{6}\pi^{\nu}}T^{3-\nu}\right)\label{equSsmall},\\
S(z_{h}\rightarrow \infty,T) &\approx& \frac{4 N_{c}^{2}}{45\pi} \frac{{\rm e}^{-\frac{3}{\sqrt{6}}\mu^{\nu}(r\,T)^{\frac{\nu}{\nu-1}}}}{(r\,T)^{\frac{3}{\nu-1}}}\label{equSlarge},
\end{eqnarray}
where $r=4\pi/\nu\mu^{\nu}$. As $z_{h}\rightarrow 0$, the entropy goes to the conformal behavior; however, the entropy of the unstable black hole (\ref{equSlarge}) exponentially decreases with rising temperatures.
From the temperature and entropy expression, the velocity of sound becomes
\begin{eqnarray}
v_{s}^{2} &=& \frac{S(z_h)T'(z_h)}{S'(z_h)T(z_h)} \nonumber\\
&=& \frac{2\sqrt[\nu]{\frac{9}{4}} \,\nu\mu^{4}z_{h}^{4} {{\rm e}^{\frac{3}{\sqrt{6}}\mu^{\nu}z_{h}^{\nu}}}}{(6 + 6\nu\mu^{\nu}z_{h}^{\nu})\left(1-\Gamma\left(\frac{4}{\nu},-\frac{3}{\sqrt{6}}\mu^{\nu}z_{h}^{\nu}\right)\right)} - \frac{6+\sqrt{6} \nu \mu^{\nu}z_{h}^{\nu}}{6+6 \nu \mu^{\nu}z_{h}^{\nu}}.
\end{eqnarray}
For the stable black hole, we see that $v_{s}$ reduces to 
\begin{eqnarray}
v_{s}^{2}(z_{h}\rightarrow 0) &\approx& \frac{1}{3} + \frac{(3-\nu)\sqrt{6} -2 \nu}{6}\mu^{\nu}z_{h}^{\nu}\\
&\approx& \frac{1}{3} +\frac{(3-\nu)\sqrt{6} -2 \nu}{6}\left(\frac{\mu}{\pi T}\right)^{\nu}.
\end{eqnarray}
For the unstable black hole,
\begin{eqnarray}
v_{s}^{2}(z_{h}\rightarrow\infty) &\approx&  \frac{1}{3 (\mu z_{h})^{\nu-1}}-\frac{1}{\sqrt{6}}\\
&\approx& \frac{4\pi T}{\nu\,\mu}-\frac{1}{\sqrt{6}},
\end{eqnarray}
the speed of sound grows without bound as the temperature increases, an unphysical result. 
From the Bekenstein entropy (\ref{equSsmall}) and (\ref{equSlarge}), we calculate the free energy as 
\begin{eqnarray}\label{equFenergyr}
\mathcal{F}(z_{h}\rightarrow 0,T) &=&  -\frac{\pi^{2}N_{c}^{2}}{45}T^{4} +\frac{12 \pi^{2-\nu}N_{c}^{2}\mu^{\nu}}{45\sqrt{6}}\frac{T^{4-\nu}}{4-\nu},\\
\mathcal{F}(z_{h}\rightarrow\infty,T) &=& -\frac{\pi^{2}N_{c}^{2}}{45}\frac{\nu-1}{\nu r}\Gamma\left(\frac{\nu-4}{\nu},(r T)^{\frac{\nu}{\nu-1}}\right).
\end{eqnarray}
The stable black hole again shows the conformal limit as $T$ increases, and the unstable black hole has vanishing free energy with increasing temperature.
It is easily seen that the thermodynamics reduce to the conformal limit as $z_{h}\rightarrow 0$ and $T\rightarrow \infty$, and the unstable black-hole results appear to be of little use for describing gauge-field thermodynamics.

\subsection{Phase Transition}

The analysis of the gravity dual actions yields information about the Hawking-Page transition \cite{Hawking:1982dh, Witten:1998zw}. When two solutions exist for a given action, this transition occurs as one solution becomes energetically favored over the other. According to the correspondence dictionary, the Hawking-Page transition marks confinement temperature, $T_{c}$, where the phase transition from hadronic matter to a QGP state occurs. To calculate the transition, one must consider the on-shell action using the metric (\ref{equ5dzmetric}) corresponding to a confined state below $T_{c}$ (thermal AdS), and then the metric (\ref{equbhmetric}) corresponding to the state above $T_{c}$ (black-hole AdS). A simple calculation of the phase transition in an AdS/QCD model was investigate in \cite{Herzog:2006ra}, using the free energy as defined in \cite{Gursoy:2008za},  
\begin{equation}\label{equfenergybk}
\beta\mathcal{F}= \lim_{R_{0}\rightarrow 0}\left[\mathcal{S}_{bh}(R_{0}) - \mathcal{S}_{th}(R_{0})\right],
\end{equation}
where again $T=1/\beta$.
The transition occurs when $\mathcal{F}=0$. We review the confinement temperature calculation done in a hard- and soft-wall phenomenological model \cite{Herzog:2006ra}.

In the hard-wall model, we analyze the action
\begin{equation}
\mathcal{S}_{hw} = -\frac{1}{16\pi G_{5}}\int{d^{5}x \sqrt{-g}\left(\mathcal{R}+\frac{12}{R^{2}}\right)}, 
\end{equation}
where the curvature of either the thermal or black-hole metric is $\mathcal{R}=-20/R^{2}$. In thermal AdS, the on-shell action can be expressed as 
\begin{equation}
\mathcal{S}_{hw,th} = \frac{1}{16\pi G_{5}}\int d^{3}x \int_{0}^{\pi z_{h}}d\tau \int_{R_{0}}^{R_{1}}\frac{8 R^{3}}{z^{5}},
\end{equation}
where $R_{0}$ ($R_{1}$) is the UV (IR) brane position and $\int d^{3}x\equiv V$. In order to match the two metrics at the boundary of AdS, we redefine $dt\rightarrow\sqrt{f(R_{0})}dt$. The action becomes
\begin{eqnarray}
\mathcal{S}_{hw,th} &=& -\frac{R^{3}V}{8G_{5}}z_{h}\sqrt{f(\epsilon)}\left(\frac{1}{R_{1}^{4}}-\frac{1}{R_{0}^{4}}\right)\nonumber\\
&=& -\frac{R^{3}V}{8G_{5}}z_{h}\left(1-\frac{R_{0}^{4}}{2 z_{h}^{4}}\right)\left(\frac{1}{R_{1}^{4}}-\frac{1}{R_{0}^{4}}\right)\nonumber\\
&=& -\frac{R^{3}V}{8G_{5}}z_{h}\left(\frac{1}{R_{1}^{4}}-\frac{1}{R_{0}^{4}}-\frac{R_{0}^{4}}{2 z_{h}^{4}}+\frac{1}{2 z_{h}}\right).\label{equHerzoghw0}
\end{eqnarray}
Switching to the black-hole action, we must now adjust the limits of the $dt$ integration. The black hole no longer terminates the space at $R_{0}$ but at $z_{h}$, so that the black-hole action becomes
\begin{eqnarray}
\mathcal{S}_{hw,bh} &=& \frac{R^{3}V}{2\pi G_{5}}\int_{0}^{\pi z_{h}} d\tau\, \int_{R_{0}}^{z_h}dz\, \frac{1}{z^{5}}\nonumber\\
&=& -\frac{R^{3}V}{8G_{5}}z_{h}\left(\frac{1}{z_{h}^{4}}-\frac{1}{R_{0}^{4}}\right).
\end{eqnarray}
Using (\ref{equfenergybk}), we see that the free energy is expressed as 
\begin{equation}
\mathcal{F} = -\frac{R_{3}V}{8\pi G_{5}}\left(\frac{1}{2 z_{h}^{4}}-\frac{1}{R_{1}^{4}}\right),
\end{equation}
where the transition occurs when $2 z_{h}^{4}=R_{1}^{4}$. Using the hard-wall relation between temperature and horizon position (\ref{equconformaltemp}), we find the critical temperature to be \cite{Herzog:2006ra}
\begin{equation}
T_{c} = \frac{2^{\frac{1}{4}}}{\pi R_{1}}.
\end{equation}

In the soft-wall model, we again follow the analysis done in \cite{Herzog:2006ra} and work with a string frame action where $\Phi = \mu^{2}z^{2}$,
\begin{equation}
\mathcal{S}_{sw} = -\frac{1}{16\pi G_{5}}\int d^{5}x \sqrt{-g}{\rm e}^{-\Phi} \left(\mathcal{R} + \frac{12}{R^{2}}\right).
\end{equation}
As in the hard wall, the temperature at which the free energy of the system vanishes marks the transition point,
\begin{eqnarray}
\beta \mathcal{F} &=& \lim_{R_{0}\rightarrow 0}\left(\mathcal{S}_{sw,bh}(R_{0}) - \mathcal{S}_{sw,th}(R_{0})\right) \nonumber\\
 &=& \lim_{R_{0}\rightarrow 0}\Bigg[-\frac{1}{16\pi G_{5}}\Bigg(\int d^{3}x \int_{0}^{\pi z_{h}} dt\int_{R_{0}}^{z_h} dz\,{\rm e}^{-\Phi} \frac{-8 R^{3}}{z^{5}} \nonumber\\
&&\quad\quad- \int d^{3}x\int_{0}^{\pi z_{h}} dt\,\sqrt{f(R_{0})} \int_{R_{0}}^{\infty}dz\, {\rm e}^{-\Phi} \frac{-8 R^{3}}{z^{5}} \Bigg)\Bigg] \nonumber\\
&=& \lim_{R_{0}\rightarrow 0}\left[-\frac{R^{3}V}{2\pi G_{5}}\pi z_{h}\left(\int_{R_{0}}^{z_{h}} dz\frac{{\rm e}^{-\Phi}}{z^{5}} - \int_{R_{0}}^{\infty}dz\,\frac{{\rm e}^{-\Phi}}{z^{5}}\left(1-\frac{R_{0}^{4}}{2 z_h^{4}}\right)\right)\right]\nonumber\\
&=& \lim_{R_{0}\rightarrow 0}\left[\frac{R^{3}V}{2 G_{5}}z_{h}\left(\int_{z_{h}}^{\infty}dz\, \frac{{\rm e}^{-\mu^{2}z^{2}}}{z^{5}} - \frac{R_{0}^{4}}{2z_{h}^{4}}\int_{R_{0}}^{\infty}dz\,\frac{{\rm e}^{-\mu^{2}z^{2}}}{z^{5}}\right)\right]\nonumber\\
&=& \frac{R^{3}V}{8 G_{5}z_{h}^{3}}\left({\rm e}^{-\mu^{2}z_{h}^{2}}(\mu^{2}z_{h}^{2}-1)+ \mu^{4}z_{h}^{4}{\rm Ei}(-\mu^{2}z_{h}^{2}) + \frac{1}{2}\right).
\end{eqnarray}
where Ei$(x)$ is the exponential integral function.
The free energy is then
\begin{equation}
\mathcal{F} =  \frac{R^{3}\pi V}{8 G_{5}z_{h}^{2}}\left({\rm e}^{-\mu^{2}z_{h}^{2}}(\mu^{2}z_{h}^{2}-1)+ {\rm Ei}(-\mu^{2}z_{h}^{2}) + \frac{1}{2}\right).
\end{equation}
The Hawking-Page transition occurs when
\begin{equation}\label{equbkTranCondition}
{\rm e}^{-x}(x-1) + \frac{1}{2} + x^{2} {\rm Ei}(x) = 0
\end{equation}
is satisfied. Since (\ref{equbkTranCondition}) cannot be solved analytically, we have defined a new variable, $x=\mu^{2} z_{h}^{2}$, to ease the graphical analysis. We find that the transition temperature is \cite{Herzog:2006ra}
\begin{equation}
T_{c} = 0.491728 \mu.
\end{equation} 
Again, we see that quantities in the soft-wall model scale with the parameter $\mu$. whereas the inverse of the IR brane position $1/R_{0}$ sets the scale in the hard-wall.

The above transition temperatures found above are purely phenomenological. The solution used in \cite{Herzog:2006ra} was not dynamically produced. Considering the soft-wall model in the complete manner, one would need to add a boundary term to the on-shell action called the Gibbons-Hawking (GH) term. It was shown in \cite{Evans:2008tu} that including the GH term in the calculation above produces a transition temperature at $T=0$, describing no confined phase. In Chapter \ref{chthermo}, we consider a dynamic soft-wall solution in an attempt to rigorously calculate a transition temperature.

\section{Conclusion}

 In this chapter, we reviewed the concepts needed to understand the original work in Chapters \ref{chzero} and \ref{chthermo}. We summarized much of the relevant literature involving meson mass spectra in the scalar, pseudoscalar, vector, and axial-vector sectors in both the hard-wall and soft-wall AdS/QCD. Next, we outlined the thermodynamics of AdS/QCD used to describe the deconfined state of QCD matter, finding the temperature, entropy, speed of sound, and free energy of the thermal fluid. We concluded with showing how to calculate the transition temperature from the on-shell actions.

\chapter{Zero Temperature Dynamics of AdS/QCD}\label{chzero}

\begin{flushright}
``Sometimes the questions are complicated and the answers are simple.''\\
Dr. Seuss
\end{flushright}

\section{Introduction} 
\label{secIntro}

 This chapter works through the dynamics of an AdS/QCD model at zero temperature. We modify the existing soft-wall version of the AdS/QCD model \cite{Karch:2006pv} in order to incorporate several phenomenological features of QCD. First, we incorporate separate sources for spontaneous and explicit chiral symmetry breaking (CSB). Second, we show this model does not restore chiral symmetry in highly excited resonances \cite{Klebanov:1999tb, Shifman:2007xn}. Finally, we produce a new dilaton background profile and retain the linear mass trajectories in the scalar, pseudoscalar, vector, and axial-vector sectors that make this phenomenological model attractive and fertile ground for further research. 

This chapter is separated into five sections. In section \ref{secDual}, we introduce the modified soft-wall Lagrangian. We describe how the new dilaton profile and higher-order terms lead to a model describing spontaneous and explicit CSB. In Section \ref{secSpectra}, we explore the QCD meson mass spectrum in the scalar, pseudoscalar, vector, and axial-vector sectors of the model. In Section \ref{secDynamics}, we determine other dynamics derived from this theory, such as the pion decay constant, the Gell-Mann-Oakes-Renner relation, the coupling of the vector mesons to the pions, and the pion electromagnetic form factor.  We conclude with a discussion in Section \ref{secDiscuss4}.

\section{The Dual Model} 
\label{secDual}

We consider a modified version of the soft-wall AdS/QCD model first introduced in 
\cite{Karch:2006pv} and further investigated in \cite{Sui:2009xe,Kwee:2007nq,Evans:2006ea, Colangelo:2008us,Cherman:2008eh,Evans:2006dj,Zuo:2009dz,Huang:2007fv, Grigoryan:2007my, Batell:2008zm,Karch:2010eg}. 
In the string frame, the background geometry is not dynamically generated but assumed to be five-dimensional AdS space with a metric
\begin{equation}\label{equstringmetric}
ds^{2}=g_{MN}dx^{M}dx^{N}=a(z)^{2}\left( \eta_{\mu\nu}dx^{\mu}dx^{\nu}+dz^{2}\right),
\end{equation}
where $a(z)=R/z$ is the warping factor and the Minkowski metric, 
\begin{equation}
\eta_{\mu\nu}=\left( \begin{array}{cccc} 
 -1 & 0 & 0 & 0\\
  0 & 1 & 0 & 0\\
  0 & 0 & 1 & 0\\
  0 & 0 & 0 & 1
\end{array} \right).
\end{equation} The coordinate $z$ has a range $0\leq z < \infty$. 
The action follows from the standard AdS/QCD dictionary. We begin with the basic action incorporating CSB with a scalar field $X$ dual to the dimension-3 $q\bar{q}$ operator introduced in Section \ref{secHadrons}. The primary shortcoming of the action (\ref{action0simple}), however, is the explicit and spontaneous CSB entanglement. No chiral limit exists in this theory. As suggested by \cite{Karch:2006pv}, we add a quartic term to the action, whose strength is controlled by the parameter $\kappa$. In section \ref{secPseudo}, we show that $X$ actually becomes a complex field to incorporate the pseudoscalar field $\pi$. 
The five-dimensional action takes the form,
\begin{equation} \label{action1}
S_{5}=-\int d^{5}x \sqrt{-g}\,{\rm e}^{-\phi(z)}{\rm Tr}\left[|D X|^{2}+ m_{X}^{2} |X|^{2}-\kappa |X|^{4}+\frac{1}{2 g_{5}^{2}}(F_{V}^{2}+F_{A}^{2})\right],
\end{equation}
where $\kappa$ is a constant that will be determined and $g_5^2=12\pi^2/N_c$ as shown in Section \ref{secHadrons}. 

To obtain linear mass trajectories from this action, we assume a soft-wall background dilaton, $\phi$, with the asymptotic behavior
\begin{equation} \label{dilatonlz}
\phi(z\rightarrow \infty) \simeq \mu^{2} z^2,
\end{equation}
where $\mu$ sets the mass scale for the meson spectrum. The $z$-dependent dilaton field also ensures that conformal symmetry is broken. In Section \ref{secVEV}, we show that assuming that if chiral symmetry remains broken as excitation number increases, as evident through the behavior of $v(z)$, then the quadratic behavior of the dilaton arises naturally.

\subsection{Bulk Scalar VEV Solution} 
\label{secVEV}

The field $X$, which is dual to the operator $\bar{q} q$, obtains a $z$-dependent vacuum expectation value (VEV), 
\begin{equation}
\label{equxvev}
\langle X \rangle \equiv \frac{v(z)}{2}
\left( \begin{array}{cc} 
  1 & 0 \\
  0 & 1  
\end{array} \right).
\end{equation}
Specifying the elements of the bifundamental field fixes the number of flavors $N_{f}$ that we consider. In the case of (\ref{equxvev}), we have two flavors of degenerate quarks. The VEV breaks the chiral symmetry SU(2)$_L\times$ SU(2)$_R \rightarrow$ SU(2)$_V$ by coupling to the axial-vector field. Assuming (\ref{equxvev}), we obtain a nonlinear equation for the VEV $v(z)$ using the variation principle
\begin{eqnarray}
\delta S&=& -2{\rm e}^{-\phi}\sqrt{-g}\,Tr\Big(g^{MN}\partial_{M}X\partial_{N}\delta X + \{A,X\}\{A,\delta X\}+[V,X][V,\delta X] \nonumber\\
&&\quad\quad\quad + m_{X}^{2}X\delta X - 2\kappa X^{\dagger}X|X|\delta X\Big)=0.
\label{equXvariation}
\end{eqnarray}
Taking the trace, integrating by parts, and using $|X|=v(z)+S(x,z)$, we find the a simplified expression for the variation,
\begin{eqnarray}
\frac{\delta S}{\delta X}\Big|_{X\rightarrow v(z)} =  \frac{1}{2}\left(\partial_{z}\left({\rm e}^{-\phi}\sqrt{-g}g^{zz}\partial_{z}v\right) + {\rm e}^{-\phi}\sqrt{-g}\,m_{X}^{2} v + {\rm e}^{-\phi}\sqrt{-g}\frac{\kappa}{2}v^{3}\right).
\end{eqnarray}
Using the metric (\ref{equstringmetric}), we find the equation of motion for $v(z)$,
\begin{equation}
\partial_z(a^3 {\rm e}^{-\phi} \partial_z v(z))-  a^5 {\rm e}^{-\phi} \left(m_X^2 v(z)-\frac{\kappa}{2} v(z)^{3}\right)=0,
\end{equation}
which simplifies to
\begin{equation}\label{equvsimple}
v'' - \left(\phi' + \frac{3}{z}\right)v' - m_{X}^{2} v(z) + \frac{\kappa}{2}v^3 = 0,
\end{equation}
where ($'$) denotes a derivative with respect to $z$.

 As noted in \cite{Shifman:2007xn}, highly excited mesons exhibit parallel Regge trajectories indicating that chiral symmetry is not restored with increasing $n$. In order to incorporate this behavior the scalar VEV, $v(z)$ must behave linearly as $z$ becomes large,
\begin{equation}
v(z\rightarrow\infty) \sim z,
\end{equation}
causing the mass difference between vector and axial-vector resonances to approach a constant as $z\rightarrow\infty$ as shown in Section \ref{secbkAxial}. By including a quartic term and requiring linear asymptotic behavior for $v$, we aim to incorporate these QCD-like characteristics into the soft-wall model.

 Instead of solving for $v(z)$ directly from (\ref{equvsimple}), we assume the VEV asymptotically behaves as expected, namely
\begin{eqnarray}
v(z\rightarrow 0) &=& \frac{m_{q} \zeta }{R} z + \frac{\Sigma}{\zeta R} z^{3}, \label{smallV} \\
v(z\rightarrow\infty) &=& \Gamma z. \label{largeV}
\end{eqnarray}
Solving for the dilaton $\phi(z)$, we see
\begin{equation}
\label{phieqn}
\phi'(z)=\frac{1}{a^3 v'(z)}\left[\partial_z(a^3 v'(z))-a^5 (m_X^2 v(z)- \frac{\kappa}{2} v^3(z))\right],
\end{equation}
where the prime ($'$) denotes the derivative with respect to $z$. Given the required behavior (\ref{smallV}) and (\ref{largeV}), we can uniquely determine the dilaton profile up to a constant. With this procedure the two sources of chiral symmetry breaking decouple while simultaneously allowing for linear trajectories in the meson spectrum.

A particularly simple parametrized form for $v(z)$ that satisfies (\ref{smallV}) and (\ref{largeV}) is
\begin{equation}
v(z) = \frac{z}{L}(A + B \tanh{C z^2}), \label{arcv}
\end{equation} 
where $A$, $B$, and $C$ are all positive coefficients dependent upon $m_{q}$, $\sigma$, $N_{c}$, and $\kappa$. Expanding (\ref{arcv}) at small and large $z$ leads to the asymptotic forms
\begin{eqnarray}
v(z\rightarrow 0)L &=&  A z + B C z^3+{\cal O}(z^5), \label{arcvsmall} \\
v(z\rightarrow \infty)L &=&  (A+B) z. \label{arcvlarge}
\end{eqnarray}
When $A=0$, corresponding to a zero quark mass, the coefficients of the cubic term in (\ref{arcvsmall}) and of the linear term in (\ref{arcvlarge}) are nonzero, implying a nonzero chiral condensate and non-restoration of chiral symmetry. Alternatively, when $B=0$ (or $C=0$), corresponding to a zero chiral condensate, the coefficients of the linear terms in (\ref{arcvsmall}) and (\ref{arcvlarge}) are both nonzero, implying a nonzero quark mass and non restoration of chiral symmetry. Thus the parametrized form in (\ref{arcv}) allows the sources of spontaneous and explicit chiral symmetry breaking to remain independent. 
Substituting (\ref{arcv}) into (\ref{phieqn}) leads to the following asymptotic behavior for the dilaton,
\begin{eqnarray}
\phi(z\rightarrow 0) &=& \frac{\kappa}{4} A^2 z^2 + \mathcal{O}(z^6), \label{arcphi0} \\
\phi(z\rightarrow\infty) &=& \frac{\kappa}{4} (A+B)^2 z^2, \label{arcphiinf}
\end{eqnarray}
where we have chosen the boundary condition $\phi(0)=0$ to recover a pure AdS metric in the $z\rightarrow 0$ limit. To reproduce the limits (\ref{arcvsmall}) and (\ref{arcvlarge}) the dilaton profile at small $z$ (\ref{arcphi0}) must differ from that at large $z$ (\ref{arcphiinf}). This does not sacrifice the linear trajectories which (as will be shown) depend on the dilaton having the asymptotic form (\ref{dilatonlz}).  Note that the quartic term with strength $\kappa$ is necessary to obtain the required behavior.  Therefore, modifying the dilaton and introducing quartic interaction terms in the Lagrangian is necessary to improve the soft-wall version of the AdS/QCD model.

\begin{figure}[h!]
\begin{center}
\includegraphics[scale=0.49,angle=90]{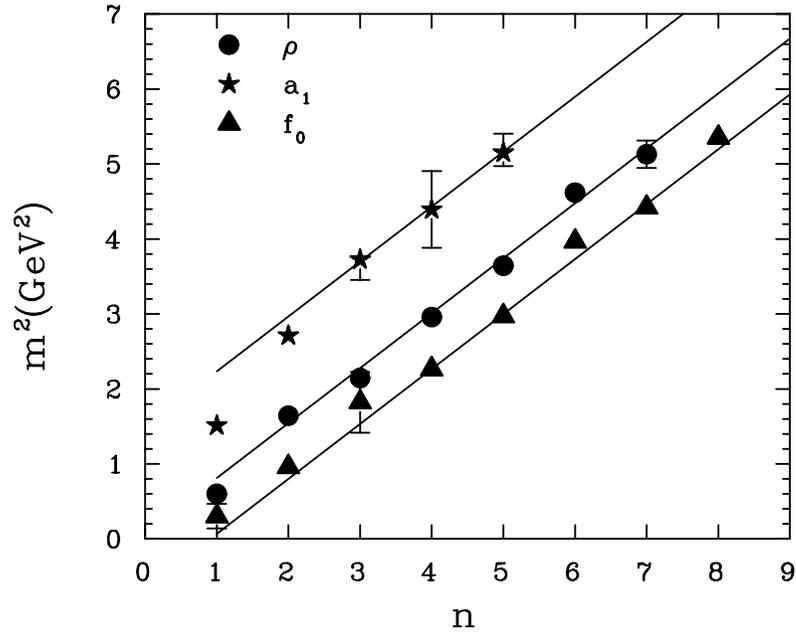}
\caption{A straight-line fit to the measured scalar, vector and axial-vector mass spectra for $n \ge 3$ used to determine the dilaton mass parameter
$\mu$. 
}
\label{combinedplot}
\end{center}
\end{figure}

From (\ref{equzeta}), the normalization parameter, $\zeta$, is set conveniently to
\begin{equation}
\zeta=\frac{\sqrt{N_c}}{(2\pi)}=\frac{\sqrt{3}}{g_5}, 
\end{equation}
where for this phenomenological model, we take $N_{c}=3$. Then the parameters $\gamma$, $A$, $B$ and $C$ can be expressed in terms of the input parameters $m_q,\sigma,\mu,\kappa$, 
\begin{eqnarray}
\Gamma &=& \sqrt{\frac{4\,\mu^{2}}{\kappa}}, \label{padeC}\\
A &=& \frac{\sqrt{3}m_q}{g_5}, \label{arcA} \\
B &=& \Gamma - A, \label{arcB} \\
C &=& \frac{g_5 \sigma}{\sqrt{3}B}. \label{arcC}
\end{eqnarray}
The input parameters are determined as follows. In order to find $\mu$, a straight line is fitted to the $m^2$ versus $n$ plot for $n \ge 3$ for the scalar, vector and axial-vector mesons, assuming the same slope $4\mu^2$ but different intercepts, namely $m^2_n = 4\mu^{2} n + m^2_0$. The fit was limited to the higher excitation modes because Regge behavior is not observed in the lower modes. Figure \ref{combinedplot} shows the plot, which gives, in units of GeV$^2$,
\begin{eqnarray}
\mu^{2} = 0.1831 \pm 0.0059,\nonumber\\
m_{S,0}^2 = -0.6634 \pm 0.0038,\nonumber\\
m_{V,0}^2 = 0.0806 \pm 0.0104,\nonumber\\
m_{A,0}^2 = 1.5023 \pm 0.0366.
\end{eqnarray}
The quark mass, quark condensate, pion decay constant, and pion mass are all related by the Gell-Mann-Oakes-Renner relation, $f_{\pi}^2 m_{\pi}^2 = 2 m_q \sigma$.  This relation holds in this model as a natural consequence of chiral symmetry \cite{Erlich:2005qh}; see also Section \ref{secDynamics}.  We use the measured values of $f_{\pi}=92.4$ MeV and $m_{\pi}=139.6$ MeV, and adjust the quark mass to reproduce the input value of $f_{\pi}$ from a solution to the axial-vector field equation in Section \ref{secDynamics} for a given value of $\kappa$.  The parameter $\kappa$ essentially controls the mass splitting between the vector and axial-vector mesons.  It is determined to be $\kappa = 15$ by a best fit to the radial spectra of the axial-vector mesons.  This results in $m_q = 9.75$ MeV and $\Sigma =(204.5$ MeV)$^3$.  The inferred value of the quark mass is consistent with an average of the up and down quark masses as summarized in the Review of Particle Physics \cite{pdg} as appropriate at the hadronic energy scale.

\begin{figure}[h!]
\begin{center}
\includegraphics[scale=0.49,angle=90]{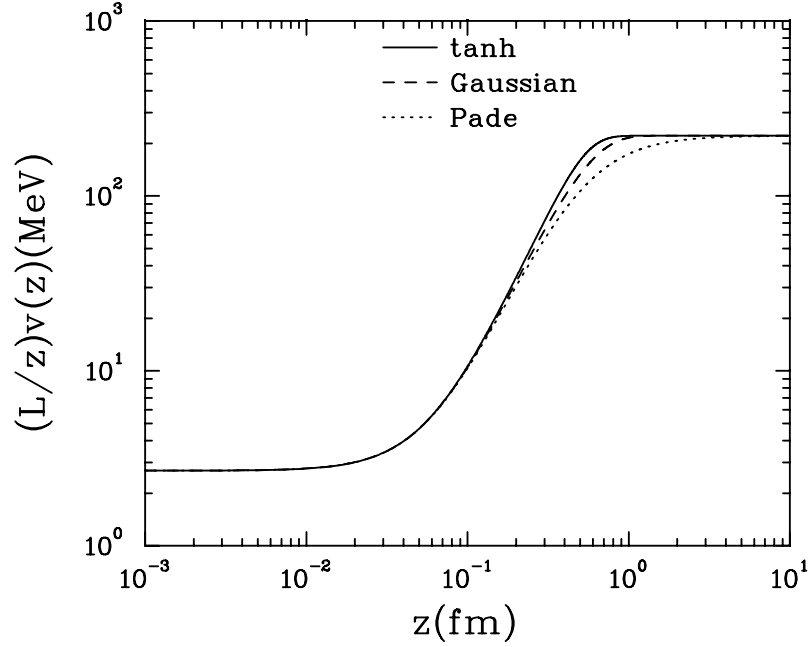}
\caption{A plot of $v(z)/z$ for various parameterizations fitted to the mass spectra. The best fit to the mass spectra is obtained with (\ref{arcv}).}
\label{vevfig}
\end{center}
\end{figure}

\begin{figure}[h!]
\begin{center}
\includegraphics[scale=0.49,angle=90]{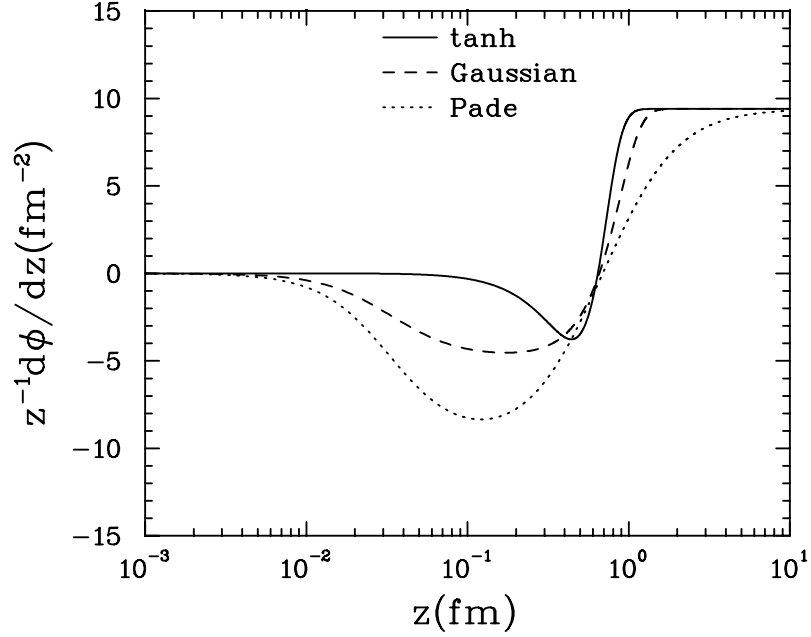}
\caption{A plot of $\phi'(z)/z$ derived from the various parameterizations of $v(z)$. The best fit to the mass spectra is obtained with the $\tanh$ form (\ref{arcv}).
}
\label{figdotdilaton}
\end{center}
\end{figure}

\begin{figure}[h!]
\begin{center}
\includegraphics[scale=0.49,angle=90]{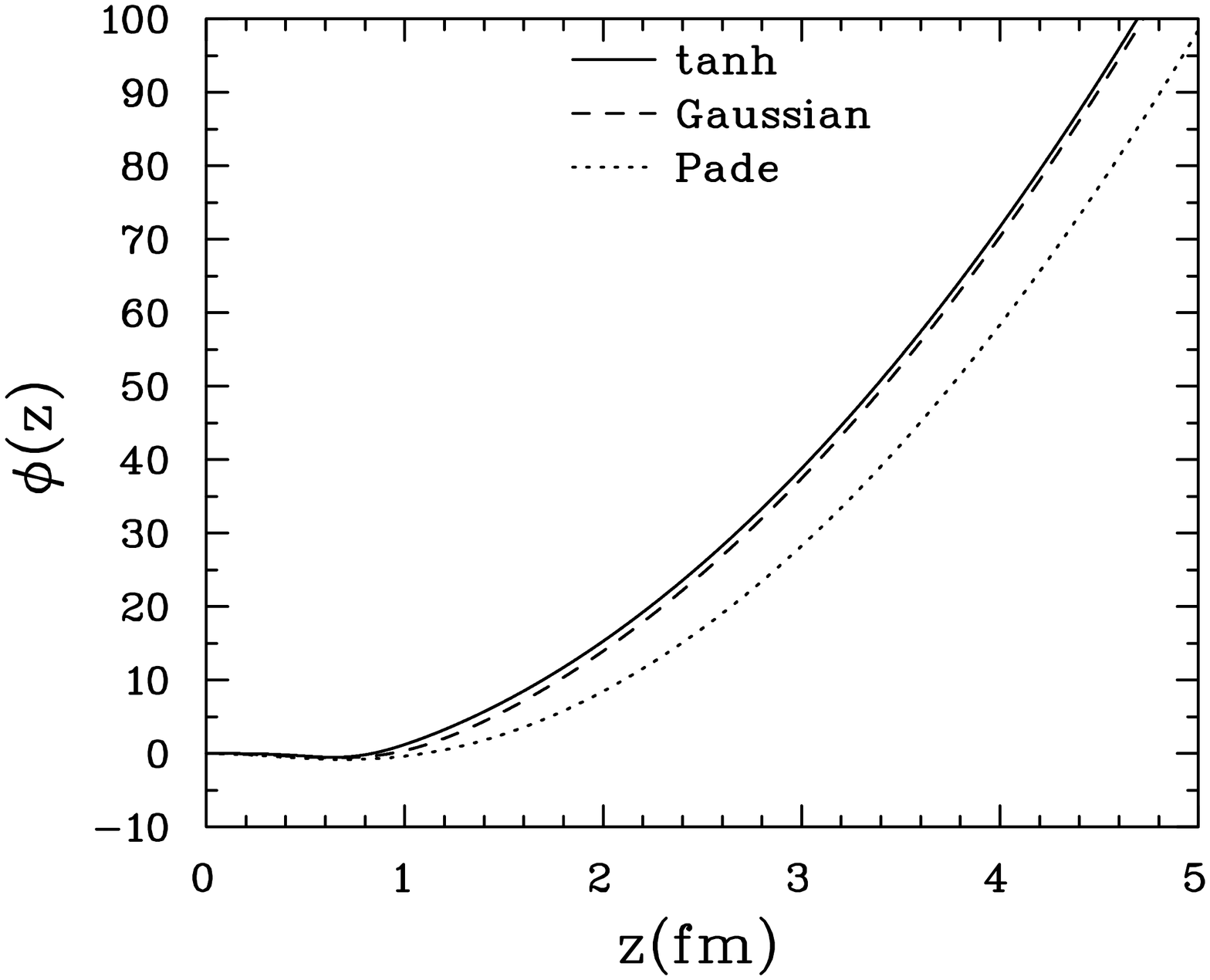}
\caption{The dilaton profile $\phi(z)$ resulting from the various parameterizations of $v(z)$. The best fit to the
meson spectra occurs with the $\tanh$ parameterization (\ref{arcv}). For $z\lesssim1$ the behavior deviates from the quadratic asymptotic form (\ref{dilatonlz}).
}
\label{figdilaton}
\end{center}
\end{figure}

Other parameterizations for $v(z)$ that lead to qualitatively similar behavior and asymptotic limits (\ref{smallV}) and (\ref{largeV}) exist. Among many possible alternatives, $v(z)$ could take the forms of a Pad\'e approximant or a Gaussian,
\begin{eqnarray}
v_{P} &=& (a_1 + a_2 z^3)/(1 + a_3 z^2)\label{equpadein0}, \\
v_{G} &=& b_1 z- b_2 z \exp(-b_3 z^2).\label{equgaussin0}
\end{eqnarray}
We explore the alternative parametrizations in Appendix \ref{appvev}. These forms were all studied but the best results were found using (\ref{arcv}).  The parameterizations (\ref{arcv}), (\ref{equpadein0}), and (equgaussin0) are compared in Figure \ref{vevfig}.  The resulting plots of $d\phi/dz$ and $\phi(z)$ are shown in Figures \ref{figdotdilaton} and \ref{figdilaton}. It becomes quite evident from the figures that a small change in $v(z)$ parametrization leads to drastic change in the behavior of the dilaton $\phi(z)$.

\section{Meson Mass Spectra}\label{secSpectra}

Our soft wall model can be used to fit the meson mass spectra. It is interesting to see how well this simple five-dimensional model matches real data.  The scalar, pseudoscalar, vector, and axial-vector mass resonances used in our fits are given in Tables \ref{scalarmasses}, \ref{tblmass}, \ref{vectormasses}, and \ref{avmasses}, respectively. Before proceeding, we give a short justification for including selective radial excitations in our analysis. All but one resonance of the included states are listed in the Review of Particle Physics \cite{pdg}.

The scalar mesons are the most ambiguous particles in this analysis. Uncertainty lies in the scalar mesons since mixing is expected among light quark mesons, four quark states, $s\bar{s}$ mesons, and glueballs.  This could shift the masses of the ``pure'' radial excitations of the lightest scalar meson by ${\mathcal O}$(100 MeV). There is evidence suggesting that $f_{0} (600)$ and $f_{0}(980)$ contain 4-quark states and that $f_0 (1370)$, $f_0 (1500)$, $f_0 (1710)$, and $f_0 (1770/1790)$ (most recently referred to as $f_{0} (1790)$) are all superpositions of $q\bar{q}$ and glueballs \cite{Bugg:2004xu,Bugg:2006uk}. However, the 2-quark character of $f_{0} (980)$ appears dominant \cite{Fariborz:2006xq}. In addition, $f_0(1500)$ and  $f_0 (1710)$ most likely have significant glueball contributions because of their absent $\gamma\gamma$ coupling, narrow width, enhanced production at low transverse momentum, and the relative close proximity to the state $f_{0} (1370)$ and $f_{0}(1790)$, respectively \cite{Bugg:2004xu, Bugg:2008eb}. Experiments suggest that $f_{0} (1790)$ makes a natural radial excitation of $f_{0} (1370)$. Almost nothing is said of the strength of the case for the $f_0 (2020)$, $f_0 (2100)$, $f_0 (2200)$, or the $f_0 (2330)$. However, $f_0(2103 \pm 8)$ and $f_0(2314 \pm 25)$ in \cite{pdg} agree well with $f_0(2102 \pm 13)$ and $f_0(2337 \pm 14)$ \cite{Bugg:2004xu}, respectively. Therefore, $f_{0}(2100)$ and $f_{0}(2330)$ complete our choices for the scalar mesons.

The pseudoscalar mesons do not have as many intricacies as the scalars. The ground-state pion has been known for decades. The excitation modes of the $\pi$ have been well established as $\pi(1300)$ and $\pi(1800)$ \cite{pdg}. We will also include some unconfirmed mass states for further comparison.

Three vector mesons are confirmed and require little debate: $\rho(775)$, $\rho(1450)$ and $\rho(1700)$ \cite{pdg}. However, it has long been known that the $\rho(1465)$ is too massive to be the first radial excitation of the $\rho(775)$ \cite{Bugg:2004xu}. Studies of the process $p + \bar{p} \rightarrow 2\pi^+ + 2\pi^-$ suggests the first radial excitation of the $\rho$ to be 1282$\pm$37 MeV, which is the value used in our fits. There is evidence of a $\rho(1570)$, but this is believed to be a OZI-suppressed decay mode of $\rho(1700)$. Based on the rating system in the review \cite{Bugg:2004xu}, we include $\rho(1900)$ as a confirmed mass resonances, and $\rho(2150)$ and $\rho(2265)$ are included as unconfirmed resonances. 

Luckily, the axial-vector mesons are limited in number. While the only confirmed axial-vector resonance is $a_{1}(1260)$, there exists a clear radial progression: $a_{1}(1640)$, $a_{1}(1930)$, and $a_{1}(2095)$ \cite{pdg}. Two unconfirmed resonances continue this progression, the $a_{1}(2270)$ and $a_{1}(2340)$, which are rather close together assuming Regge trajectories. However, it is clear that $a_{1}(2270)$ is backed with more experimental evidence \cite{pdg}.

\subsection{Scalar sector} 
\label{secScalar}

Introducing a quartic term in the Lagrangian causes the scalar excitations to couple with their own VEV, giving a modified equation of motion unlike those in \cite{DaRold:2005zs, Colangelo:2008us}. Assuming $X(x,z)\equiv (v(z)/2+S(x,z)){\rm e}^{ i P}$, where $P$ will be defined in Section \ref{secPseudo}, and a KK decomposition,
\begin{equation}
 S(x,z)=\sum_{n=0}^{\infty}{\cal S}_n(x) S_n(z), 
\end{equation}
we obtain by variation of the action (\ref{action1}),
\begin{equation}
\partial_z(a^3 {\rm e}^{-\phi}\partial_{z}S_n(z))- a^5 {\rm e}^{-\phi}(m_X^{2} -\frac{3}{2}\kappa v^2(z)) S_n(z)= -a^3 {\rm e}^{-\phi} m_{S_n}^2 S_n(z), \label{scalarequ}
\end{equation}
where $S_{n}(z)$ are the Kaluza-Klein modes. Note that by ignoring the nonlinear terms in (\ref{scalarequ}) we are assuming infinitesimally small amplitudes $S_{n}$. Because of the $z$-dependent mass term, (\ref{scalarequ}) is difficult to solve analytically for the parametrized solution of $v(z)$; however, we implement a shooting method discussed in Appendix \ref{appShooting}. 

\begin{figure}[h!]
\begin{center}
\includegraphics[scale=0.45,angle=90]{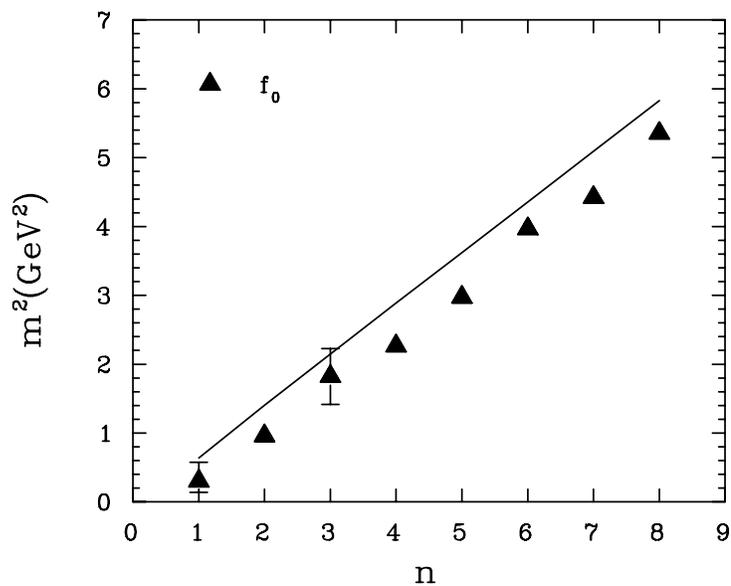}
\caption{Comparison of the predicted scalar mass eigenvalues using the $\tanh$ form (\ref{arcv}) of $v(z)$ (solid) with the QCD $f_{0}$ scalar mass spectrum \cite{pdg}. 
}
\label{ScalarMasses}
\end{center}
\end{figure}

The scalar equation of motion (\ref{scalarequ}) can be brought into a Schr\"{o}dinger-like form using the process in Appendix \ref{appSchTransform} with the substitution 
\begin{equation}
S_n(z)={\rm e}^{\omega_s/2}s_n(z),\quad\quad\quad \omega_s = \phi(z) + 3\log{z}.
\end{equation} 
The eigenvalue equation becomes
\begin{equation} \label{schroS}
-\partial_z^{2}s_n(z)+\left(\frac{1}{4}\omega_s'^2-\frac{1}{2}\omega_s''-\frac{3}{2}\frac{L^2}{z^2} \kappa v^2(z)-\frac{3}{z^2}\right)s_n(z)=m_{S_n}^{2}s_n(z),
\end{equation}
where the boundary conditions are
\begin{equation}
\lim_{z_0\rightarrow 0} s_n(z_0)=0,\quad\quad\quad \partial_z s_{n}(z\rightarrow\infty)=0.
\end{equation}
We find the scalar mass spectra listed in Table~\ref{scalarmasses}, and displayed in Figure \ref{ScalarMasses}. The reproduction of the experimentally measured masses is reasonable, apart from an overall constant.  This could be a failure of the model in describing this sector.  However, considering that these light quark/antiquark radial excitations mix with scalar $s\bar{s}$ excitations, scalar glueballs, and possible four quark states, it may be that either the lowest or first radially excited state has been misidentified.  Removing either the $f_0(550)$ or the $f_0(980)$ would shift all the higher masses to the left by one unit of $n$, resulting in a much better fit to the model.  An obvious extension of this work would be to include strange quarks and glueballs and to determine the mixing among the resulting scalar states.

\begin{table}[h!]
\begin{center}
\begin{tabular}{|c||c|c|}
\hline
$n$ & $f_0$ experimental (MeV) & $f_0$ model (MeV) \\
\hline
\hline
1 & $550^{+250}_{-150}$ & $799$ \\
\hline
2 & $980 \pm 10$ & $1184$ \\
\hline
3 & $1350 \pm 150$ & $1466$ \\
\hline
4 & $1505 \pm 6$ & $1699$ \\
\hline
5 & $1724 \pm 7$ & $1903$ \\
\hline
6 & $1992 \pm 16$ & $2087$ \\
\hline
7 & $2103 \pm 8$ & $2257$ \\
\hline 
8 & $2314 \pm 25$ & $2414$ \\
\hline 
\end{tabular}
\caption{The experimental and predicted values of the scalar meson masses.}
\label{scalarmasses}
\end{center}
\end{table}

\subsection{Pseudoscalar Sector}
\label{secPseudo}

Analyzing the pseudoscalar sector is particularly intriguing because of the pseudoscalar representation used in \cite{DaRold:2005zs, DaRold:2005vr} {\it appears} to be incompatible with the representations used in \cite{Erlich:2005qh, Grigoryan:2008cc, Kwee:2007nq, DeFazio:2008mb}. Both are in the exponential form 
\begin{equation}
X = S {\rm e}^{i P},
\end{equation}
where $P$ is the pseudoscalar part of a complex field $X$. The former papers define $P$ using the vacuum expectation value $v(z)$ of the scalar field, 
$S(x,z)$, as $P = \pi^a(x,z)t^a/v(z)$, while the latter papers specify $P = 2\pi^{a}(x,z)t^a$. The two representations produce seemingly different equations of motion and {\it potentially} different mass eigenvalues. Alternatively, one may use a linear representation, $X = X_{0} + i \pi^{a}t^a$, as in \cite{Kaplan:2009kr}. Any physical observables calculated from this model should, of course, be independent of the choice of representation.  

In this section, we derive and explore the equations of motion of the two representations, showing that the sets of equations are, indeed, equivalent. One can simplify the equations of motion in two different ways: (i) by eliminating one field as in \cite{Sui:2009xe}, or (ii) by transforming the full system of second-order differential equations into Schr\"odinger-like form. Numerically calculating the mass eigenvalues, we find that the eigenvalues found from one method do not match those of the other in our case.  Apparently, \cite{Sui:2009xe} has found the eigenvalues not of the pseudoscalars but of a derivative field, and they are \textit{not necessarily} equivalent. However, it should be noted that the original authors of \cite{Sui:2009xe} again assert that methods (i) and (ii) give equivalent eigenvalues \cite{Sui:2010ay}; we cannot confirm this. 

Concentrating on the pseudoscalar sector, we explore two representations for the field $X$,
\begin{eqnarray}
X_e &=& (v(z)/2+S(x,z)) \, \emph{1} \, {\rm e}^{2i \pi_{e}^{a}(x,z)t^{a}} \label{equXe},\\ 
X_l &=& (v(z)/2+S(x,z)) \, \emph{1} + i \pi_{l}^{a}(x,z)t^{a} \label{equXl},
\end{eqnarray}
where $\emph{1}$ is the $2\times 2$ identity matrix. We refer to $X_{e}$ as the exponential representation and $X_{l}$ as the linear representation. The exponential representation is used in \cite{Sui:2009xe,Kwee:2007nq, Erlich:2005qh, DeFazio:2008mb,Grigoryan:2008cc}, where it is assumed to be the canonically normalized pion field, $\pi_{e} = \tilde{\pi}/f_{\pi}$. The linear representation has been specified before in \cite{Kaplan:2009kr}, where the $\pi$ field carries the same dimensions as other fields in the Lagrangian. We already see an apparent difference between representations; (\ref{equXl}) allows for an explicit quartic term in $\pi$ when substituted into the action (\ref{action1}), whereas there is no such term in the case of (\ref{equXe}). The consequence of such quartic terms in $\pi$ will not be addressed here; however, the quartic term strength impacts the pseudoscalar mass spectrum through the parameter $\kappa$. We only consider field terms up to quadratic order.

\subsubsection{Representations} 
\label{secRep}

In this section, we derive the equations of motion arising from the two representations (\ref{equXe}) and (\ref{equXl}). The pseudoscalar and longitudinal components of the axial-vector field mix in the Lagrangian; therefore, we find two coupled differential equations for each representation. This makes the numerical work more involved than for the scalar, vector, and axial-vector sections, which were already studied in \cite{Gherghetta:2009ac}.  The last part of this section shows that the two sets of differential equations are equivalent.

\paragraph{Exponential Representation}
Let us take (\ref{equXe}) and substitute it into (\ref{action1}), where we focus only on the terms involving the field $\pi(x,z)$,
\begin{eqnarray}
\mathcal{L}_e &=&  -\sqrt{-g} \,{\rm e}^{-\phi(z)} \frac{1}{2}\delta^{ab}\Big(g^{MN} \left(v^{2}\,\partial_{M}\pi\partial_{N}\pi + v^{2} A_{M} A_{N}-2v^{2}\partial_{M}\pi A_{N}\right)\nonumber\\
&&+ \frac{g^{MP}g^{NR}}{g_{5}^{2}}\left(\partial_{M} A_{N}\partial_{P} A_{R} - \partial_{M} A_{N}\partial_{R}A_{P}\right) \Big).\label{start}
\end{eqnarray}
We work in the axial gauge, $A_{z} = 0$, and define $A_{\mu} = A_{\mu\perp} + \partial_{\mu}\varphi$, where $\partial_{\mu}A_{\perp}^{\mu}=0$. Separating (\ref{start}) explicitly into four-dimensional components and bulk terms, we obtain
\begin{eqnarray}
\mathcal{L}_e &=& -\frac{1}{2} {\rm e}^{-\phi(z)}\Big[\sqrt{-g}\,g^{\mu\nu}\left(v^{2}\partial_{\mu}\pi\partial_{\nu}\pi+v^{2}A_{\mu}A_{\nu} - 2v^{2}\partial_{\mu}\pi A_{\nu}\right)\nonumber \\
&&+\sqrt{-g}g^{zz}v^{2}\partial_{z}\pi\partial_{z}\pi\ + \frac{\sqrt{-g}g^{\mu\nu}g^{\rho\sigma}}{g_{5}^{2}}(\partial_{\mu} A_{\rho} \partial_{\nu} A_{\sigma} - \partial_{\mu} A_{\rho} \partial_{\sigma} A_{\nu}) \nonumber\\
&&+ \frac{\sqrt{-g}g^{zz}g^{\mu\nu}}{g_{5}^{2}}(\partial_{z} A_{\mu}\partial_{z} A_{\nu})\Big].\label{equLbase}
\end{eqnarray}
Keeping only terms of the longitudinal part of $A_{\mu}$ gives
\begin{eqnarray}
\mathcal{L}_{e} &=& -\frac{1}{2} {\rm e}^{-\phi(z)}\Big[\sqrt{-g}\,g^{\mu\nu}(v^{2}\partial_{\mu}\pi\partial_{\nu}\pi+v^{2}\partial_{\mu}\varphi\partial_{\nu}\varphi - 2v^{2}\partial_{\mu}\pi \partial_{\nu}\varphi) \nonumber\\
&& + \sqrt{-g}g^{zz}v^{2}\partial_{z}\pi\partial_{z}\pi\ 
+ \frac{\sqrt{-g}g^{zz}g^{\mu\nu}}{g_{5}^{2}}(\partial_{z} \partial_{\mu}\varphi\partial_{z} \,\partial_{\nu}\varphi)\Big].\label{equLproper}
\end{eqnarray}
Varying (\ref{equLproper}) with respect to $\pi$ gives
\begin{equation}
\delta\mathcal{L}_e = \partial_{z} {\rm e}^{-\phi}\sqrt{-g}\,g^{zz}v^{2}\partial_{z}\pi \delta\pi + {\rm e}^{-\phi}\sqrt{-g}\,v^{2}g^{\mu\nu}\partial_{\nu}\partial_{\mu}(\pi - \varphi)\delta\pi. \nonumber\\
\end{equation}
Using a Kaluza-Klein decomposition,
\begin{eqnarray}
\pi(x,z) &=& \sum_{n}\Pi_{n}(x) \pi_{n}(z) \label{equKKpi}, \\
\varphi(x,z) &=& \sum_{n}\Phi_{n}(x)\varphi_{n}(z) \label{equKKphi},
\end{eqnarray}
and
\begin{equation}
\partial^{2} \Pi_{n}(x) = m_{n}^{2}\Pi_{n}(x) \, , \quad\quad \partial^{2}\Phi_{n}(x) = m_{n}^{2}\Phi_{n}(x) \,, 
\end{equation}
we can express the system of equations in terms of its $z$-dependent parts. We obtain the first equation of motion,
\begin{equation}
\frac{{\rm e}^{\phi}}{v^{2} a^{3}}\partial_{z}\left({\rm e}^{-\phi}v^{2}a^{3}\partial_{z}\pi_{n}\right) + m_{n}^{2}(\pi_{n} -\varphi_{n}) = 0. \label{equOne}
\end{equation}
Varying (\ref{equLproper}) with respect to $\phi$ and breaking it down into KK modes gives the second equation of motion,
\begin{equation} \label{equTwo}
{\rm e}^{\phi}\partial_{z}\left(\frac{{\rm e}^{-\phi}}{z}\partial_{z}\varphi_{n}\right) + \frac{g_{5}^{2}R^{2}v^{2}}{z^{3}}(\pi_{n}-\varphi_{n}) = 0. 
\end{equation}

Alternatively, we can express (\ref{equOne}) and (\ref{equTwo}) in a Schr\"odinger-like form. We rewrite them in the same form as (\ref{equphiAppend}) and (\ref{equpiAppend}) in the Appendix by substituting
\begin{eqnarray}
\pi_{n} &=& {\rm e}^{f(z)}\tilde\pi_{n}, \quad\quad\quad f(z) = \phi(z) + \log{\frac{z^{3}}{v(z)^{2}}}, \\
\varphi_{n} &=& {\rm e}^{g(z)}\tilde\varphi_{n}, \quad\quad\quad g(z) = \phi(z)+ \log{z},
\end{eqnarray}
which eliminate terms involving the first derivative of the fields, $\pi_{n}'$ and $\varphi_{n}'$. After reverting back to the notation $\tilde\pi\rightarrow\pi$ and $\tilde\varphi\rightarrow\varphi$ the equations of motion become
\begin{eqnarray}
&& -\pi_{n}'' + \left(\frac{\phi'^{2}}{4} - \frac{\phi''}{2}-\frac{\phi'v'}{v} + \frac{3\phi'}{2z} + \frac{15}{4z^{2}}-\frac{3 v'}{v z}+\frac{v''}{v}-m_{n}^{2}\right)\pi_{n}\nonumber\\
&&\quad\quad = -m_{n}^{2}\frac{v^{2}R^{2}}{z^{2}}\varphi_{n}, \label{equSchexppi}\\
&& -\varphi_{n}'' + \left(\frac{\phi'^{2}}{4} - \frac{\phi''}{2}+\frac{\phi'}{2z} + \frac{3}{4 z^{2}} + \frac{g_{5}^{2}v^{2}R^{2}}{z^{2}}\right)\varphi_{n} = g_{5}^{2} \pi_{n}, \label{equSchexpphi}
\end{eqnarray}
where ($'$) indicates the derivative with respect to $z$.

\paragraph{Linear Representation}
 
When considering the linear representation of the pseudoscalar field (\ref{equXl}), we find quadratic and quartic $\pi$ terms that were not explicitly present in the exponential representation. After making the appropriate substitutions, we find that
\begin{eqnarray}
\mathcal{L}_l &=& -\frac{1}{2} {\rm e}^{-\phi}\sqrt{-g}\Big(g^{\mu\nu}\partial_{\mu}\pi\partial_{\nu}\pi + g^{zz}\partial_{z}\pi\partial_{z}\pi - 2 v g^{\mu\nu}\partial_{\mu}\pi\partial_{\nu}\varphi + m_{X}^{2}\pi^{2} -\frac{\kappa}{2}v^{2}\pi^{2} \nonumber\\
&+& g^{\mu\nu}v^{2} \partial_{\mu}\varphi\partial_{\nu}\varphi + \frac{g^{\mu\nu} g^{zz}}{g_{5}^{2}}\partial_{z}\partial_{\mu}\varphi\partial_{z}\partial_{\nu}\varphi\Big).
\end{eqnarray} 
Once again, we derive two coupled equations. Varying with respect to $\phi$ produces a result similar to $X_{e}$ with the exception of factors of the VEV in the mixing term, giving
\begin{equation}\label{equphi}
{\rm e}^{\phi}\partial_{z}\left(\frac{{\rm e}^{-\phi}}{z}\partial_{z}\varphi_{n}\right) + \frac{g_{5}^{2}R^{2} v}{z^{3}}\left(\pi_{n} - v \varphi_{n}\right) = 0.
\end{equation}
Varying with respect to $\pi$ gives the second equation of the linear representation,
\begin{equation}\label{equpi}
z^{3}{\rm e}^{\phi}\partial_{z}\left(\frac{{\rm e}^{-\phi}}{z^{3}}\partial_{z}\pi_{n}\right) - \left(\frac{m_{X}^{2}}{z^{2}} - \frac{\kappa R^{2} v^{2}}{2 z^{2}}\right)\pi_{n}+m_{n}^{2}\pi_{n} = m_{n}^{2}v\varphi_{n}.
\end{equation}
We can express (\ref{equphi}) and (\ref{equpi}) in a Schr\"odinger-like form with the substitutions,
\begin{eqnarray}
\pi_{n} &=& {\rm e}^{f}\tilde{\pi_{n}},\quad\quad\quad f = \frac{\phi}{2} + \frac{3}{2} \log{\frac{z}{R}}, \\
\varphi_{n} &=& {\rm e}^{g}\tilde{\varphi_{n}},\quad\quad\quad g = \frac{\phi}{2} + \frac{1}{2} \log{\frac{z}{R}}.
\end{eqnarray}
Simplifying the equations and reverting back to the notation $\tilde{\pi_{n}}\rightarrow\pi_{n}$ and $\tilde{\varphi_{n}}\rightarrow\varphi_{n}$ for simplicity, we find
\begin{eqnarray}
&&-\varphi_{n}'' + \left(\frac{\phi'^{2}}{4} - \frac{\phi''}{2} + \frac{3}{4z^{2}} + \frac{\phi'}{2 z} + \frac{g_{5}^{2}R^{2}v^{2}}{z^{2}}\right)\varphi_{n} = \frac{g_{5}^{2}R v}{z}\pi_{n},\label{equSchphi} \\
&&-\pi_{n}'' + \left(\frac{\phi'^{2}}{4} - \frac{\phi''}{2} + \frac{3}{4z^{2}} + \frac{3\phi'}{2 z} - \frac{\kappa R^{2}v^{2}}{2z^{2}} - m_{n}^{2} \right)\pi_{n} = -m_{n}^{2}\frac{v\,R}{z}\varphi_{n}.  \label{equSchpi}
\end{eqnarray}

\paragraph{Representation Equivalence}
 
The pseudoscalar field representation should not affect the physical results obtained from the model. Examining the two sets of coupled equations in each representation, we see that neither (\ref{equOne}) nor (\ref{equTwo}) contains an explicit dependence on $\kappa$, whereas (\ref{equpi}) does. Although $\kappa$ does not appear explicitly in the exponential representation, the dependence enters through the function $v(z)$.

We begin by expanding $X_{e}$,
\begin{eqnarray}
X_{e} &=& \left(\frac{v}{2} + S\right)(1 + 2 i \pi_{e} + \ldots)\nonumber \\
&=& \frac{v}{2} + S + i\pi_{e} v. \label{equXexpand}
\end{eqnarray}
Comparing (\ref{equXexpand}) to (\ref{equXl}), we surmise that $\pi_{e} v(z)\rightarrow \pi_{l}$ relates the two representations. Let us substitute $\pi_{e}\rightarrow \pi_{l}/v(z)$ into the equations of motion of the exponential representation and attempt to obtain the equations of motion of the linear representation. The substitution into (\ref{equTwo}) is trivial; it yields
\begin{equation}
{\rm e}^{\phi}\partial_{z}\left(\frac{{\rm e}^{-\phi}}{z}\partial_{z}\varphi\right) + \frac{g_{5}^{2}v}{z^{3}}(\pi_{l}- v\varphi) = 0,
\end{equation}
which is equivalent to (\ref{equphi}) as expected. Showing the equivalence of the other two equations requires a bit more work. First, we substitute for $\pi_{e}$ in (\ref{equOne}) and then simplify the expression,
\begin{equation} 
\frac{z^{3}{\rm e}^{\phi}}{v}\partial_{z}\left(\frac{{\rm e}^{-\phi}v^{2}}{z^{3}}\left(\frac{\pi_{l}'}{v} - \frac{\pi_{l} v}{v^{2}}\right)\right) + m_{n}^{2}(\pi_l -v\varphi) = 0,
\end{equation}
which becomes
\begin{equation}
\pi_{l}'' - \left(\phi' + \frac{3}{z}\right)\pi_{l}' - \frac{\pi_l}{v}\left(v'' - \phi'v' - \frac{3}{z}v'\right) + m_{n}^{2}(\pi_l -v\varphi) = 0. \label{equOnemid}
\end{equation}
Substituting (\ref{equvsimple}), which does not depend on the representation, into (\ref{equOnemid}), we find
\begin{equation}
\pi_{l}'' - \left(\frac{3}{z}+\phi'\right)\pi_{l}' + \left(\frac{3}{z^{2}} + \frac{\kappa R^{2}v^{2}}{2 z^{2}}\right)\pi_{l} + m_{n}^{2}\left(\pi_{l} -  v\varphi\right) = 0,
\end{equation}
which is the same as the equation of motion of the linear representation (\ref{equpi}). In a similar way, this equivalence can be shown by starting with the linear representation and substituting $\pi_{l} = v(z)\pi_{e}$.

\subsubsection{Pseudoscalar Mass Eigenvalues}
\label{secPME}

We investigate two ways to calculate the pseudoscalar eigenvalues $m_{n}^{2}$. Rearranging and eliminating the longitudinal component $\phi$ is one strategy outlined in \cite{Sui:2009xe} and is briefly presented here. Alternatively, Appendix \ref{appMatrix} contains a numerical routine we use to solve the set of coupled equations. Using this method, we find that $\pi_{e}$, the ratio of $\pi_{l}$ and $v(z)$, is extremely sensitive to boundary conditions, the reason being that $v(z)$ goes to zero as $z$ goes to zero.  Resolving the eigenvalues is difficult and subject to significant numerical error. Fortunately, we have shown explicitly that physical results do not depend on the particular representation of the pseudoscalars. Therefore we determine the eigenvalues in the linear representation. 

First, we follow the method of \cite{Sui:2009xe}. We manipulate (\ref{equOne}) and (\ref{equTwo}) to eliminate the $\varphi$ field. Adding (\ref{equOne}) and (\ref{equTwo}) yields
\begin{equation}\label{equOneMod}
g_{5}^{2}a^{2}v^{2}\partial_{z}\pi_{n} = m_{n}^{2}\partial_{z}\varphi_{n} \, .
\end{equation}
We then use (\ref{equOneMod}) to replace the first term in (\ref{equTwo}) and solve for $\varphi$,
\begin{eqnarray} 
\varphi_{n} &=& \frac{1}{g_{5}^{2}h(z)}\partial_{z}\left[a \,{\rm e}^{-\phi}\left(-\frac{g_{5}^{2}}{q^{2}}a^{2}v^{2}\partial_{z}\pi_{n}\right)\right] + \pi_{n},\nonumber \\
m_{n}^{2}\partial_{z}\varphi_{n} &=& \partial_{z}\left(h(z)^{-1}\partial_{z}(h(z)\partial_{z}\pi_{n})\right) + m_{n}^{2}\partial_{z}\pi_{n} 
\end{eqnarray}
where $h(z) = a(z)^{3}v^{2} {\rm e}^{-\phi}$. Using (\ref{equOneMod}) again, we denote $\partial_{z}\pi\rightarrow \tilde\pi$ and rearrange to find the eigenvalue equation
\begin{equation}\label{equTildepi}
-\partial_{z}[h^{-1}\partial_{z}(h \tilde\pi_{n})] + g_{5}^{2}a ^{2}v^{2}\tilde\pi_{n} = m_{n}^{2} \tilde\pi_{n} \,.
\end{equation}
This can be put into the Schr\"odinger-like form by substituting $\Pi = \tilde\pi/\sqrt{h(z)}$.  Then (\ref{equTildepi}) becomes
\begin{equation}\label{equMassEigen}
-\Pi_{n}'' + V(z)\Pi_{n} = m_{n}^{2} \Pi_{n},
\end{equation}
where the potential takes the form,
\begin{eqnarray}
V(z) &=& \frac{3h'^{2}}{h^{2}}-\frac{h''}{2h}+\frac{g_{5}^{2}R^{2}v^{2}}{z^{2}}\nonumber \\
&=& \frac{3}{4 z^{2}} -\frac{3 v'}{z v} + 2\frac{v'^{2}}{v^{2}} + \frac{3\phi'}{2 z} - \frac{v'\phi'}{v} + \frac{\phi'^{2}}{4} - \frac{v''}{v} + \frac{\phi''}{2} + \frac{g_{5}^{2}R^{2}v^{2}}{z^{2}}. \label{equPot}
\end{eqnarray}
Solving (\ref{equMassEigen}) using a standard shooting method gives the mass spectrum shown in Table \ref{tblmass}. There is no light pseudo-Goldstone boson and no large mass gap between the first two eigenvalues. What we have done is taken two second-order differential equations and reduced them to a third-order differential equation and found the eigenvalues of $\partial_{z}\pi$. We do not recover mass values that match the experimental mass spectra using this method. 

\begin{figure}[h!]
\begin{center}
\includegraphics[scale=0.42]{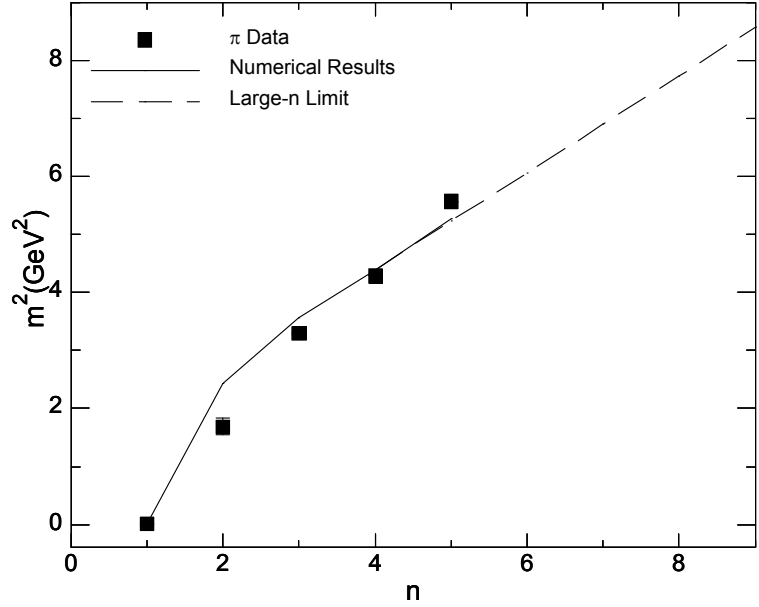}
\caption{The mass spectrum calculated in the AdS/QCD model is plotted along with the experimental data \cite{pdg}. The eigenvalues display two characteristics matching the QCD pion spectrum: (1) light ground state and (2) a large gap between the ground state and the first excited state. The large-$n$ mass trajectory clearly follows our calculated eigenvalues from $n\approx 4$ when our numerical routine inadequately follows the oscillations of the higher eigenfunctions.}
\label{linEigen}
\end{center}
\end{figure}

The numerical routine described in Appendix \ref{appMatrix} calculates the mass eigenvalues of (\ref{equSchphi}) and (\ref{equSchpi}), which are then plotted in Figure \ref{linEigen} and listed in Table \ref{tblmass}. Solving the set of equations directly produces a mass spectrum with a light pseudo--Nambu-Goldstone boson and mass eigenstates that match well with the observed radial pion excitations. These results show that eliminating one of the fields from the system of second-order differential equations also eliminates information from the mass spectrum. Thus, the eigenvalues of $\partial_{z}\pi_{e}$ do not match those of $\pi_{l}$ or $\pi_{e}$, at least not in this model. The validity of equating the two sets of eigenvalues in \cite{Sui:2009xe} may need to be reassessed.

\begin{table}[h!]
\begin{center}
\begin{tabular}{| c || c | c | c | c |}
\hline
n & $\pi$ Data (MeV) 	& $\pi_{l}$ (MeV) 		& Large-$n$ $\pi_{l}$ 	& $\partial_{z}\pi_{e}$ (MeV) \\
\hline
\hline
1 & 140  		& 143  				& -			& 1440\\
\hline
2 & 1300 $\pm$ 100 	& 1557  			& -			& 1706 \\
\hline
3 & 1816 $\pm$ 14 	& 1887  			& -			& 1925\\
\hline
4 & 2070* 		& 2095 				& - 			& 2117\\
\hline
5 & 2360* 		& 2298 				& 2245 			& 2290\\
\hline
6 &  -   		& - 				& 2403 			& 2451\\
\hline
7 &  -   		& -				& 2551 			& 2601\\
\hline
\end{tabular}
\caption{The observed masses \cite{pdg} and calculated masses using the linear representations. The large-$n$ limit solutions are valid from $n\approx 4$. From that point onward, the numerical method used becomes increasingly inaccurate and tends to skew the $\pi_{l}$ eigenvalues to larger values than are expected from the linear behavior. The eigenvalues found using the method of \cite{Sui:2009xe} are also shown. *=Appears strictly in the further states of \cite{pdg}.}
\label{tblmass}
\end{center}
\end{table}

For large-$n$ excitations, the technique described in Appendix \ref{appMatrix} runs into boundary condition problems for $n\geq 4$. As the number of oscillations in the eigenfunctions increases for higher $n$ modes, the routine finds eigenvalues that are skewed to larger values. To uncover the correct asymptotic behavior for large $n$, we take the large-$z$ limit of (\ref{equSchphi}) and (\ref{equSchpi}). As $n$ increases, the eigenfunction is largely determined by the behavior of the effective potential at large $z$. At large $z$, the dilaton and tachyon behave as
\begin{eqnarray}
v(z) &=&  \Gamma z,\\
\phi(z) &=& \mu^{2} z^{2}.
\end{eqnarray}
To take the large-$z$ limit of both representations, we introduce a new dimensionless parameter, $\xi=\mu z$, and expand in $\xi$. In the exponential representation, we find that (\ref{equSchexppi}) and (\ref{equSchexpphi}) become
\begin{eqnarray}
-\pi_{k}'' + \xi^{2}\pi_{k}  &=& \frac{m_{k}^{2}}{\mu^{2}}\left(\pi_{k} - \Gamma\varphi_{k}\right), \label{equpiExpSHO}\\
-\varphi_{k}'' + \xi^{2}\varphi_{k} &=& \frac{g_{5}^{2}}{\mu^{2}}\left(\pi_{k} - \Gamma\varphi_{k}\right). \label{equphiExpSHO}
\end{eqnarray}
Similarly, in the linear representation the expansion of (\ref{equSchphi}) and (\ref{equSchpi}) at large $\xi$ yields the eigenvalue equations
\begin{eqnarray}
-\pi_{k}'' + \xi^{2} \pi_{k}&=& \left(\frac{\kappa\Gamma^{2}}{2 \mu^{2}} -2 + \frac{m_{k}^{2}}{\mu^{2}}\right)\pi_{k} - \frac{m_{k}^{2}\Gamma}{\mu^{2}}\varphi_{k}, \label{equpiSHO} \\
-\varphi_{k}'' + \xi^{2} \varphi_{k} &=& \frac{g_{5}^{2}\Gamma}{\mu^{2}}\left(\pi_{k} - \Gamma\varphi_{k}\right),\label{equphiSHO}
\end{eqnarray}
where ($'$) indicates differentiation with respect to $\xi$. Each set of equations appears to describe a pair of simple harmonic oscillators, the equations of motion of which are
\begin{eqnarray}
-\varphi_{k}'' + \xi^{2} \varphi_{k} &=& (2k+1)\varphi_{k}, \label{equphiSHOG} \\
-\pi_{k}'' + \xi^{2}\pi_{k} &=& (2k+1)\pi_{k}\quad\quad k=0,1,\ldots. \label{equpiSHOG}
\end{eqnarray}
It is a reasonable assumption that $\varphi_{k} = c_{k}\pi_{k}$; it ensures that (\ref{equpiExpSHO}), (\ref{equphiExpSHO}), (\ref{equpiSHO}), and (\ref{equphiSHO}) have solutions. Using the form of (\ref{equphiSHOG}) and (\ref{equpiSHOG}) to solve for $c_k$ and $m_{k}^{2}$ in both representations, we find
\begin{eqnarray}
c_{k} &=& \frac{g_{5}^{2}}{g_{5}^{2}\Gamma^{2} +(2k+1)\mu^{2}}, \label{equexpcn}\\
m_{k}^{2} &=& (2k+1)\mu^{2} + g_{5}^{2}\Gamma^{2}
\end{eqnarray}
for the exponential representation and 
\begin{eqnarray}
c_{k} &=& \frac{g_{5}^{2}\Gamma}{g_{5}^{2}\Gamma^{2} +(2k+1)\mu^{2}}, \label{equlinearcn}\\
m_{k}^{2} &=& \frac{\left((2k+3)\mu^{2} -\frac{1}{2}\kappa\Gamma^{2}\right)\left(g_{5}^{2}\Gamma^{2} + (2k+1)\mu^{2}\right)}{(2k+1)\mu^{2}} \label{equlinearmn}
\end{eqnarray}
for the linear representation. 

So far, we have neglected an important fact:  $z\geq 0$. The eigenfunctions $\phi$ and $\pi$ describe \textit{half}-harmonic oscillators and contain only half the modes that full harmonic oscillators do; therefore, we must take $k\rightarrow 2k$. The mass eigenvalues for large $n$, where $n=k+1$, in both representations then become 
\begin{equation} \label{equMass4Both}
m_{n}^{2} = (4 n-3)\mu^{2} + g_{5}^{2}\Gamma^{2}\quad\quad n=4,5\ldots,
\end{equation}
which are also listed in Table \ref{tblmass} and plotted in Figure \ref{linEigen}. Combining (\ref{equMass4Both}) and the numerical technique, we obtain all the pseudoscalar eigenvalues. By simple investigation, we find that the large-$n$ eigenvalues should be trusted over the ones found with the numerical routine for $n \geq 4$.

\subsection{Vector sector} 
\label{secVector}

The soft-wall model with the quadratic dilaton describes the $\rho$ meson spectrum surprisingly well \cite{Karch:2006pv}. In fact, since the scalar field VEV does not couple to the vector sector, any dilaton with the behavior (\ref{dilatonlz}) causes the vector mass spectrum to exhibit linear trajectories for the higher resonances. Examining the QCD experimental data, one sees that the $\rho$ mass spectrum exhibits linear behavior around 
$\rho(1465)$ or $\rho(1720)$; therefore, one expects the appropriately modified dilaton will 
only affect lower lying resonances as higher eigenfunctions localize towards the IR and are less dependent on 
small-$z$ behavior.  

From the action (\ref{action1}), the vector equation of motion is derived exactly as in Section \ref{secbkVector}. Using the KK decomposition, $V_{\mu}(x,z) = \sum_{n=0}^{\infty}V_{n}(z)\mathcal{V}_{\mu,n}(x)$ and (\ref{equProca}), we obtain 
\begin{equation}\label{equVnonSch}
V_{n}'' -\phi' V_{n} - \frac{1}{z} V_{n} + m_{n}^{2} V_{n} = 0,
\end{equation}
using the axial gauge $V_5=0$. Using Appendix \ref{appSchTransform}, we transform (\ref{equVnonSch}) to a Schr\"{o}dinger form
\begin{equation} \label{vschroedinger}
-\partial_z^{2}v_{n}+\left(\frac{1}{4}\omega'^{2} - \frac{1}{2}\omega''\right) v_{n}=m_{V_n}^{2}v_{n}, 
\end{equation}
where 
\begin{equation}
V_{n}={\rm e}^{\omega/2} v_{n}, \quad\quad\quad \omega=\phi(z)+\log{z}.
\end{equation}
Fully simplified, (\ref{vschroedinger}) becomes
\begin{equation}\label{equVgeneral}
-v_{n}'' +\left(\frac{1}{4}\phi'^{2} + \frac{1}{2 z}\phi' - \phi'' + \frac{3}{4z^{2}}\right)v_{n} = m_{V_n}^{2}v_{n},
\end{equation} 
with the boundary conditions,
\begin{eqnarray}
\lim_{z_0\rightarrow 0} v_n(z_0)=0 \\
\partial_z v_{n}(z\rightarrow\infty)=0.
\end{eqnarray}

\begin{figure}[h!]
\begin{center}
\includegraphics[scale=0.42,angle=90]{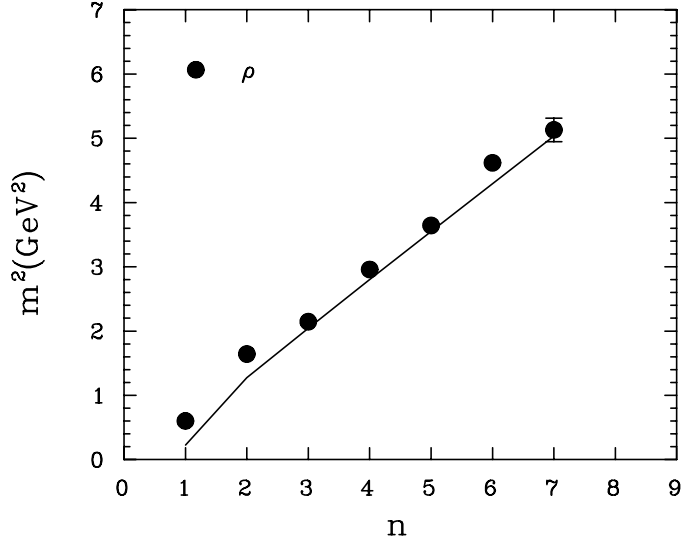}
\caption{Comparison of the predicted vector mass eigenvalues using the $\tanh$ form (\ref{arcv}) of $v(z)$ (solid) with the QCD $\rho$ mass spectrum \cite{pdg}. }
\label{VectorMasses}
\end{center}
\end{figure}

\begin{table}[h!]
\begin{center}
\begin{tabular}{|c||c|c|}
\hline
$n$ & $\rho$ experimental (MeV) & $\rho$ model (MeV) \\
\hline
\hline
1 & $775.5 \pm 1$ & $475$ \\
\hline
2 & $1282 \pm 37$ & $1129$ \\
\hline
3 & $1465 \pm 25$ & $1429$ \\
\hline
4 & $1720 \pm 20$ & $1674$ \\
\hline
5 & $1909 \pm 30$ & $1884$ \\
\hline
6 & $2149 \pm 17$ & $2072$ \\
\hline
7 & $2265 \pm 40$ & $2243$ \\
\hline
\end{tabular}
\caption{The experimental and predicted values of the vector meson masses.}
\label{vectormasses}
\end{center}
\end{table}
Since the dilaton specified in (\ref{phieqn}) is modified for small $z$, there is a change in the slope of the mass spectrum around $n=2$ which matches the behavior of the experimental data. The numerical vector mass spectrum is compared to the experimental data in Figure~\ref{VectorMasses} and listed in Table \ref{vectormasses}. While the prediction for the $\rho(775)$ mass is low, the rest of the vector meson masses are in reasonable agreement with experiment.  Most likely the agreement with the $\rho(775)$ could be improved upon by using a parameterization of $v(z)$ which rises more rapidly to its asymptotic value at large $z$.  Nevertheless, the purpose of this analysis is to explore the consequences of the improved CSB in soft-wall AdS/QCD models, not just to fit data.

\subsection{Axial-vector sector} 
\label{secAxial}

Unlike the vector field, the axial-vector couples to the scalar field VEV, producing a $z$-dependent mass term in its equation of motion. Similarly to the vector field case, the equation of motion assuming $A_\mu(x,z)={\cal A}_\mu^n(x) A_n(z)$ using the axial gauge $A_5=0$ is given by 
\begin{equation} \label{swaxial}
-\partial_z^{2}A_n+\omega'\partial_z A_n+g_5^2 \frac{R^2}{z^2} v^2(z) A_n=m_{A_n}^{2}A_n.
\end{equation}
Using the same transformation in Appendix \ref{appSchTransform} as for the vector field, where $A_n={\rm e}^{\omega/2} a_n$, one can express (\ref{swaxial}) as
\begin{eqnarray} 
-\partial_z^{2}a_{n}+\left(\frac{1}{4}\omega'^2 - \frac{1}{2}\omega'' +g_5^2 \frac{R^2}{z^2} v^2(z) \right)a_n &=& m_{A_n}^2 a_n,\nonumber\\
 -a_{n}'' + \left(\frac{1}{4}\phi'^{2} + \frac{1}{2z}\phi' - \phi'' + \frac{3}{4 z^{2}} + g_5^2 \frac{R^2}{z^2} v^2(z) \right)a_n &=&  m_{A_n}^2 a_n.\label{swaxialtrans}
\end{eqnarray}
The expression (\ref{swaxialtrans}) for the axial-vector field matches that of the vector field except for the 
additional term, $g_5^2 v^2(z) R^2/z^2$. Because of this $z$-dependent mass term, equation (\ref{swaxialtrans}) is difficult to solve analytically and again requires a numerical solution. Using the shooting method with the boundary conditions,
\begin{eqnarray}
 \lim_{z_0\rightarrow 0} a_n(z_0)=0, \\
 \partial_z a_{n}(z\rightarrow\infty)=0,
\end{eqnarray}
 the axial-vector meson spectrum is obtained for fixed values of $\mu$, $m_q$, $\sigma$, and $\kappa$.

\begin{figure}[h!]
\begin{center}
\includegraphics[scale=0.42,angle=90]{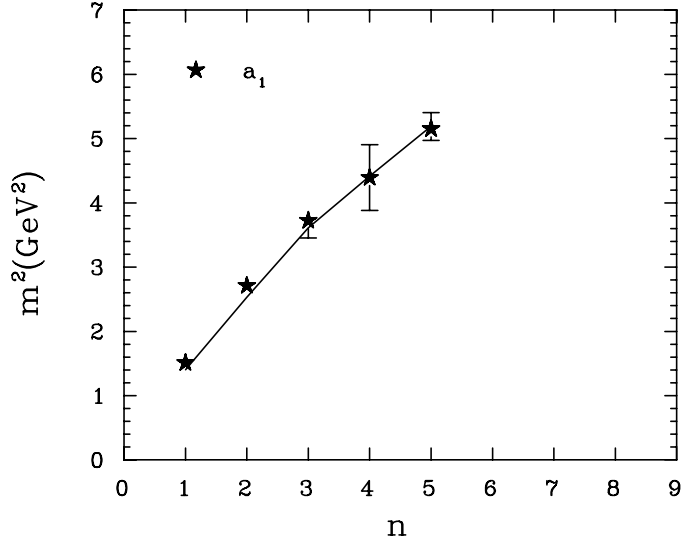}
\caption{Comparison of the numerical results for the axial-vector mass eigenvalues using the $\tanh$ form (\ref{arcv}) of $v(z)$ (solid)
with the QCD $a_1$ mass spectrum \cite{pdg}.
}
\label{AxialMasses}
\end{center}
\end{figure}

The linear behavior of $v(z)$ as $z\rightarrow\infty$ leads to a constant shift between the vector and axial-vector spectra at high mass values.  Comparing the equations of motion (\ref{equVgeneral}) and (\ref{swaxialtrans})
for these fields one finds the asymptotic behavior
\begin{equation}\label{deltam2}
\Delta m^2 \equiv \left(m_{A_n}^2 - m_{V_n}^2\right)_{n \rightarrow \infty}
= g_5^2 \frac{R^2}{z^2} v^2(z \rightarrow \infty) = \frac{4g_5^2 \mu^{2}}{\kappa}.
\end{equation}
Together with the slope $\mu^{2}$, the mass difference, $\Delta m^{2}$, indicates that $\kappa\sim 30$, although (\ref{deltam2}) is only valid as $n\rightarrow\infty$. The best global fit to all the data suggests $\kappa = 15$, which is used in this work. The results of our analysis are plotted in Figure \ref{AxialMasses} and displayed in Table~\ref{avmasses}. The $a_1(1260)$ resonance is predicted to within 5$\%$ and there is good agreement with the higher resonances of $a_1$. 

Note that from (\ref{deltam2}), $\Delta m^2 >0$ implies that $\kappa >0$, which means
that the potential in (\ref{action1}) is unbounded from below.  To address the stability of the gravity-dilaton background requires a complete fluctuation analysis generalizing the work in \cite{Breitenlohner:1982bm}. Even though this is beyond the scope of the present work it does suggest that higher-order terms may be needed for stability.

\begin{table}[h!]
\begin{center}
\begin{tabular}{|c||c|c|}
\hline
$n$ & $a_1$ experimental (MeV) & $a_1$ model (MeV) \\
\hline
\hline
1 & $1230 \pm 40$ & $1185$ \\
\hline
2 & $1647 \pm 22$ & $1591$ \\
\hline
3 & $1930^{+30}_{-70}$ & $1900$ \\
\hline
4 & $2096 \pm 122$ & $2101$ \\
\hline
5 & $2270^{+55}_{-40}$ & $2279$ \\
\hline
\end{tabular}
\caption{The experimental and predicted values of the axial-vector meson masses.}
\label{avmasses}
\end{center}
\end{table}

\section{Other Dynamics} \label{secDynamics}

To further confirm the correspondence between our AdS/QCD model and large-$N_{c}$ QCD, we take a closer look at the Gell-Mann--Oakes--Renner (GOR) relation, the coupling $g_{\rho_{n}\pi\pi}$, and the form factor $F_{\pi}(q)$. We find that our model satisfies the GOR relation naturally and matches $F_{\pi}$ data quite well, but produces values for $g_{\rho_{n}\pi\pi}$ smaller than some soft-wall models and the experimental value. In both Sections \ref{secGOR} and \ref{secPion}, we need the axial-pion mixing equations that come from Section \ref{secPseudo},
\begin{eqnarray}
\left[ {\rm e}^{\phi} \partial_{z}\left(\frac{{\rm e}^{-\phi}}{z}\partial_{z}A_{\mu}(q,z)\right)-\frac{q^{2}}{z}A_{\mu}(q,z)-\frac{g_{5}^{2} L^2 v^2(z)}{z^{3}}A_{\mu}(q,z)\right]_{\perp}=0,\label{Aperp} \\
{\rm e}^{\phi} \partial_{z}\left(\frac{{\rm e}^{-\phi}}{z}\partial_{z}\varphi(z)\right)
+\frac{g_{5}^{2} R^2 v^2(z)}{z^{3}}(\pi_{e}(z) - \varphi(z))=0, \label{mixed1}\\
q^{2}\partial_{z}\varphi(z) + \frac{g_{5}^2 R^2 v(z)^{2}}{z^{2}} \partial_{z}\pi_{e}(z) =0,\label{mixed2}
\end{eqnarray}
where we have switched to momentum space and separated out the source from the axial-vector field, so that
\begin{equation}
A(0,R_{0})\Big|_{R_{0}\rightarrow 0} = 1.
\end{equation}
Setting $q=0$, we find several facts,
\begin{eqnarray}
A(0,z) = \varphi(z) - \pi(z), \\
\pi(z) =  \text{constant},
\end{eqnarray}
which become useful in our work.

\subsection{Gell-Mann--Oakes--Renner Relation} 
\label{secGOR}

Using the established equivalence between the exponential and linear representations, $\pi_e = \pi_l/v(z)$, and inserting it into (\ref{equOneMod}), we obtain
\begin{equation}
\frac{g_5^2L^2v^2}{z^2} \partial_z\left(\frac{\pi_l}{v}\right) = m_\pi^2\partial_z\phi \,.
\end{equation}
Following the method of \cite{Erlich:2005qh}, we construct a perturbative solution in $m_{\pi}$ where $\phi(z) = A(0,z)-1$ and use 
\begin{equation}\label{equ:fpi}
f_\pi^2=\left. -R \frac{\partial_z A(0,z)}{g_5^2z}\right |_{z\rightarrow0}. 
\end{equation}
 From this relation, it follows that
\begin{equation}
\pi(z)=m_\pi^2\,v(z)\int_0^z du\, \frac{u^3}{v^2(u)} \frac{\partial_z A(0,u)}{g_5^2u} \,.
\end{equation}
The function $u^3/v^2(u)$ is significant only at small values of $u\sim\sqrt{m_q/\sigma}$, where we may use (\ref{equ:fpi}) to relate the derivative of $A(0,u)$ to the pion decay constant, so that
\begin{equation} \label{equPreGOR}
\frac{\pi_l}{v}=-\frac{m_\pi^2 f_\pi^2}{2m_q\sigma} \,.
\end{equation}
We find that $\pi_l=-v(z)$ solves the axial-vector field equations (\ref{Aperp}), (\ref{mixed1}), and (\ref{mixed2}) in the region of small $z$ and as $q\rightarrow 0$. As a result, (\ref{equPreGOR}) becomes the expected Gell-Mann--Oakes--Renner (GOR) relation,
\begin{equation}
2m_q\sigma=m_\pi^2 f_\pi^2 \,.
\end{equation}
This perturbative behavior for $\pi_{e}$ and $\pi_l$ justifies the use of Neumann and Dirichlet boundary conditions, respectively, such that
\begin{equation}
\pi_{e}(0) = -1 \,, \quad\quad\quad \pi_{l}(0) = v(0) = 0 \,.
\end{equation}

\begin{figure}[h!]
\begin{center}
\includegraphics[scale=0.42]{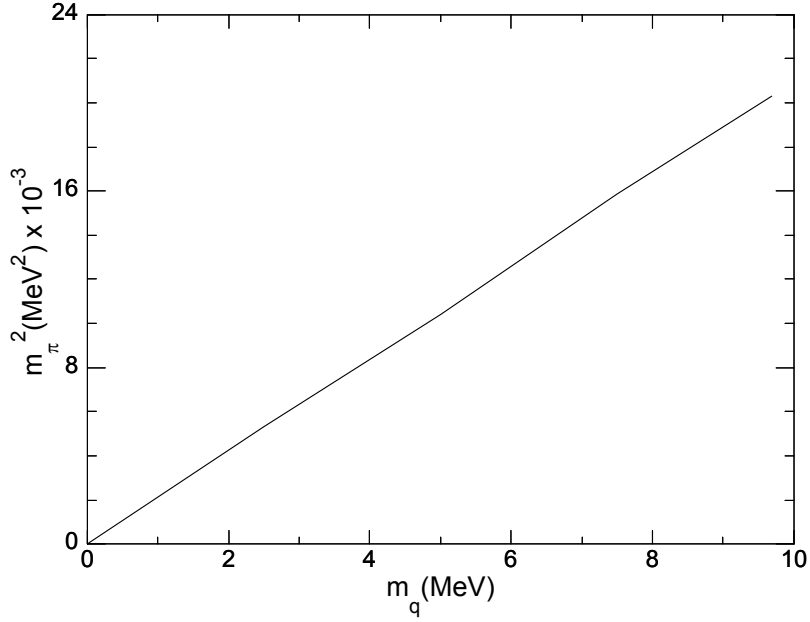}
\caption{Plot of $m_{\pi}^{2}$ versus $m_{q}$ produces a straight line from which the pion decay constant $f_{\pi}$ is calculated using (\ref{equGORform2}).}
\label{figGOR}
\end{center}
\end{figure}

The ratio of $m_{q}$ and $m_{\pi}^{2}$ should be a constant following the GOR relation,
\begin{equation} \label{equGORform2}
\frac{m_{q}}{m_{\pi}^{2}} = \frac{f_{\pi}^{2}}{2\sigma}.
\end{equation}
We solve the pair of coupled differential equations (\ref{equSchphi}) and (\ref{equSchpi}) for the ground-state pseudoscalar mass $m_{\pi}$ using differing values of $m_{q}$ to ensure that the numerical routine of Appendix \ref{appMatrix} respects the GOR relation. The results are plotted in Figure \ref{figGOR}. We see a linear relationship between $m_{q}$ and $m_{\pi}^{2}$, indicating that (\ref{equGORform2}) is indeed satisfied. The slope of the line in Figure \ref{figGOR} implies $f_{\pi} \approx 90$ MeV, a result consistent with the input parameters. 

\subsection{Pion Coupling}\label{secPion}

\begin{figure}[h!]
\begin{center}
\includegraphics[scale=0.45,angle=90]{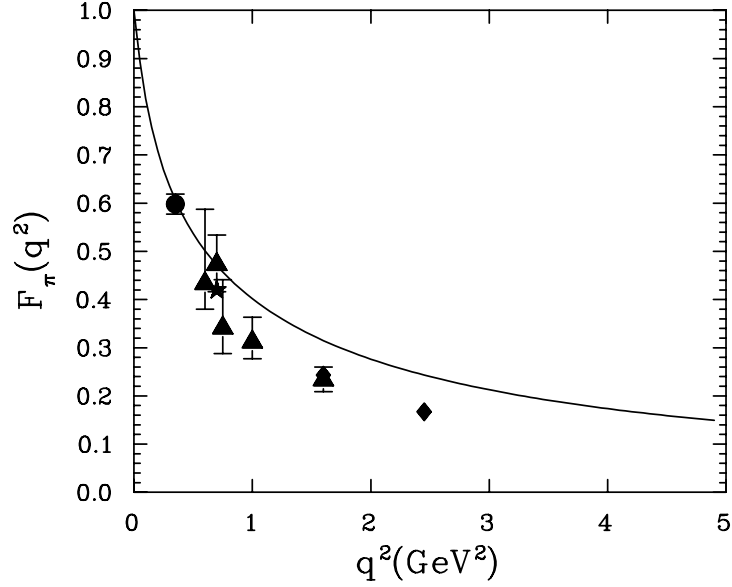}
\caption{The line shows the predicted space-like behavior of the pion form factor $F_{\pi}(q^2)$ which is compared to the experimental data obtained from \cite{Kwee:2007nq}. The triangles are data from DESY, reanalyzed by \cite{Tadevosyan:2007yd}. The diamonds are data from Jefferson Lab \cite{Horn:2006tm}. The circles \cite{Ackermann:1977rp} as well as the star \cite{Brauel:1977ra} are also data obtained from DESY.}
\label{Fpiplot}
\end{center}
\end{figure}

To investigate aspects of the pion coupling, we need the value of $f_{\pi}$ from (\ref{equ:fpi}). For our model where $m_q = 9.75$ MeV, $\kappa = 15$, and $\mu^{2} = 0.1831$ GeV$^2$, we find that $f_\pi=92.4$ MeV. In order to calculate the vector-pion-pion coupling $g_{\rho_{n}\pi\pi}$, we need the $V\pi\pi$ interaction terms. Using our action (\ref{action1}) and the steps and definitions in \cite{Kwee:2007nq}, we find that 
\begin{equation} \label{grpp}
g_{\rho_{n}\pi\pi} =  \frac{1}{f_{\pi}^2} \int dz \,V_n(z) {\rm e}^{-\phi(z)} \left(\frac{1}{g_{5} z}(\partial_{z}\varphi(z))^{2} + \frac{g_5 R^2 v^2(z)}{z^{3}}(\pi(z)-\varphi(z))^{2}\right)~,
\end{equation}
where $V_n$ are the KK modes of the $\rho$ meson. They are normalized as
\begin{equation}
\int dz\, \frac{{\rm e}^{-\phi(z)}}{z} V_n(z) V_m(z) = \delta_{mn}.
\end{equation}
The functions $\pi(z)$ and $\varphi(z)$ must be determined from the system of equations for the axial-vector and pion as given in (\ref{Aperp}), (\ref{mixed1}), and (\ref{mixed2}) \cite{Kwee:2007nq}. The evaluation is made easier because we can approximate by setting $\varphi(z) = A_{\mu}(0,z) - 1$ to first order. Previous soft-wall models \cite{Kwee:2007nq} have obtained values a factor of two smaller than the experimental result of $g_{\rho\pi\pi}\approx 6$. Similarly, our calculations produce a small value $g_{\rho\pi\pi} =2.89$.  Once we calculate $g_{\rho_{n}\pi\pi}$, the space-like pion form factor can easily be determined from a sum over vector meson poles,
\begin{equation} \label{Fpisum}
F_{\pi}(q^{2}) = \sum_{n=1}^{\infty}\frac{f_{n}g_{\rho_{n}\pi\pi}}{q^{2}+m_{V_n}^{2}},
\end{equation} 
where $f_{n}$ are the decay constants of the vector modes. However, (\ref{Fpisum}) converges much too slowly and numerically it is much better to use the expression in terms of the vector and axial-vector bulk-to-boundary propagators as in \cite{Kwee:2007nq}
\begin{equation}
F_{\pi}(q^{2}) =\int{dz\, {\rm e}^{-\phi(z)} \frac{V(q,z)}{f_{\pi}^{2}}\left(\frac{1}{g_5^2 z} (\partial_{z}\varphi(z))^{2} +\frac{v^2(z)}{z^{3}}(\pi(z)-\varphi(z))^2\right)}.
\end{equation}
The results of our $F_{\pi}(q^{2})$ calculation are plotted in Figure \ref{Fpiplot}, and shows a slight improvement in matching the experimental values compared to that obtained in \cite{Kwee:2007nq}. It is apparent that the QCD behavior is mimicked reasonably well, beyond that expected from a simple soft-wall AdS/QCD model.

\section{Conclusion} \label{secDiscuss4}

In this chapter, we have shown that higher-order terms in the action improves upon the soft-wall AdS/QCD model. We correctly incorporate chiral symmetry breaking decoupling the sources for spontaneous and explicit breaking. This is achieved by introducing a quartic term in the potential for the bulk scalar field dual to the quark bilinear operator ${\bar q} q$. This changes the dilaton profile for small $z$, while simultaneously maintaining the large $z$ quadratic behavior and therefore linear trajectories for the radially excited states. In addition, our model is built from the assumption of preserving chiral symmetry for highly excited states, which is supported by the experimental values of the QCD mass spectrum. This enables us to obtain reasonable agreement within 10$\%$ of the QCD meson mass spectra for scalar, vector and axial-vector fields, although the lowest lying $\rho$ and $f_0$ predictions are not as good.

Even though our modification of the soft-wall version of the AdS/QCD model is simple and predictive,
any further progress must recognize the limitations of this type of phenomenological model. Genuine
stringy behavior is most likely required to fully describe the characteristics of QCD. Nevertheless, some
features such as masses and couplings seem to agree better than expected and it would be worth 
using the modified dilaton profile to study further details of the meson spectrum. On the theoretical side,
it would be interesting to further understand the soft-wall model from the top-down including finding a 
dynamical solution of the features exhibited in our model along the lines considered in \cite{Batell:2008zm}. 
In addition, the stability of the scalar potential will most likely require higher-order terms that can only be 
studied from the top-down. It is interesting that the simple five-dimensional model contains QCD-like features and suggests 
that a further understanding of QCD can be obtained from the gauge/gravity correspondence.

\chapter{Finite Temperature Thermodynamics}\label{chthermo}

\begin{flushright}
``Nothing in life is certain except death, taxes and the second law of thermodynamics.''\\
-Seth Lloyd
\end{flushright}

\section{Introduction}

In this chapter, we focus on two descriptions of our QCD-like gauge theory. Confined hadronic matter where baryons and mesons are the primary degrees of freedom is characterized by the AdS metric similar to the one from Chapter \ref{chzero}. Deconfined plasma where quarks and gluons associate freely is modeled by extending the metric to an AdS-Schwarzschild solution, describing an extra-dimensional generalization of a black hole. The thermodynamic description of black holes is well established \cite{Bekenstein:1973,Bekenstein:1974,Hawking:1974sw}. The correspondence connects the black-hole thermodynamics with the thermodynamic properties of a deconfined gauge theory. The transition between these two phases occurs at a critical temperature $T_{c}$ and corresponds to the Hawking-Page transition in the gravity dual. Experimental evidence confirms that the QGP is strongly coupled \cite{Adams:2005dq, Adcox:2004mh, Arsene:2004fa, Back:2004je}, lending support to the theoretical picture obtained via a five-dimensional dual gravity description.

In this chapter, we find a finite-temperature description of a strongly coupled gauge theory with a soft-wall geometry based on the model in \cite{Batell:2008zm}. The five-dimensional Einstein equations involve two scalar fields. The black-hole metric is asymptotically AdS with an event horizon located at a finite value of the bulk coordinate, $z$. In the string frame, one scalar field plays the role of a string-theory dilaton, while the other scalar field is similar to a string-theory tachyon. In terms of the correspondence, we begin with the assumption that the dilaton field is dual to the dimension-4 gluonic operator, Tr$(F_{\mu\nu}F^{\mu\nu})$, while the second tachyon field is dual to the dimension-3 chiral operator $q\bar{q}$. The true field/operator correspondence is more nuanced, as we will discuss.

The solution outlined in this chapter is an expansion only valid in the limit of $z, z_h<1$. Unfortunately, this is the price we pay to study the soft-wall thermodynamics; the equations of motion have no known closed-form solution nor a reasonable numerical solution. We explore a purely phenomenological model that does not dynamically generate the soft-wall geometry in Appendix \ref{appPhenom}. We attempt to introduce rigor into the soft-wall thermodynamics, beyond what has been done in \cite{Herzog:2006ra, Colangelo:2009ra}.

The thermodynamics resulting from our solution leads to interesting consequences. We show that a nonzero thermal condensate function $\mathcal{G}$S induces a phase transition \cite{Gursoy:2008za, Gursoy:2008bu, Gubser:2008ny}. Furthermore, $\mathcal{G}$ contributes leading-order terms that are absent in the lattice results to the soft-wall thermodynamics \cite{Boyd:1996bx, Miller:2006hr}. However, qualitatively the two agree. The entropy has the expected behavior at high temperatures, scaling as $T^3$, and the speed of sound through the thermal plasma is consistent with the conformal value of $1/3$, agreeing with the upper bound advocated in \cite{Cherman:2009tw}. 

Our analysis begins in Section \ref{secDynamicAct}, where we present the thermal AdS and black-hole AdS solution and compute the on-shell action. In Section \ref{secThermo}, we study the thermodynamics of our solution, including a general expression for the entropy and squared speed of sound. We then calculate the free energy difference by carefully matching our two solutions at the AdS boundary. This enables us to compute the transition temperature between the confined and deconfined phases of the gauge theory. Concluding remarks are given in Section \ref{secDiscuss}.

\section{Finite-Temperature Action} 
\label{secDynamicAct}

We begin by specifying the five-dimensional action in the string frame and transforming it to the Einstein frame. The equations of motion become simpler in the Einstein frame. We are extending the AdS solution with two scalar fields first investigated by \cite{Batell:2008zm}. 

\subsection{5D Lagrangian}
We start with the string-frame action inspired by the dimensionally-reduced type IIB supergravity action first introduced in Section \ref{secbktempaction},
\begin{eqnarray} 
\label{equString}
\mathcal{S}_{string} &=&-\frac{1}{16 \pi G_5} \int d^5 x \sqrt{-g} \Bigg[{\rm e}^{-2\Phi} 
\Bigg(R + 4\,g^{MN}\partial_M\Phi\partial_N\Phi - \frac{1}{2}\,g^{MN}\partial_M \chi\partial_N \chi 
- \nonumber\\ 
&&\qquad\qquad\qquad\qquad V_S(\Phi, \chi)\Bigg) +\, {\rm e}^{-\Phi}\mathcal{L}_{{\rm meson}}\Bigg] + S_{GH}^{(s)},
\end{eqnarray}
where $\mathcal{L}_{{\rm meson}}$ contains all the mass terms not considered in this chapter, $V_S$ is the string-frame scalar potential, and the indices $M,N = (t,x_{1},x_{2},x_{3},z)$. We are working with two scalar fields $\Phi$ and $\chi$ where previous models have only focused on the dilaton \cite{BallonBayona:2007vp, Gursoy:2008za, Gursoy:2008bu, Alanen:2009na, Alanen:2010tg, Gubser:2008ny, Franco:2009if}. The inclusion of a Gibbons-Hawking term $\mathcal{S}_{GH}^{(s)}$ is an attempt to be more rigorous than previous models. The string-frame metric is assumed to have an AdS-Schwarzschild form
\begin{equation} \label{equStringMetric}
ds^2 = g_{MN} dx^M dx^N= \frac{R^{2}}{z^{2}}\left(-f(z) dt^{2} + d\vec{x}^{2} + \frac{dz^{2}}{f(z)}\right), 
\end{equation}
where $f(z)$ determines the location of the black-hole horizon. Furthermore, the string-frame action (\ref{equString}) contains similarities with noncritical string theory. The $\Phi$ scalar field behaves like a dilaton, while the $\chi$ scalar field behaves like a closed-string tachyon field. This suggests that our setup could be the low-energy limit of some underlying string theory.

While the string-frame action provides a suitable starting point, it is more practical to do calculations in the Einstein frame. Switching to the Einstein frame involves a simple conformal transformation,
\begin{equation} \label{equconformal}
g_{MN}^{(s)} = {\rm e}^{\frac{4}{3}\Phi}g_{MN}^{(E)}.
\end{equation}
The gravity-dilaton-tachyon action in the Einstein frame then becomes
\begin{equation} 
\mathcal{S}_E =-\frac{1}{16 \pi G_{5}} \int d^5x \sqrt{-g}\left(R - \frac{1}{2}(\partial\phi)^2 - \frac{1}{2}(\partial\chi)^2- V(\phi, \chi)\right)
+ S_{GH},
\label{equEinstein}
\end{equation}
where $\phi = \sqrt{8/3} \Phi$ and $V = V_S\,{\rm e}^{\frac{4}{3}\Phi}$. We explicitly define the Gibbons-Hawking term $S_{GH}$ as
\begin{eqnarray}
\mathcal{S}_{GH} &=& \frac{1}{8 \pi G_5}\int d^4 x \sqrt{-\gamma} K \nonumber\\
&=& \frac{2N_c^2}{45\pi^2 R^3} \int d^3 x \int_0^\beta d\tau \sqrt{-\gamma} K, \label{GHaction}
\end{eqnarray}
where $K=\gamma^{\mu\nu}K_{\mu\nu}$ and $\gamma$ is the four-dimensional induced metric at the AdS boundary. The extrinsic curvature $K_{\mu\nu}$ is defined by
\begin{equation}\label{defnintcurv}
      K_{\mu\nu} = \nabla_\mu n_\nu = \frac{1}{2} n^M \partial_M \gamma_{\mu\nu}~,
\end{equation}
 where the vector $n_{M}$ is the outward directed normal to the boundary, normalized by
\begin{equation}
g_{MN}\,n^{M}n^{N} = 1.
\end{equation}
The boundary of the AdS$_{5}$ space considered here is the $z=0$ plane, meaning the normal vector is 
\begin{equation}
      n^M = -\frac{1}{\sqrt{g_{zz}}} \left(\frac{\partial}{\partial z}\right)^M = 
      \frac{\delta^M_z}{\sqrt{g_{zz}}}.
      \label{defnormal}
\end{equation}
The Gibbons-Hawking term does not affect the equations of motion, but will have consequences when considering the free energy and deconfinement temperature.

To convert to a description at finite temperature, we compactify the Euclidean time coordinate, 
$\tau\equiv i t\rightarrow it+ \beta$, where $\beta=1/T$ is the inverse temperature. The 
finite-temperature metric in the Einstein frame then becomes
\begin{equation} 
\label{equEinsteinMetric}
ds^{2}=a(z)^{2}\left(f(z)d\tau^2+d\vec{x}^{2}+\frac{dz^{2}}{f(z)}\right),
\end{equation}
with the finite-temperature action given by
\begin{eqnarray}
\mathcal{S}_E(\beta)&=&-\frac{1}{16 \pi G_{5}} \int d^4 x \int_0^{\beta} d\tau \int dz \sqrt{-g}\left(R - \frac{1}{2}(\partial\phi)^2 - \frac{1}{2}(\partial\chi)^2 - V(\phi, \chi)
\right)\nonumber \\
&&\qquad\qquad +\frac{1}{8 \pi G_5}\int d^3 x \int_0^{\beta} d\tau \sqrt{-\gamma} K.
\label{equEinstein1}
\end{eqnarray}
The Einstein frame metric (\ref{equEinsteinMetric}) and action (\ref{equEinstein1}) will be 
used to solve the Einstein's equations in two realms, thermal AdS (thAdS) and black-hole AdS (bhAdS). All quantities with a subscript $_{0}$ are associated with the thAdS solution.

\subsection{Thermal AdS Solution}
\label{secThermalAdS}

We begin by obtaining a thAdS solution that is effectively equivalent to the zero-temperature case explored in Chapter \ref{chzero}, where $f(z)=1$.  Assuming the scalar fields are a function of only the $z$ coordinate, the action can be expressed as
\begin{eqnarray} 
\label{equAction1}
\mathcal{S}_0(\delta) &=& -\frac{N_c^2}{45\pi^2}\frac{V}{R^3 T} \int_{\delta}^{\infty}{dz \sqrt{-g}\left(R-\frac{1}{2}
g^{55}\phi'^{2} - \frac{1}{2}g^{55}\chi'^{2} - V(\phi,\chi)\right)}\nonumber\\
 && \qquad\qquad+S_{0,GH},
\end{eqnarray}
where $V$ is the spatial three-volume and a UV cutoff at $z=\delta$ has been introduced to regularize 
any singular behavior at the AdS boundary. 

 The metric associated with the thAdS solution is
\begin{equation} 
\label{equEmetric}
      ds^2 = a(z)^{2}(d\tau^2 + d\vec{x}^2 + dz^2) \equiv {\rm e}^{-2 c\,\phi(z)}\frac{R^2}{z^2}(d\tau^2 + d\vec{x}^2 + dz^2),
\end{equation}
where the constant $c$ depends on the conformal transformation. In our case, $c=1/\sqrt{6}$. Two equations come from the five-dimensional Einsteins equations,
\begin{eqnarray}
12 \frac{a'^{2}}{a^{2}} - 6 \frac{a''}{a} &=& \phi'^{2} + \chi'^{2}, \label{equGeneralField} \\
6\frac{a'^{2}}{a^{2}} + 3 \frac{a''}{a} &=& -a^{2} V(z), \label{equGeneralV}
\end{eqnarray}
and two more equations come from the varying the action,
\begin{eqnarray}
a^{2}\frac{\partial V}{\partial \phi} &=& \phi'' + 3 \phi' \frac{a'}{a}, \label{equVphi} \\ 
a^{2}\frac{\partial V}{\partial \chi} &=& \chi'' + 3 \chi' \frac{a'}{a}, \label{equVchi}
\end{eqnarray}
where prime $(')$ denotes derivatives with respect to $z$. 
Using (\ref{equEmetric}), we express (\ref{equGeneralField}), (\ref{equGeneralV}), (\ref{equVphi}), and (\ref{equVchi}) all in terms of $\phi$, $\chi$, and $V\left(\phi(z),\chi(z)\right)$,
\begin{eqnarray}
\chi'^{2} &=& \frac{2\sqrt{6}}{z}\phi' + \sqrt{6} \phi'', \label{equGeneralField2}\\
V(z) &=& \frac{{\rm e}^{\frac{2}{\sqrt{6}} \phi}}{R^2}\left(-12 - 3\sqrt{6} z \phi' - \frac{3z^{2}}{2} \phi'^{2} + \sqrt{\frac{3}{2}} z^{2} \phi''\right), \label{equGeneralV2}\\
\frac{\partial V}{\partial \phi} &=& \frac{z^{2}{\rm e}^{\frac{2}{\sqrt{6}}\phi}}{R^{2}} \left(\phi'' - \sqrt{\frac{3}{2}} \phi'^{2} - \frac{3}{z} \phi'\right),\label{equVphi2}\\
\frac{\partial V}{\partial \chi} &=& \frac{z^{2}{\rm e}^{\frac{2}{\sqrt{6}}\phi}}{R^{2}} \left(\chi'' - \sqrt{\frac{3}{2}} \phi'\,\chi' - \frac{3}{z} \chi'\right),\label{equVchi2}
\end{eqnarray}
where we see that the nonlinear term $\phi'^{2}$ has conveniently cancelled in (\ref{equGeneralField2}). Of the four equations, we find that three are independent.

As shown in \cite{Batell:2008zm}, the solution in the soft-wall model with a quadratic dilaton gives
\begin{eqnarray}
\phi(z) &=& \sqrt{\frac{8}{3}} \mu^2 z^2, \label{equphisol}\\
\chi(z) &=& 2 \sqrt{6} \mu z, \label{equchisol}\\
V(z) &=& \frac{{\rm e}^{\frac{4}{3} \mu^2 z^2}}{R^2} \left(-12 - 20 \mu^2 z^2 - 16 \mu^4 z^4\right), \label{equpotsol}
\end{eqnarray}
where $\mu$ sets the hadronic mass scale. The quadratic behavior of the $\phi$ solution (\ref{equphisol}) leads to a Regge-like hadron mass spectrum $(m_n^2\sim \mu^2 n)$. The linear behavior of $\chi$ indicates similar chiral symmetry consequences as the scalar VEV function $v(z)$ did in Chapter \ref{chzero}. The potential is a function of the $z$ coordinate with no unique solution for $V(\phi,\chi)$. However, a potential was suggested in \cite{Batell:2008zm},
\begin{equation}\label{equbatellV}
V(\phi,\chi) = \frac{\chi}{2}{\rm e}^{\frac{\chi^{2}}{18}} + 2 \phi^{2}{\rm e}^{\frac{2}{\sqrt{6}} \phi} - 12 \left(3 {\rm e}^{\frac{\chi}{36}}-2\left(1-\frac{2}{\sqrt{6}}\right){\rm e}^{\frac{\phi}{\sqrt{6}}}\right)^{2}.
\end{equation}
We can define an alternative potential that still produces (\ref{equphisol}), (\ref{equchisol}), and (\ref{equpotsol}),
\begin{equation}\label{equkelleyV}
V_{\rm alt}(\phi,\chi) = \frac{{\rm e}^{\frac{2}{\sqrt{6}}\phi}}{R^{2}}\left(-12 + 4\sqrt{6} \phi - \frac{3}{2}\chi^{2} - 4\phi^{2} + \frac{7}{3\sqrt{6}}\phi\chi^{2}-\frac{2}{27}\chi^{4}\right).
\end{equation}
More examples of consistent and well-defined potentials written in terms of two scalar fields can be found in \cite{Kapusta:2010mf}.

Part of the the free energy expression is found by substituting the thAdS solution into (\ref{equAction1}) and finding the Gibbons-Hawking term. In general, we find the Ricci scalar is 
\begin{equation} \label{equRicciS1}
R = -\frac{8 a''}{a^{3}} - \frac{4 a'^{2}}{a^{4}}.
\end{equation}
For the thermal AdS solution, we find the extrinsic curvature takes the form of 
\begin{equation}\label{equKcurvature}
K_{\mu\nu}  =\frac{1}{2}\frac{\delta_{M}^{z}}{\sqrt{g_{zz}}}\eta_{\mu\nu}\partial_{M}a_{0}(z)^{2}=a_{0}'(z).
\end{equation}
Calculating the trace, we find
\begin{equation}
    \gamma^{\mu\nu} K_{\mu\nu} \equiv K = 4\frac{a_0'}{a_0^2}.
\end{equation}
Substituting this value into (\ref{GHaction}) gives the on-shell Gibbons-Hawking term for the
thermal AdS solution,
\begin{equation} \label{equGHterm0}
\mathcal{S}_{0,GH} = \frac{N_c^2}{45\pi^2} \frac{V}{R^3 T} \left( 8 a_0^2 a_0'\right).
\end{equation}
The on-shell action (\ref{equAction1}) then becomes
\begin{equation}
\mathcal{S}_0(\delta) = -\frac{N_c^2}{45\pi^2} \frac{V}{R^3 T}\left( 2 a'(\delta) a^2(\delta)\right) + S_{0,GH} = 
\frac{N_c^2\,V}{15\pi^2} \frac{2 a'(\delta) a^2(\delta)}{R^3 T}.
\label{equS1}
\end{equation}
The on-shell action is a pure boundary term and strictly depends on the AdS boundary conditions, where $\delta\rightarrow 0$.

\subsubsection{Field/Operator Correspondence}\label{secFOcorr}

According to the AdS/CFT dictionary, the scalar field masses determine the corresponding operator dimensions in the dual gauge theory via (\ref{equSmass}). Expanding the potential (\ref{equbatellV}) from \cite{Batell:2008zm}, we see that 
\begin{equation}\label{equAcase}
m_{\phi}^{2}R^{2} = -4, \quad\quad\quad m_{\chi}^{2}R^{2} = -3,
\end{equation}
suggesting that $\phi$ is dual to a dimension-2 operator, and $\chi$ is dual to a dimension-3 operator. When compared to the standard form (\ref{equstandardform}), the behavior of the solutions (\ref{equphisol}) and (\ref{equchisol}) confirm the operator dimensions suggested by (\ref{equAcase}).  
 However, the potential (\ref{equkelleyV}) gives the masses as
\begin{equation}\label{equGcase}
m_{\phi}^{2}R^{2} = 0, \quad\quad\quad m_{\chi}^{2}R^{2} = -3,
\end{equation}
indicating that $\phi$ and $\chi$ are dual to a dimension-4 and dimension-3 operator, respectively. It is fairly clear that the chiral operator, $q\bar{q}$, and the gluonic operator, Tr$\left[F^{2}\right]$, are the dimension-3 and dimension-4 operators, but the dimension-2 operator is much less clear. 

The most likely dimension-2 operator candidate is $A_{\mu}^{2}$, which becomes a local expression in the Laudau gauge, $\partial^{\mu}A_{\mu}=0$. Coupling a source term to $A_{\mu}^{2}$, however, makes the theory nonrenormalizable at the quantum level. A quadratic source term can be added to remedy this obstacle, though, this ruins the energy interpretation of the effective action \cite{Vercauteren:2010rk}. In the context of the AdS/CFT correspondence, $A_{\mu}^{2}$ is often understood to convey information about the topological defects in the gravity dual \cite{Gubarev:2000eu, Gubarev:2000nz}. Much more work concerning $A_{\mu}^{2}$ has been conducted in \cite{Dudal:2009tq, Vercauteren:2010cg, Vercauteren:2011ze}.

In the current soft-wall case, the field/operator correspondence appears complicated. The ambiguity stems from the fact that the original AdS/CFT dictionary was formulated considering purely free scalar fields. The potentials (\ref{equbatellV}) and (\ref{equkelleyV}) clearly have interaction terms. Resolving the issue of whether interaction terms affect the field/operator correspondence and determining the interpretation of the dimension-2 operator is ultimately beyond the scope of this thesis. In that spirit, we expand upon the published potential (\ref{equbatellV}) and assume that the fields $\chi$ and $\phi$ correspond to the operators $q\bar{q}$ and Tr$\left[F^{2}\right]$, which is the general consensus. The significance of our work relies on the fact that the dilaton is dual to \emph{some} temperature-dependent operator. The identity of that operator is a topic for further research.

\subsection{Black-Hole AdS solution}
\label{secbhAdS}

Next, we consider the black-hole AdS solution that describes a deconfined phase mimicking a free gluon plasma. 
Assuming the solutions are only a function of the $z$ coordinate, the five-dimensional action associated with the black-hole AdS solution simplifies to
\begin{eqnarray} 
\label{equAction2}
\mathcal{S}_{bh}(\delta) &=& -\frac{N_{c}^{2}}{45 \pi^{2}}\frac{V}{R^3 T(z_{h})} \int_{\delta}^{z_h}{dz \sqrt{-g}\left(R-\frac{1}{2}g^{55}\phi'^{2} - \frac{1}{2}g^{55}\chi'^{2} - V(\phi,\chi)\right)} \nonumber\\
&&\qquad\qquad\qquad\qquad+S_{bh,GH},
\end{eqnarray}
where $z_h$ is the location of the black-hole horizon. 
We will see that $z_h$ is directly related to the temperature of the gauge theory. We begin with the black-hole metric (\ref{equEinsteinMetric}) and find four independent equations,
\begin{eqnarray}
f''(z) &=& -3 f'(z)\frac{a'(z)}{a(z)}, \label{equbkf}\\
\phi'(z)^{2} + \chi'(z)^{2} &=& 12 \frac{a'(z)^{2}}{a(z)^{2}} - 6 \frac{a''(z)}{a(z)},\label{equbkfields}\\
a(z)^{2}\frac{\partial V}{\partial \phi} &=& f(z)\phi''(z) + f'(z)\phi'(z) + 3 f(z)\phi'\frac{a'(z)}{a(z)}, \label{equbkVphi}\\
a(z)^{2}\frac{\partial V}{\partial \chi} &=& f(z)\chi''(z) + f'(z)\chi'(z) + 3 f(z)\chi'\frac{a'(z)}{a(z)} \label{equbkVchi}.
\end{eqnarray}
Unlike in the thAdS case, the potential is already determined. We must use the potential (\ref{equbatellV}) to connect the soft-wall action to the free energy investigated in Section \ref{secThermo}. With four independent equations and four unknown functions, $f(z)$, $a(z)$, $\phi(z)$, and $\chi(z)$, we no longer can assume a fixed relation between the warp factor $a(z)$ and the dilaton $\phi(z)$. These quantities must evolve independently as the temperature varies. The system of equations associated with bhAdS are difficult to solve but for the simplest cases. The conformal case discussed in Chapter \ref{chApplications} is the only known exact solution. 

Using the series expansions, we construct another solution. We use the thAdS solution of Section \ref{secThermalAdS} as the starting points for these series expansions,
\begin{eqnarray}
a(z) &=& \frac{R}{z}{\rm e}^{-\frac{\phi}{\sqrt{6}} + \sum_{n=2}^{\infty} m_{n}(\mathcal{G},z_h) z^{n}}, \label{equseriesmetric} \\
\phi(z) &=& \sqrt{\frac{8}{3}}\mu^{2}z^{2} + \sum_{n=2}^{\infty} p_{n}({\cal G},z_h) z^n, \label{equseriesphi} \\
\chi(z) &=& \sum_{n=1}^{\infty} c_{n}(\mathcal{G},z_h) z^{n},\label{equserieschi}\\
f(z) &=& 1 + \sum_{n=4}^{\infty} f_{n}(\mathcal{G},z_{h}) z^{n},\label{equseriesf}
\end{eqnarray}
where we see that one of the black-hole conditions, $f(0)=1$, is automatically satisfied.
The condensate function $\mathcal{G}(z_h)$ plays an important role in the free energy and phase transition of the thermal plasma.
By solving (\ref{equbkf}), (\ref{equbkfields}), (\ref{equbkVphi}), and (\ref{equbkVchi}) in successive powers of $z$, we find the coefficients up to $n=8$ in terms of $f_{4}(\mathcal{G},z_h)$. the other black-hole condition, $f(z_h)=0$, determines the final unknown coefficient. We construct a solution where the majority of the coefficients are zero. We calculate the non-zero coefficients for the metric,
\begin{eqnarray}
m_{2}  &=& -\frac{\sqrt{6}}{4\mu^{2}}\mathcal{G}(z_h),\\
m_{6}  &=& \frac{1}{8\mu^{4} + 3\sqrt{6}\mathcal{G}(z_h)}\Bigg( -\frac{8\mu^{6}}{21}f_{4} + \frac{3\sqrt{3}\mu^{2}}{7\sqrt{2}}f_{4}\mathcal{G}(z_h) + \frac{32\sqrt{2}\mu^{6}}{21\sqrt{3}}\mathcal{G}(z_h) \nonumber\\
&& \quad+ \frac{45}{56\mu^{2}}f_{4}\mathcal{G}(z_h)^{2} + \frac{131\mu^{2}}{42}\mathcal{G}(z_h)^{2} +\frac{9\sqrt{3}}{8\sqrt{2}\mu^{2}}\mathcal{G}(z_h)^{3}\nonumber\\
&&\quad - \frac{93}{112\mu^{6}}\mathcal{G}(z_h)^{4} \Bigg),\\ 
m_{8}  &=& \frac{1}{8\mu^{4}+3\sqrt{6}\mathcal{G}(z_h)}\Bigg(-\frac{4\mu^{8}}{9}f_{4} + \frac{11\mu^{4}}{21\sqrt{6}}f_{4}\mathcal{G}(z_h) +\frac{184\sqrt{2}\mu^{8}}{243\sqrt{3}}\mathcal{G}(z_h) \nonumber\\
&&\quad + \frac{71}{56}f_{4}\mathcal{G}(z_h)^{2} + \frac{2543\mu^{4}}{1701}\mathcal{G}(z_h)^{2} + \frac{1507}{378\sqrt{6}}\mathcal{G}^{3} +\frac{117\sqrt{3}}{224\sqrt{2}\mu^{4}}f_{4}\mathcal{G}(z_h)^{3}\nonumber\\
&&\quad + \frac{203}{432\mu^{4}}\mathcal{G}(z_h)^{4} - \frac{895}{896\sqrt{6}\mu^{8}}\mathcal{G}(z_h)^{5} \Bigg),
\end{eqnarray}
the field $\phi$,
\begin{eqnarray}
p_{4}  &=& -\mathcal{G}(z_h),\\ 
p_{6}  &=& -\frac{\mu^{2}}{\sqrt{6}}f_{4} - \frac{\mu^{2}}{3}\mathcal{G}(z_h) + \frac{11}{8\sqrt{6}\mu^{2}}\mathcal{G}(z_h)^{2},\\
p_{8}  &=& \frac{1}{8\mu^{4} + 3\sqrt{6}\mathcal{G}}\Bigg(-\frac{80\sqrt{2}\mu^{8}}{21\sqrt{3}}f_{4} - \frac{12\mu^{4}}{7}f_{4}\mathcal{G}(z_h) - \frac{1408\mu^{8}}{567}\mathcal{G}(z_h) \nonumber\\
&&\quad + \frac{3\sqrt{6}}{7}f_{4}\mathcal{G}(z_h)^{2} -\frac{107\sqrt{2}\mu^{4}}{63\sqrt{3}}\mathcal{G}(z_h)^{2}-\frac{29}{27}\mathcal{G}(z_h)^{3}\nonumber\\
&&\quad-\frac{3083}{1008\sqrt{6}\mu^{4}}\mathcal{G}(z_h)^{4}\Bigg),
\end{eqnarray}
the field $\chi$,
\begin{eqnarray}
c_{1}  &=& \sqrt{24\mu^{2}+\frac{9\sqrt{6}}{\mu^{2}}\mathcal{G}(z_h)},\\
c_{3}  &=& \frac{\frac{\sqrt{3}}{2\mu^{4}}\mathcal{G}(z_h)^{2}-2\sqrt{2}\mathcal{G}(z_h)}{\sqrt{8\mu^{2}+\frac{3\sqrt{6}}{\mu^{2}}\mathcal{G}(z_h)}},\\
c_{5}  &=& \frac{1}{\sqrt{8\mu^{4}+3\sqrt{6}\mathcal{G}(z_h)}}\Bigg(-\sqrt{3}\mu^{3}f_{4} - \frac{9\sqrt{2}}{8\mu}f_{4}\mathcal{G}(z_h) - 3\sqrt{2}\mu^{3}\mathcal{G}(z_h)\nonumber\\
&&\quad - \frac{5\sqrt{3}}{4\mu}\mathcal{G}(z_h)^{2} + \frac{9\sqrt{2}}{8\mu^{5}}\mathcal{G}(z_h)\Bigg),\\
c_{7}  &=& \frac{1}{\sqrt{8\mu^{4}+3\sqrt{6}\mathcal{G}(z_h)}}\Bigg(-\frac{20\mu^{5}}{7\sqrt{3}}f_{4} - \frac{39\mu}{14\sqrt{2}}f_{4}\mathcal{G}(z_h) - \frac{568\sqrt{2}\mu^{5}}{189}\mathcal{G}(z_h)\nonumber\\
&&\quad - \frac{111\sqrt{3}}{112\mu^{3}}f_{4}\mathcal{G}(z_{4})^{2} - \frac{1123\mu}{126\sqrt{3}}\mathcal{G}(z_h)^{2} - \frac{67}{24\sqrt{2}\mu^{3}}\mathcal{G}(z_h)^{3} \nonumber\\
&&\quad+ \frac{151\sqrt{3}}{224\mu^{7}}\mathcal{G}(z_h)^{4} \Bigg),
\end{eqnarray}
and $f(z)$,
\begin{eqnarray}
f_{4} &=& \frac{-1}{z_{h}^{4}\left(1+\frac{4\mu^{2}}{3}z_{h}^{2}+\mu^{4}z_{h}^{4} + \left(\frac{\sqrt{6}}{2\mu^{2}} + \frac{\sqrt{6}}{2}\right)z_{h}^{2}\mathcal{G}(z_h) + \frac{27}{32\mu^{4}}z_{h}^{4}\mathcal{G}(z_h)^{2}\right)},\\
f_{6} &=& \frac{4\mu^{2}}{3}f_{4} + \frac{3}{\sqrt{6}\mu^{2}}f_{4}\mathcal{G}(z_h), \\
f_{8} &=& \mu^{4} f_{4} + \frac{\sqrt{6}}{2}f_{4}\mathcal{G}(z_h) + \frac{27}{32\mu^{4}}f_{4}\mathcal{G}(z_h)^{2}.
\end{eqnarray}

In general, an arbitrary power $z^n$ multiplying a factor of ${\cal G}(z_h)$ in the coefficients $p_{n}$ and $m_{n}$ has direct consequences for the free energy. For $0<n<4$, the free energy contains divergences of the power $n$, while for $n>4$ the function ${\cal G}(z_h)$ does not affect the free energy. Only the $n=4$ coefficients affect the free energy expression. As we will see in Section~\ref{secThermo}, the condensate function ${\cal G}$ then plays a crucial role in giving rise to a finite transition temperature.

The calculation of the Gibbons-Hawking term is more complicated in the bhAdS solution. The induced metric in this case is
\begin{equation}\label{equinducedbh}
\gamma = a(z)^{2}\left( f(z)d\tau^{2} + d\vec{x}^{2} \right).
\end{equation}
To obtain the trace of the extrinsic curvature, we need to separate the $\tau$ components from the rest,
\begin{equation}
K = \gamma^{00}K_{00} + \gamma^{ij}K_{ij}.
\end{equation} 
Using (\ref{equKcurvature}), we find that
\begin{eqnarray}
\gamma^{00}K_{00} &=& \frac{f'}{2\sqrt{f}a} + \frac{\sqrt{f}a'}{a^{2}},\\
\gamma^{ij}K_{ij} &=& \frac{3\sqrt{f}a'}{a^{2}}.
\end{eqnarray} 
Given that $\sqrt{\gamma} = a^{4}\sqrt{f}$, the Gibbons-Hawking term becomes
\begin{equation}\label{equSbhgh}
\mathcal{S}_{bh,GH} = \frac{N_{c}V}{45\pi^{2}R^{3}T}\left(8 f \,a^{2}a' + f' \,a^{3}\right).
\end{equation} 

Finally, an expression for the on-shell action can be obtained by substituting the solutions back into
(\ref{equAction2}). Using the fact that the Ricci tensor can be expressed as 
\begin{equation}
R = -\frac{4 f a'^{2}}{a^{4}} - \frac{8 f a''}{a^{3}} - \frac{8 f'a'}{a^{3}} - \frac{f''}{a^{2}},
\end{equation}
the on-shell action can then be written as
\begin{eqnarray}
\mathcal{S}_{bh}(\delta) 
&=& -\frac{N_c^2}{45\pi^2}\frac{V}{R^3 T}2 f(\delta)a'(\delta) a^2(\delta) +S_{bh,GH} \nonumber\\
&=& \frac{N_c^2}{45\pi^2}\frac{V}{R^3 T} \left(6 f(\delta)a'(\delta) a^2(\delta) +f'(\delta) a^3(\delta) \right).
\label{bhonshell}
\end{eqnarray}
Again, the on-shell action is a pure boundary term and depends only on the AdS boundary conditions.
This action will be used to compute the free energy in Section~\ref{secThermo}.

\subsubsection{Field/Operator Correspondence at Finite Temperature}

The scalar field solutions (\ref{equseriesphi}) and (\ref{equserieschi}) correspond to turning on thermal condensates in the gauge theory. To show the relation between ${\cal G}(z_h)$ and the operator condensates, we assume the kinetic term of a canonically normalized bulk scalar field fluctuation, 
denoted $\omega$, to be 
\begin{equation} 
\label{equgenbulk}
\mathcal{S}_{{\rm bulk}} = \frac{1}{2 L^3}\int d^5 x \sqrt{-g}\, \partial_M \omega\partial^M \omega.
\end{equation}
Let us first consider the scalar field $\phi$, where the coupling to the four-dimensional boundary operator is
\begin{equation} \label{equgenboundary}
\mathcal{S}_{{\rm boundary}} = \int{d^{4}x \,\omega_{\phi}\, {\rm Tr}F^2}.
\end{equation}
In terms of the 't Hooft coupling $\lambda = {\rm e}^{\Phi}$, the Yang-Mills field strength is 
\begin{equation}\label{equYMcouple}
\mathcal{S}_{{\rm boundary}} = -\int{d^{4}x\, \frac{1}{4\lambda} {\rm Tr} F^{2} },
\end{equation}
making the dilaton fluctuation, 
\begin{equation}\label{equfluxphi}
\delta \mathcal{S}_{{\rm boundary}}  = \frac{1}{4} \int{ d^4x\, \delta\Phi \,{\rm e}^{-\Phi} {\rm Tr}(F^{2})}.
\end{equation}
We are only interested in the fluctuation $\delta\Phi = \Phi - \Phi_{0}$, which allows one to compute the difference between thermal and vacuum values of $\langle {\rm Tr}F^{2} \rangle$ \cite{Gursoy:2008za}. From (\ref{equseriesphi}), we find that $\delta\Phi=-\sqrt{3/8}{\cal G}(z_h) z^4$. Recall that the relation between the expectation value of a $\Delta$-dimensional operator in $d$-dimensional space and the field is
\begin{equation}
\frac{\omega}{z^{\Delta}} \rightarrow \frac{\langle \mathcal{O} \rangle}{2\Delta - d}.
\end{equation}
The function ${\cal G}(z_h)$ then relates directly to the gluon condensate,
\begin{equation}
\langle {\rm Tr} (F^{a})^{2} \rangle -  \langle {\rm Tr} (F^{a})^{2} \rangle_{0} = 
-\sqrt{\frac{2}{3}} \left(\frac{32 N_c^2\lambda}{45\pi^2}\right) {\mathcal G}(z_h). 
\end{equation}

A similar correspondence can be obtained for the scalar field $\chi$, where it is tempting to
relate $\chi$ to the three-dimensional operator $q \bar{q}$. In this case, the coefficient of the 
$z^3$ term in (\ref{equserieschi}) corresponds to the renormalized chiral condensate, $\langle q \bar{q}\rangle -\langle q \bar{q}\rangle_{0} $. 
For this correspondence to hold, $\chi$ would need to be a bifundamental field in the gauge theory. This can be done by 
promoting $\chi$ to a bifundamental field $\chi^{ab}$, where $a,b$ are group indices and writing, for example
\begin{equation}
\langle \chi^{ab} \rangle = \chi(z) 
 \left( 
 \begin{array}{cc} 
 1 & 0 \\
 0 & 1
\end{array}\right)
\label{bifundveveqn}
\end{equation}
for an $SU(2)\times SU(2)$ symmetry. The bulk $\chi$ solution (\ref{equserieschi}) is then identified with the vacuum expectation value $\chi(z)$ in (\ref{bifundveveqn}). 

Under this assumption, we can follow the same procedure as for the $\phi$ field, and derive an operator correspondence for $\chi$. Assuming a bulk kinetic term of the form (\ref{equgenbulk}) and boundary operator coupling
\begin{equation}
\mathcal{S}_{boundary} = \int{d^{4}x \,\omega_{\chi}\,\bar{q}q},
\end{equation}
we find that
\begin{equation}
\langle \bar{q} q\rangle - \langle \bar{q} q \rangle_{0} =
-\frac{2 N_{c}^{2}}{27\pi^{2}}\left(\frac{\mathcal{G}(z_h)}{\mu}-\frac{5}{8}\frac{\mathcal{G}(z_h)^{2}}{\mu^{5}}\right).
\end{equation}
Thus, we see that the function ${\cal G}(z_h)$ is intimately related to the thermal condensates 
$\langle \bar{q}q \rangle$ and $\langle {\rm Tr} F^2 \rangle$. In fact, the ratio of the gluon condensate and the
chiral condensate is given roughly by
\begin{equation}
        \frac{\langle {\rm Tr} (F^{a})^{2} \rangle -  \langle {\rm Tr} (F^{a})^{2} \rangle_{0}}
        {\langle \bar{q} q\rangle - \langle \bar{q} q \rangle_{0}} \approx \frac{16}{5} \sqrt{6} \lambda \mu,
\end{equation}
and is consistent with that obtained in perturbation theory to first order \cite{Shifman:1978bx}.

\section{Thermodynamics} 
\label{secThermo} 

Having determined the black-hole solution of the five-dimensional gravity-dilaton-tachyon system we now
investigate the thermodynamics. We begin
with Hawking's black-hole thermodynamics. An expression for the temperature $T$ was derived in Section \ref{secThermodynamics}
\begin{eqnarray}
T(z_h) &=& -\frac{\partial_{z}f(z)}{4\pi}\Big|_{z=z_h} = \frac{1}{4\pi \mathcal{P}(z_h)\,a(z_h)^{3}}\nonumber\\
&=& \frac{2}{\pi z_{h}}\left(1-\frac{1}{\frac{1}{2} - 3\mu^{2}z_{h}^{2} - \frac{9\sqrt{3}z_{h}^{2}}{4\sqrt{2}\mu^{2}} + \frac{6\mu^{2}(\sqrt{6}z_{h}^{4}\mathcal{G}(z_h)-5)}{12\mu^{2} + 8\mu^{4}z_{h}^{2} + 3\sqrt{6}z_{h}^{2}\mathcal{G}(z_h)} }\right). \label{equexplicitT}
\end{eqnarray}
Of course, all thermodynamic quantities depend on the condensate function, whose behavior we address in Section \ref{secCond}.

As stated earlier, the graph of $T(z_h)$ contains information about the stability of the black-hole solution. The specific heat $c_{H}$ of the system is found using standard thermodynamics,
\begin{equation}
c_{H} = \frac{dE}{dT} = \frac{dE/dz_h}{dT/dz_{h}}.
\end{equation}
In general, $E'(z_h)<0$; therefore, the zeros of $T'(z_h)$ indicate where the specific heat switches sign. Whenever $T'(z_h)<0$, we find a stable AdS black hole. The behavior of the temperature as a function of $z_h$ with and without a condensate function is shown in Figure \ref{figTemp}. We use (\ref{equTinfty}) to indicate the general the large-$z_h$ behavior.

\subsection{Entropy} \label{subsecEntropy}
The entropy is found in the usual way from black-hole thermodynamics, a subject extensively covered in \cite{Bekenstein:1973,Bekenstein:1974,Hawking:1974sw}. As expected for a relativistic gas at high temperature, the entropy density, $s(T)$, behaves like $T^3$. 
In our model, we can also calculate the subleading temperature behavior and check that it is consistent.
We compute the entropy density from the area of the black-hole horizon using the induced metric $\gamma$. Using the metric (\ref{equseriesmetric}), we obtain
\begin{equation} \label{equBHarea}
A_{bh} = \int d^3x \,\sqrt{\gamma} = \frac{V\,R^3}{z_h^3}{\rm e}^{-3\left( \frac{\phi}{\sqrt{6}} + m2\,z_{h}^{2} + m6\,z_{h}^{6} + m8\,z_{h}^{8}\right)} ,
\end{equation}
where $V$ is the spatial volume. Using the first few terms in our metric expansion, the entropy density is then given by
\begin{equation}
s = \frac{4N_{c}^{2}V}{45\pi}\left(\frac{1}{z_{h}^{3}} - \frac{2\mu^{2}}{z_h} - \frac{3\sqrt{3}\mathcal{G}}{2\sqrt{2}\mu^{2}z_h} + 2\mu^{4}z_{h} + 2\sqrt{6} z_{h} \mathcal{G} + \frac{27 z_{h}\mathcal{G}^{2}}{16\mu^{4}}\right).
\end{equation}

\subsection{Speed of Sound} \label{secvsound}
With the temperature and entropy known, we also determine the speed of sound through the deconfined medium of the gauge theory. The speed of sound $v_s$ characterizes the hydrodynamic evolution of the deconfined, strongly coupled plasma. It has been suggested that $v_s^2$ in holographic models obeys an upper limit of 1/3 \cite{Cherman:2009tw}.
This can be checked for our solution using the relation
\begin{eqnarray} 
v_s^2 &=& \frac{s \frac{dT}{dz_h}}{T \frac{ds}{dz_h}} = \frac{d\log{T}}{d\log s} \nonumber\\
&=& -1 - \frac{1}{3}\frac{d\log{\mathcal{P}}}{d\log{a}}.
\end{eqnarray}
In our model, the exact form of $v_{s}^{2}$ greatly depends on the form of $\mathcal{G}$. However, we do confirm that 
\begin{equation} 
\frac{d\log{\mathcal{P}}}{d\log{a}} \rightarrow -4,
\end{equation}
in the high temperature limit, recovering the conformal limit with an upper bound of $v_s^2=1/3$.

\subsection{Free Energy} \label{secFEnergy}

The free energy of the deconfined phase is calculated in two ways. We can use thermodynamic identities to define the free energy as
\begin{equation}
\mathcal{F} = -\int{S\, dT}= \mathcal{F}_{{\rm min}}-\int{s\, V \frac{dT}{dz_{h}}dz_{h}}. \label{equfreeEdefine}
\end{equation}
where we have found the Bekenstein entropy and have an expression for $T(z_h)$. We have explicitly written the integration constant $\mathcal{F}_{{\rm min}}$ since the problem with using (\ref{equfreeEdefine}) is common among energy definitions, setting the zero point. Because $z_h(T)$ is a multi-valued function, the integral is non-trivial. Calculating the value $\mathcal{F}_{{\rm min}}$ generally requires the expression of entropy in the large-$z_h$ region, exactly where our expanded solutions are invalid. While the integral in (\ref{equfreeEdefine}) is completely calculable, we find the free energy by using the on-shell action. Finding the zero-point energy using actions is much easier; one merely subtracts the background free energy as defined by the thAdS action,
\begin{equation}\label{equfreeEaction}
\mathcal{F} = {\rm lim}_{\delta\rightarrow 0} T\left[\mathcal{S}_{bh}(\delta) - \mathcal{S}_0(\delta)\right].
\end{equation}
Both $\mathcal{S}_{bh}$ and $\mathcal{S}_{0}$ have been computed earlier. 

Before evaluating (\ref{equfreeEaction}), we must properly match the thermal AdS and black-hole AdS metrics at the boundary, $\delta\rightarrow 0$ \cite{Witten:1998zw,Gursoy:2008za}. This requires matching the intrinsic geometry of the two solutions at the boundary cut-off, where
\begin{eqnarray}
a_0(\delta) &=& a(\delta)\sqrt{f(\delta)}, \nonumber\\
V_{0}a_0(\delta)^{3} &=& V a(\delta)^{3}.\label{equmatch}
\end{eqnarray}
In order for (\ref{equmatch}) to be satisfied, we must evaluate the thAdS and bhAdS solutions at different cut-off points, $\tilde{\delta}$ and $\delta$ respectively. We find that 
\begin{equation}
\tilde{\delta} = \frac{\sqrt{8\mu^{2}\delta^{2} - 2\sqrt{6}\mathcal{G}\delta^{4} + \frac{3\sqrt{6}\mathcal{G}}{\mu^{2}}\delta^{2}}}{2\sqrt{2}\mu}.
\end{equation}
Combining the matching with (\ref{equfreeEaction}), we obtain a rather simple expression for the free energy,
\begin{equation}\label{equfreeEmod2}
\mathcal{F} = \frac{N_c^2 V}{45\pi^2 R^3} {\rm lim}_{\delta\rightarrow 0} \left(6 f(\delta) a(\delta)^{2}a'(\delta) + f' a(\delta)^{3} -6 \sqrt{f(\delta)} a(\delta)^4 \frac{a_0'(\tilde{\delta})}{a_0(\tilde{\delta})^2}\right),
\end{equation}
which can be reduced to
\begin{eqnarray} \label{equfreeEfG}
\mathcal{F} = \frac{N_c^{2}V}{45\pi^2}\left(f_{4}(z_h,\mathcal{G}) + 2\sqrt{6} \mathcal{G} \right) = \frac{2N_c^{2}V}{45\sqrt{6}\pi^2}\mathcal{G} -\frac{1}{4} T S.
\end{eqnarray}
It should be noted that we also checked that the black-hole energy $E$ satisfies the thermodynamic formula 
$E={\cal F} + T S$ by computing the ADM energy in Appendix \ref{appADM}.

\section{The Condensate Function}\label{secCond}

All the thermodynamic relationships rely on the behavior of the condensate function; therefore, we need to find a solution to $\mathcal{G}$ to evaluate the temperature, entropy, and speed of sound. The free energy of the system gives us enough information to solve for $\mathcal{G}$. We only need to set (\ref{equfreeEdefine}) and (\ref{equfreeEfG}) equal to one another. Taking the derivative with respect to $z_h$ removes any unknown constants, giving
\begin{equation}\label{equdiffforG}
\frac{df_{4}}{d z_h} + 2\sqrt{6}\frac{d\mathcal{G}}{d z_h} = s\,V\,\frac{dT}{d z_h}, 
\end{equation}
or in more simplified terms,
\begin{equation} \label{equdiffG}
    \frac{d\mathcal{G}(z_h)}{dz_h} = \frac{1}{2 \sqrt{6} \mathcal{P}(z_h)} \left(\frac{a'(z_h)}{a(z_h)} + \frac{\mathcal{P}'(z_h)}{4 \mathcal{P}(z_h)}\right),
\end{equation}
where
\begin{equation}
\mathcal{P}(z) = \int_{0}^{z} dx \,a(x)^{-3}.
\end{equation}
Unfortunately, (\ref{equdiffG}) is a stiff equation. Stiff differential equations contain terms that lead to rapid variations in its solution, causing numerical instabilities. One can often find stable solutions to these equations within a certain region, but our case is barely within a region of stability. 

\begin{figure}[h!]
\begin{center}
\includegraphics[scale=0.45]{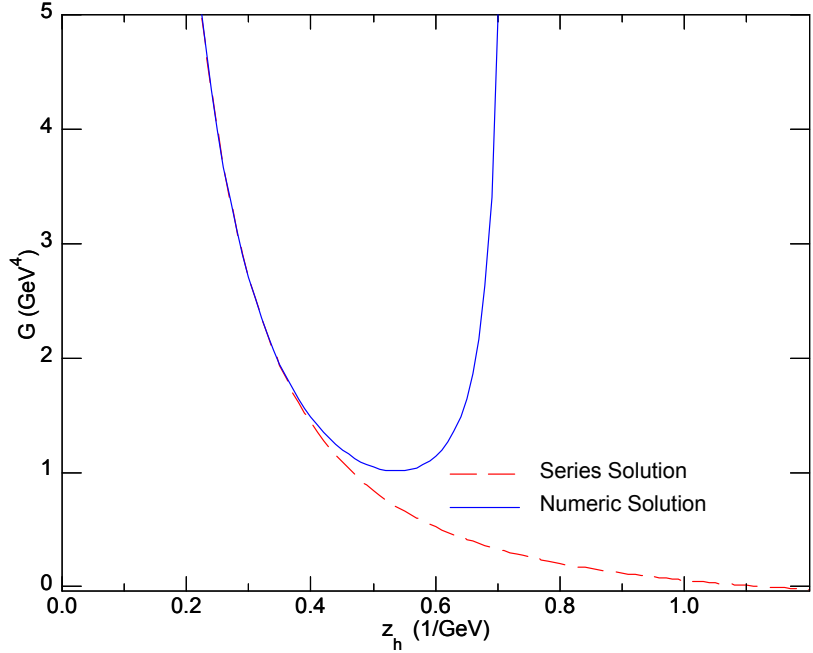}
\caption{The series and numerical solution for $\mathcal{G}$ is plotted. We use the series solution for the thermodynamics since the differential equation clearly gives a numerically unstable solution for $z\ge 0.3$ GeV$^{-1}$.}
\label{figcondensate}
\end{center}
\end{figure}

\begin{figure}[h!]
\begin{center}
\includegraphics[scale=0.45]{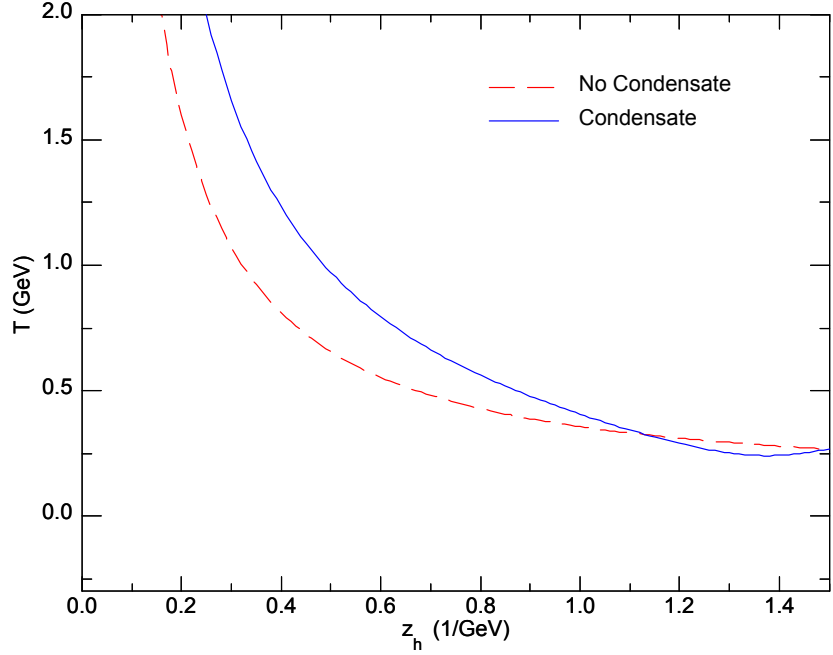}
\caption{The temperature as a function of the horizon location $z_h$ with and without a condensate 
$\mathcal{G}$. The inclusion of a condensate in this particular model precludes a second, unstable black-hole 
solution from developing.}
\label{figTemp}
\end{center}
\end{figure}

 We use two methods to solve for $\mathcal{G}$. First, we introduce a series expansion for $\mathcal{G}$,
\begin{equation}\label{equGseries}
\mathcal{G}(z_h) = \sum_{j=-\infty}^{\infty} g_{j} \mu^{j+4} z_{h}^{j},
\end{equation}
and expand (\ref{equdiffG}) in powers of $z_h$. Performing the expansion and matching coefficients, we find that the nonzero $g_{j}$'s are
\begin{eqnarray}\label{equGcoeff}
g_{-2}\equiv g = 1.43290,\nonumber\\
g_{0} = -6.09417,\nonumber\\
g_{2} = -29.7867,\nonumber\\
g_{4} = -570.637.
\end{eqnarray}
Since the nonlinear nature of (\ref{equdiffG}) spoils the series solution quite quickly, we find that only the first two terms give an accurate solution for $G$ in the range of $z_h<1$,
\begin{equation}
G(z_h) = 1.43290 \frac{\mu^{2}}{z_{h}^{2}} - 6.09417 \mu^{4}.
\end{equation}
 Using the leading order terms to inform the boundary conditions, we then numerically solve for $\mathcal{G}$ in (\ref{equdiffG}). Numerically, (\ref{equdiffG}) is difficult to solve because of its stiff nature. However, we are able to obtain a valid solution in a small range of $z_{h}$, which matches the series expansion, until $\mathcal{G}$ diverges. The two solutions are plotted in Figure \ref{figcondensate}. With a solution to $\mathcal{G}$, we can take a second look at the thermodynamics.

Given the expression (\ref{equseriesf}), the temperature can be written as 
\begin{eqnarray}
T(z_h) &\approx& \frac{2}{\pi z_h}\left(\frac{16+12\sqrt{6}g + 27g^{2}}{32+12\sqrt{6}g + 27g^{2}}\right) + \frac{\mu^{2}z_{h}^{2}}{\pi(32+12\sqrt{6}g + 27g^{2})^{2}}\times\nonumber\\
&&\qquad\Bigg(\frac{2048}{3}+516\sqrt{6}g+192 g^{2}+256\sqrt{6}g_{0} + 1728 g g_{0}\nonumber\\
&&\qquad+\, 216\sqrt{6} g^{2} g_{0}\Bigg).\label{equTexactwithG}
\end{eqnarray}
We clearly see that the condensate function contributes a finite piece to the temperature expression. Increasing the strength of the condensate produces 
\begin{equation}\label{equTlimit}
\lim_{g\rightarrow \infty}\,T(z_h)\rightarrow \frac{2}{\pi z_h},
\end{equation}
which is a factor of 2 larger than the conformal limit in (\ref{equT0}) or that found in \cite{Gursoy:2008za, Gursoy:2009kk}. We find that other work in this area has assumed that the condensate terms are suppressed logarithmically. However, in the construction of this model, the soft-wall set-up has no natural means of generating the logarithmic suppression in $\mathcal{G}$. As a result, we see that the condensate contributes to leading order behavior in the small-$z_h$/large-$T$ limit.

It will be useful to have an inverted function $z_{h}(T)$ which we use to transform functions of $z_h$ to functions of $T$. For simplicity, we use only the first-order term,
\begin{equation}\label{equTexpand}
z_{h}(T) \approx \frac{2}{\pi\,T}\left(\frac{16+12\sqrt{6}g + 27g^{2}}{32+12\sqrt{6}g + 27g^{2}}\right).
\end{equation} 

\begin{figure}[h!]
\begin{center}
\includegraphics[scale=0.45]{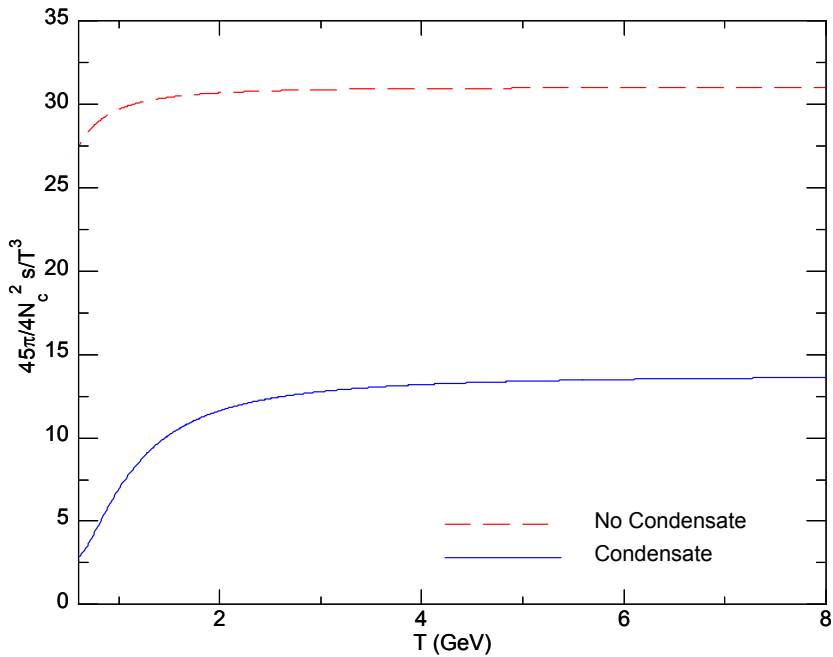}
\caption{The entropy density divided by $T^3$ as a function of temperature. At high temperatures the entropy density slowly evolves to the conformal case of $s\sim T^3$ in each case. }
\label{figEntropy}
\end{center}
\end{figure}

Analytically, the entropy of the deconfined phase can be written in terms of $z_h$, 
\begin{eqnarray}
s(z_h) &=& \frac{4N_c^2}{45\pi}\frac{1}{(32+16\sqrt{6}g+27g^{2})^{2}}\Bigg[32 -\frac{86\sqrt{6}}{7}g +\frac{279}{14}g^{2} - \frac{313\sqrt{3}}{14\sqrt{2}} \nonumber\\
&&\quad + \frac{3623}{168}g^{4} - \frac{313}{224\sqrt{6}}g^{5} + \frac{90081}{3584}g^{6}\Bigg] +\ldots\nonumber\\
&\approx& \frac{1.77}{z_{h}^{3}} - \frac{25.35\mu^{2}}{z_{h}} + 156.27 \mu^{4} z_{h} +\ldots,\label{equspecificentropy} 
\end{eqnarray}
or in terms of $T$,  
\begin{eqnarray}
s(T) &\approx& \frac{4N_{c}^{2} V}{45\pi}\frac{\pi^{3}T^{3}}{(16 + 12\sqrt{6}g + 27 g^{2})^{3}}\Big(4096+\frac{17664\sqrt{6}}{7}g + \frac{43200}{7} g^{2}  \nonumber\\ 
&&\quad- \frac{3056\sqrt{6}}{7} g^{3} - \frac{57520}{21} g^{4} - \frac{22331\sqrt{3}}{14\sqrt{2}}g^{5} +\frac{102549}{16}g^{6}  \nonumber\\
&&\quad+ \frac{992899\sqrt{3}}{112\sqrt{2}}g^{7}+12069g^{8} + \frac{9652689\sqrt{3}}{1792\sqrt{2}}g^{9} +  \frac{65669049}{28672}g^{10}\Big)\nonumber + \ldots,\\
&\approx& 13.88 T^{3} - 50.35 \mu^{2}\,T +\ldots, \nonumber\\
s(T) &=& C(T) T^{3},\label{equGEntropy}
\end{eqnarray}
where the function $\mathcal{C}(T)$ modifies the entropy behavior at low temperatures.
Our result, plotted in Figure \ref{figEntropy}, agrees qualitatively with the lattice data presented in \cite{Boyd:1996bx}. The high-temperature limit of $s(T)/T^3$, however, is shifted from lattice results because of the contribution from the condensate function.

\begin{figure}[h!]
\begin{center}
\includegraphics[scale=0.45]{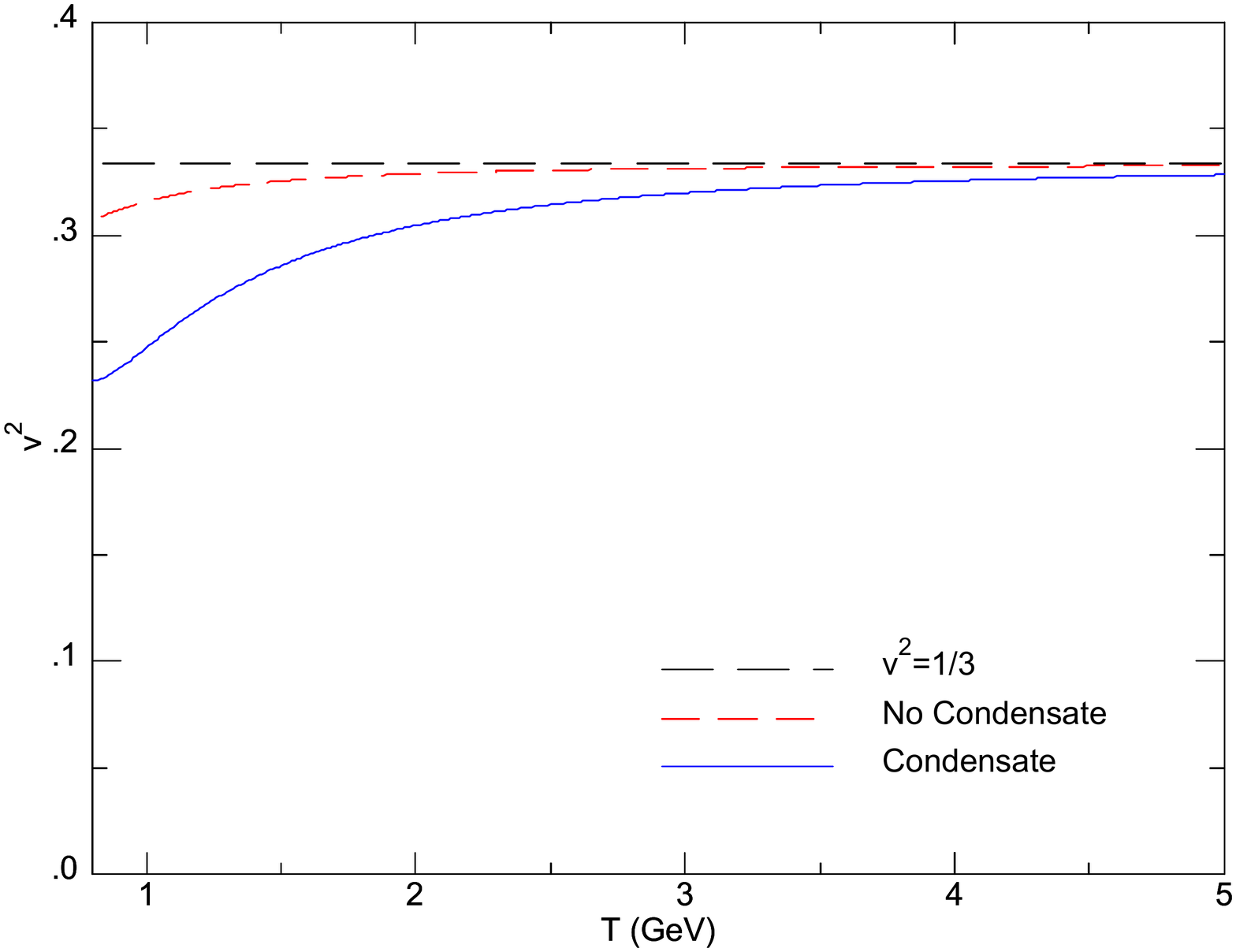}
\caption{The squared speed of sound in the strongly coupled plasma as a function of the temperature. The conformal limit of $v_s^2 = \frac{1}{3}$ is recovered at high temperatures.}
\label{figVelo}
\end{center}
\end{figure}

The speed of sound through the QGP-like thermal phase can be expressed as 
\begin{eqnarray}
v_{s}^{2}(z_h) &\approx& \frac{1}{3} - 2.718\mu^{2}z_{h}^{2} + 21.938 \mu^{4}z_{h}^{4},\label{equv2zh}\\
v_{s}^{2}(T)   &\approx& \frac{1}{3} - 0.689\frac{\mu^{2}}{T^{2}} + 1.409\frac{\mu^{4}}{T^{4}}.\label{equv2T}
\end{eqnarray} 
We clearly see that (\ref{equv2zh}) and (\ref{equv2T}) give the expected $v_{s}^{2} = 1/3$ in the small-$z$ and large-$T$ limit. The speed of sound through our QGP is plotted in Figure \ref{figVelo}. As the figure shows, $v_{s}^{2}$ deceases more rapidly with the condensate terms included; however, the difference between the two cases are not as stark as it was in the entropy case. Again, our results would match those of the no condensate scenario if the terms contributed by the condensate were logarithmically suppressed as in \cite{Gursoy:2008za, Gursoy:2009kk}.

An issue arises when we consider the behavior of $\mathcal{G}$. In Section \ref{secFOcorr}, we argued that the dual operator corresponding to the dilaton field is either a dimension-2 or a dimension-4 gluonic operator. On dimensional grounds, previous work asserts that $\mathcal{G}\sim T^{4}$. However, our model clearly shows that any self-consistent solution in the soft-wall model requires that $\mathcal{G}\sim  T^{2}$. The consequences of the temperature dependence on the field/operator duality are subjects for further research. 

\subsection{Deconfinement Temperature} 
\label{secDeTemp}

Finding the deconfinement temperature requires examining the free energy of the deconfined phase in (\ref{equfreeEfG}). The point at which $\mathcal{F}=0$ occurs when the black-hole solution becomes more energetically favored than the thermal solution. It is instructive to first consider the case without a condensate. By setting ${\cal G}(z_h)=0$ in (\ref{equfreeEfG}), the transition temperature is determined from the condition
\begin{equation}
\label{equHerzog}
f_{4}(\mathcal{G}(z_h),z_h)=0.
\end{equation} 
However, (\ref{equHerzog}) has only one possible solution, $z_{h}\rightarrow\infty$; therefore, no transition temperature exists since $z_h\rightarrow\infty$ occurs in the region of the unstable black hole. This mimics the scenario considered in \cite{Herzog:2006ra, Colangelo:2009ra}, but we include the full back-reaction of the scalar field and take the Gibbons-Hawking term into account. Therefore, we see that a nonzero condensate is needed to obtain a transition temperature in our model. 

\begin{figure}[h!]
\begin{center}
\includegraphics[scale=0.45]{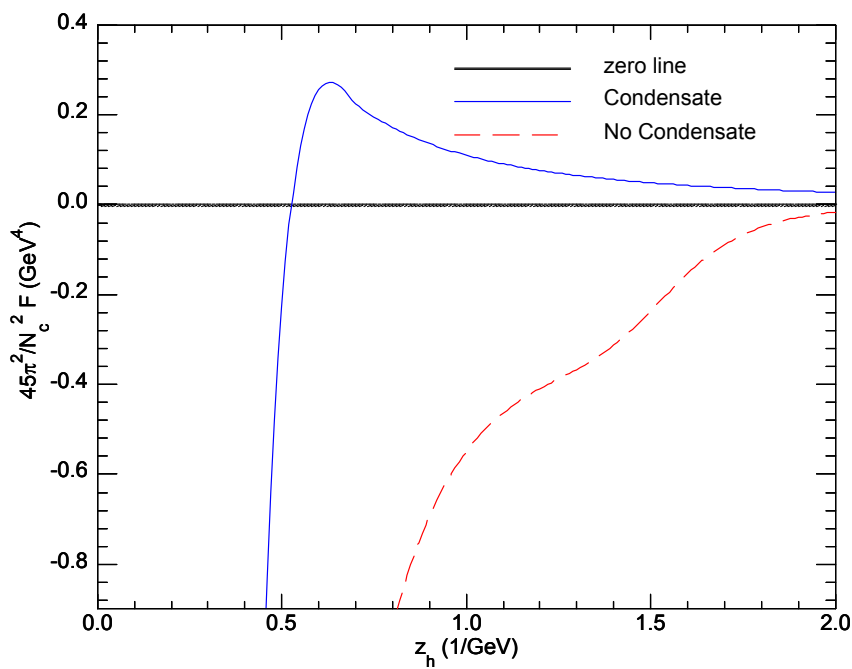}
\caption{The free energy $\cal F$ plotted as a function of $z_h$. Only when including a gluon condensate do we find a solution to $\mathcal{F} = 0$. The free energy with no gluon condensate approaches zero as $z_h\rightarrow\infty$, but never crosses the x-axis.}
\label{figEnergy}
\end{center}
\end{figure}

Using the leading behavior and the numerical solution, we find the free energy behavior and plot it in Figure \ref{figEnergy}. The phase transition occurs at a critical $z_c=0.5262$. Using (\ref{equexplicitT}) and the first two terms of the expansion for $\mathcal{G}$, this corresponds to a critical temperature of $T_{c}=919$ MeV. This is much larger than either theoretical reasoning or lattice calculations suggest \cite{Shifman:1978bx,Shifman:1978bw,Shifman:1978by, Boyd:1996bx, Miller:2006hr}. There are no current plans at the LHC for experiments that reach such high temperatures; however, experimental evidence for QGP has already been detected. The large transition temperature is most likely the product of the crude soft-wall model and the absence of any suppression terms in $\mathcal{G}$. As $z_h$ increases ($T$ decreases), we see that free energy goes to zero as in \cite{Gursoy:2008za}; however, we should only trust the free energy up to the point of validity of the bhAdS solutions, $z_h\approx 1$. Thus, the true behavior of the free energy at increasing values of $z_h$ requires more numerical work. 

\section{Conclusion}  
\label{secDiscuss}

We have presented a five-dimensional, gravity-dilaton-tachyon, black-hole, dynamical solution that represents a dual description of a strongly coupled gauge theory at finite temperature. The solution generalizes the soft-wall geometry considered in \cite{Batell:2008zm}, which generates a quadratic dilaton and Regge mass trajectory ($m_n\sim\sqrt{n}$). In the string frame, the dilaton and tachyon fields appear to be duals to a gluon and chiral operators, respectively. However, the actual field/operator correspondence is ambiguous. The black-hole solution describes a deconfined free-gluon phase. A transition from thAdS exists, predicted at a extremely high $T_{c}=919$ MeV. A condensate function is needed for any transition to occur.

 Despite being a mere series expansion, we argue that it gives insight into thermodynamics of a strongly coupled gauge theory produced by the soft-wall model. We expected to find two major contributions to the thermodynamics: the underlying conformal limit and condensate terms. Although we considered a finite number of terms in our infinite solution expansion, our calculations suggest that the condensate terms produce convergent quantities, as shown in (\ref{equTlimit}). With further study, we believe a viable closed-form, black-hole solution with a lower transition temperature will be found. A much more reasonable transition temperature is obtained by the phenomenological model in Appendix \ref{appPhenom}.

While our model has some features reminiscent of QCD at finite temperature, it still represents a crude approximation with a number of shortcomings. The bhAdS solution is only valid in the region of small $z_h$. The power-law dependence of the scalar fields and metric does not include the logarithmic corrections needed to suppress condensate terms as in \cite{Gursoy:2008za}, resulting in noticeable changes in the behaviors of the temperature, entropy, and speed of sound. Furthermore, using the AdS/CFT dictionary, the dilaton field appears to be dual to a dimension-2 operator, $(A_{\mu})_{min}^{2}$, in contradiction to the standard assumption of a dimension-4 operator, $Tr[(F^{a})^{2}]$. Nevertheless, our five-dimensional dynamical black-hole solution with two scalar fields provides a toy model to understand the nontrivial properties of strong-coupled gauge theories at finite temperature.

\chapter{Discussion and Conclusion}\label{conclusion}
\label{conclusion_chapter}

\begin{flushright}
``I've learned that people will forget what you said, people will forget what you did, but people will never forget how you made them feel.''\\
-Maya Angelou
\end{flushright}

We began this thesis with a quick overview of gauge/gravity dualities. In Chapter \ref{chbackground}, we reviewed basic concepts of AdS/CFT and its connection to AdS/QCD. We took a new perspective on the mass spectra produced by the hard-wall and soft-wall AdS/QCD correspondence in Chapter \ref{chApplications}. In the same chapter, we introduced thermodynamics into the correspondence via a black-hole metric. In Chapter \ref{chzero}, we explored the consequences of adding higher-order interaction terms in the action. We found a new way of expressing spontaneous and explicit CSB. The modified background dilaton field produces meson mass spectra that agree with experimental results better than the previous pure linear trajectories. We also edified the pseudoscalar sector that had been neglected by past soft-wall work. In Chapter \ref{chthermo}, we investigated the means of dynamically generating the soft-wall from the gravitational action. We explore an approximated series solution that produces a quadratic dilaton and linear tachyon to leading order.

The phenomenological soft-wall AdS/QCD correspondence mimics QCD in a number of aspects: (i) radial Regge trajectories, (ii) running coupling constant, (iii) Gell-Mann--Renner--Oakes relation, (iv) deconfinement temperature, and (v) high-temperature thermodynamics. In our research, we have shown that non-trivial improvements are made when adding higher-order terms. Low-mass resonances that experimentally show non-Regge behavior are reproduced. Independent sources of spontaneous and explicit chiral symmetry naturally arise. The thermodynamics of the soft-wall model is much more nuanced. Solving Einstein's equations that dynamically generate the soft-wall has proven exceedingly difficult. Predictions from such attempts give the right high-temperature behaviors, and a transition between confined matter and deconfined plasma occurs. However, results are skewed by leading order contributions from condensates that are thought to be logarithmically suppressed. Phenomenological soft-wall models appear to describe the thermodynamics of deconfined QCD much better, as shown in Appendix \ref{appPhenom}.

Perennial limitations and shortcomings plague some aspects of the soft-wall AdS/QCD correspondence. Currently, the running coupling of QCD cannot be analytically calculated at low energy. Therefore, we cannot ensure that the gauge coupling in this model,
\begin{equation}
\lambda={\rm e}^{\Phi(z)},
\end{equation}
resembles QCD at the relevant energies. The series solution in Chapter \ref{chthermo} is also problematic. This solution is only valid in the regime of $z,z_h<1$. However, small $z$ corresponds to high energy, the region in which the strong-coupling assumptions should fail.   

Our research could be expanded in a number of ways. Since the bulk coordinate $z$ is related to the renormalized energy scale, one needs to consider running parameters as functions of $z$,
\begin{equation}
m_{q}\rightarrow m_{q}(z)\qquad\qquad \sigma\rightarrow \sigma(z).
\end{equation} 
We may even incorporate the anomalous dimensions of operators by assuming the scalar mass from the field/operator correspondence is also dependent upon $z$ \cite{Vega:2010ns}. Further study must be done to sort out the field/operator correspondence of the dilaton $\phi$. Whether this scalar field is dual to $Tr(F^{2})$ or $A_{\mu}^{2}$ may have consequences to other parts of the theory. In addition, we should investigate if the $A_{\mu}^{2}$ contains relevant information for the gauge theory. The most fertile area for further research involves the thermodynamics. We presented a series solution that produces many general properties of finite-temperature QCD, but numerical analysis may reveal more richness.

In our research, we attempted to evaluate and improve the relatively simple correspondence presented by \cite{Karch:2006pv}. Further study into these topics, especially those of Chapter \ref{chthermo} will most likely require extensive numerical work and a deeper understanding of stringy effects. While AdS/QCD is not yet a precision tool for QCD calculations, it provides qualitative and quantitative insights to difficult problems involving strongly coupled gauge theories. 


\bibliography{thesis}

\appendix

\chapter{Mathematical Methods}

\section{Useful Relations}\label{appuseful}

Bessel Functions:
\begin{eqnarray}
J_{n}(x\rightarrow\infty) &\sim& \sqrt{\frac{2}{\pi x}}{\rm cos}\left(x-\frac{n\pi}{2}-\frac{\pi}{4}\right)\label{appBesselJlimit}\\
Y_{n}(x\rightarrow\infty) &\sim& \sqrt{\frac{2}{\pi x}}{\rm sin}\left(x-\frac{n\pi}{2}-\frac{\pi}{4}\right)\label{appBesselJlimit} 
\end{eqnarray}
Incomplete Gamma Functions:
\begin{eqnarray}
\Gamma(s, x\rightarrow\ 0) &\sim& \Gamma(s) - \frac{e^{-x}x^{s}}{s}\label{appgamma0} \\
\Gamma(s,x\rightarrow\infty) &\sim& -x^{s-1} e^{-x}\label{appgammainf}\\
\Gamma(s,0) &=& \Gamma(s)
\end{eqnarray}

\subsection{Common Differential Equation}\label{appdiffeq}

When solving mass eigenvalue equations in the AdS/QCD soft-wall model, we often face a differential equation of the form 
\begin{equation}
\psi''(z) + \left(z^{2} + \frac{m^{2}-\frac{1}{4}}{z^{2}}\right)\psi(z) = E \psi(z). 
\end{equation}
The eigenfunctions, $\psi_{n}$, and eigenvalues, $E$, are found in \cite{Karch:2006pv},
\begin{eqnarray}
\psi_{n}(z) &=&  e^{-\frac{z^{2}}{2}}z^{m+\frac{1}{2}} \sqrt{\frac{2 n!}{(m+n)!}}L_{n}^{m}(z^{2}),
\end{eqnarray}
where $L_{n}^{m}$ are associated Laguerre polynomials.

\section{Schr\"{o}dinger Transform}\label{appSchTransform}

This transform is used to eliminate all first order derivatives from a second-order differential equation of the form, 
\begin{equation} \label{equAppPsi}
-\Psi''(z) + \omega'(z)\Psi'(z) + W(z)\Psi(z) = m^{2}\Psi(z),
\end{equation}
where ($'$) denotes differentiation with respect to $z$ \cite{Karch:2006pv}.
One can substitute 
\begin{eqnarray}
\Psi &=& e^{\frac{\omega(z)}{2}}\psi, \\
\Psi' &=& e^{\frac{\omega(z)}{2}}\left(\psi' + \omega'\,\psi\right),\\
\Psi'' &=& e^{\frac{\omega(z)}{2}}\left(\psi''+\omega'\,\psi' + \frac{\omega''}{2}\,\psi + \frac{\omega'^{2}}{4}\,\psi\right),
\end{eqnarray}
into (\ref{equAppPsi}) so that it becomes
\begin{eqnarray}
-\psi'' + \frac{\omega'^{2}}{4}\psi - \frac{\omega''}{2}\psi + W(z)\psi = m^{2}\psi.
\end{eqnarray}
Thus, we have a Schr\"{o}dinger-like equation,
\begin{equation}\label{equAppSchlike}
-\psi'' + V(z)\psi = m^{2} \psi,
\end{equation}
where
\begin{equation}
V(z) = \frac{\omega'^{2}}{4} - \frac{\omega''}{2} + W(z).
\end{equation}

\section{Numerical Methods}

\subsection{Shooting Method}\label{appShooting}

\begin{figure}[h!]
\begin{center}
\includegraphics[scale=0.60]{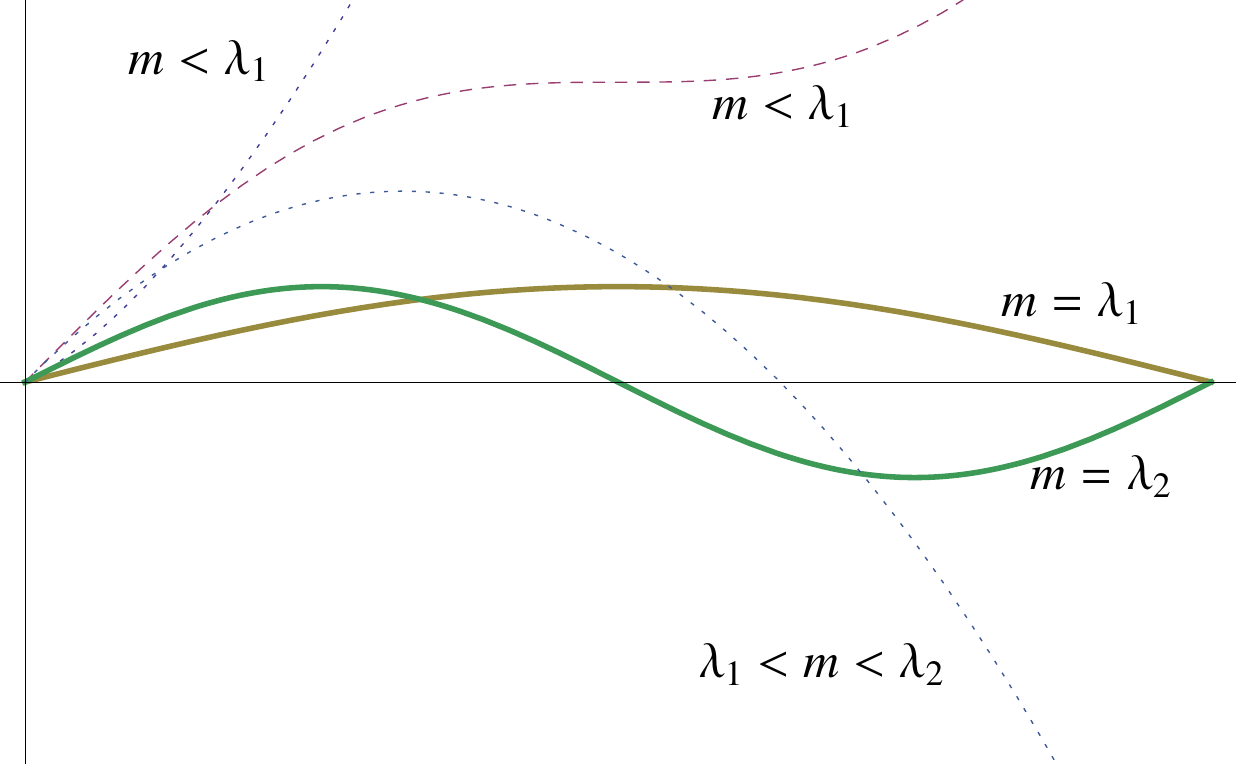}
\caption{The two dotted lines show the wavefunction as far away from any eigenvalues $\lambda_{i}$. The first dotted line has no antinodes and is below the first eigenvalue. The second dotted line has one antinode but still diverges; therefore, it is between the first and second eigenvalues. The dashed line shows a wavefunction for $m$ approaching the first eigenvalue, however, it has not changed its direction of divergence. The two solid lines represent eigenfunctions corresponding to the ground state (one antinode) and the first excited state (two antinodes).}
\label{figShooting}
\end{center}
\end{figure}

Since most eigenvalue problems involving equations similar to (\ref{equAppSchlike}), where $\psi\rightarrow\psi_{n}$ and $m\rightarrow m_{n}$, cannot be solved analytically, and thus, do not give a closed expression for $m_{n}^{2}$, a numerical shooting method was employed. This shooting method involves iterating through all values of $m_{n}$ and numerically solving for the wavefunction, $\psi_{n}$. For values of $m_{n}$ not equal to the eigenvalue,  $\psi_{n}$ diverges; the direction of the divergence is the informative piece of information. As we iterate through wavefunction solutions, we discover that if successive values of $m_{n}$ produce $\psi_{n}$'s that switch their direction of divergence, then the true eigenvalue lies between those values. The excitation mode, $n$, whose ground state begins at $n=1$, is then determined by the number of antinodes, $a$, that exist in the function where $n=a$. Figure \ref{figShooting} illustrates the shooting method concept. 

\subsection{Matrix Method}\label{appMatrix}

The equations of motion can be reduced to a set of second-order differential equations,
\begin{eqnarray}
-\phi'' + V_1(z)  \phi + f(z)\pi  = 0, \label{equphiAppend}\\
-\pi'' + V_2(z) \pi + g(z)\phi = 0, \label{equpiAppend}
\end{eqnarray}
where the eigenvalues are contained within the coefficient functions. These equations can be expressed as a system of first-order differential equations
\begin{equation} \label{equmatrixphi}
\Phi' + W(z) \Phi = 0,
\end{equation}
where $W$ is the matrix
\begin{equation}
W = \left(\begin{array}{cccc}
0     & 1     & 0     & 0 \\ 
V_1(z)& 0     &   f(z)& 0 \\
0     & 0     & 0     & 1 \\
g(z)  & 0     & V_2(z)& 0
\end{array}\right)
\end{equation}
and $\Phi$ is the vector
\begin{equation}
\Phi_{\alpha i} = \left(\begin{array}{c} \phi_{i} \\
			-\phi'_{i} \\
			\pi_{i} \\
			-\pi'_{i} \end{array}\right)		
\end{equation}
that forms an orthonormal basis of solutions. We can propagate the solution $\Phi$ between two boundary points
\begin{equation}
\Phi(z_1) = U(z, z_1, z_0, m_n^2)\Phi(z_0),
\end{equation}
where we solve (\ref{equmatrixphi}) with the appropriate boundary condition at $z_0$. The eigenvectors and eigenvalues of $U$ are then calculated. We find two large and two small eigenvalues corresponding to two nonrenormalizable and two normalizable eigenfunctions, respectively. Let us assume the eigenvectors $u_3$ and $u_4$ correspond to the small eigenvalues, $\lambda_3$ and $\lambda_4$. Then, any solution for $\Phi_{i}$ can be written as 
\begin{equation}
\Phi_{i} = \alpha u_3 + \beta u_4,
\end{equation}   
where we take the boundary condition as 
\begin{equation}
\Phi_{i}(z_0) =\left( \begin{array}{c} \phi(z_0) \\ -\phi'(z_0) \\ \pi(z_0) \\ -\pi'(z_0) \end{array}\right) \, . 
\end{equation}
In order for $\alpha$ and $\beta$ to be nontrivial, we must satisfy 
\begin{equation}\label{equNontrivial1}
\left(\begin{array}{cc} u_3^1 & u_4^1 \\ u_3^{3} & u_4^{3} \end{array}\right)\left(\begin{array}{c} \alpha \\ \beta\end{array}\right) = 0
\end{equation}
for Dirichlet or
\begin{equation}\label{equNontrivial2}
\left(\begin{array}{cc} u_3^2 & u_4^2 \\ u_3^{4} & u_4^{4} \end{array}\right)\left(\begin{array}{c} \alpha \\ \beta\end{array}\right) = 0
\end{equation}
for Neumann boundary conditions. We do this by cycling through eigenvalues $m_n^2$ that minimize the determinant of the 2$\times$2 matrix in (\ref{equNontrivial1}) or (\ref{equNontrivial2}). Practically, we find the singular points in the graph of the quantity $u_3^1 u_4^3 - u_3^3 u_4^1$ (or $u_3^2 u_4^4 - u_3^4 u_4^2$) versus $m_n^2$. An abrupt change in its behavior signals an eigenvalue. Of course, the elements chosen from the eigenvectors $u_3$ and $u_4$ are dependent upon the choice of Neumann or Dirichlet conditions on the boundary $z_0$.

\chapter{Alternative Parameterization for the Scalar VEV}
\label{appvev}

In this section, we explore two different parameterizations of the scalar VEV. While the mass spectra are modified, we see that the results are qualitatively the same. The parametrizations that we investigate take the form of a Pad{\'e} approximant and a Gaussian function,
\begin{eqnarray}
v_{P} &=& \frac{a_{1}z + a_{2}z^{3}}{1+a_{3} z^{2}}, \label{equpade}\\
v_{G} &=& b_{1} z + b_{2} z \,e^{b_{3}z^{2}} \label{equgauss}.
\end{eqnarray}

\section{Pad{\'e} Approximant}
Using the form (\ref{equpade}), we find that the desired limiting behavior of the VEV is satisfied,
\begin{eqnarray}
v_{P}(z\rightarrow 0) &=& a_{1} z +  (a_{2} - a_{1}a_{3}) z^{3},\\
v_{P}(z\rightarrow \infty) &=& \frac{a_{2}}{a_{3}} z.
\end{eqnarray}
According to the correspondence dictionary, the coefficients take the values
\begin{eqnarray}
a_{1} &=& \frac{\sqrt{3}m_{q}}{g_{5}},\\
a_{2} &=& \frac{2 g_{5}^{2}\mu\sigma}{2\sqrt{3} g_{5}\mu - 3 m_{q}\sqrt{\kappa}},\\
a_{3} &=& \frac{g_{5}^{2}\sigma\sqrt{\kappa}}{2\sqrt{3}g_{5}\mu - 3 m_{q}\sqrt{\kappa}}.
\end{eqnarray}
Given this particular form for the scalar VEV, the dilaton becomes
\begin{equation}
\frac{d\phi_{P}(z)}{dz} = \frac{16 a_{3}(a_{1}a_{3}-a_{2})z^{3} + \kappa(a_{1}+a_{2}z^{2})^{3}z}{2(1+ a_{3}z^{2})\left(a_{1}+(3 a_{2}-a_{1}a_{3})z^{2} + a_{2}a_{3}z^{4}\right)},
\end{equation}
with asymptotic behavior,
\begin{eqnarray}
\phi_{P}(z\rightarrow 0) &=& \frac{\kappa a_{1}}{4}z^{2} + \left(2 a_{3}^{2} - \frac{2 a_{2}a_{3}}{a_{1}}\right)z^{4} + \ldots \\
&=& \frac{3 \kappa m_{q}^{2}}{4 g_{5}^{2}}z^{2} + \frac{2\sigma^{2}g_{5}^{2}(3 m_{q}\kappa + 2\sqrt{3\kappa} g_{5}\mu)}{9(3m_{q}^{3}\kappa - 4 g_{5}^{2}m_{q}\mu^{2})} + \ldots,\\
\phi_{P}(z\rightarrow\infty) &=& \frac{\kappa a_{2}^{2}}{4a_{3}^{2}}z^{2} + \ldots\\
&=& \mu^{2}z^{2} + \ldots. \label{equpadelargephi}
\end{eqnarray}
In the large-$z$ limit, the leading behavior of $\phi_{P}$ is the quadratic term $\mu^{2}z^{2}$ as expected. Using the same equations of motion and methods detailed in Chapter \ref{chzero}, we find the scalar, pseudoscalar, vector, and axial-vector meson mass spectra produced by (\ref{equpade}), listed in Table \ref{tblpade}

\begin{table}[h!]
\begin{center}
\begin{tabular}{|c||c|c|c|c|}
\hline
$n$ & $m_{f_{0}}$ (MeV) & $m_{\pi}$ (MeV) & $m_{\rho}$ (MeV) & $m_{a_{1}}$ (MeV)\\
\hline
\hline
1  &  440 &  140 &  390 &  940 \\
\hline
2  &  880 & 1194 &  840 & 1290 \\
\hline
3  & 1180 & 1499 & 1150 & 1550\\
\hline
4  & 1430 & 1746 & 1400 & 1770\\
\hline
5  & 1650 & 1993 & 1620 & 1970\\
\hline
6  & 1840 & 2255 & 1820 & 2150\\
\hline
7  & 2020 & 2528 & 2000 & 2310\\
\hline
8  & 2190 & 2812 & 2170 & 2460 \\
\hline
\end{tabular}
\end{center}
\caption{Meson mass spectra produced by the Pad{\'e} Approximant.}
\label{tblpade}
\end{table}

\section{Gaussian Function}
Using the form (\ref{equgauss}), we again find the desired limiting behavior of the VEV,
\begin{eqnarray}
v_{G}(z\rightarrow 0) &=& (b_{1}+b_{2}) z +  b_{2}b_{3} z^{3},\\
v_{G}(z\rightarrow \infty) &=& b_{1} z.
\end{eqnarray}
According to the correspondence dictionary, the coefficients become
\begin{eqnarray}
b_{1} &=& \sqrt{\frac{4 \mu^{2}}{\kappa}},\\
b_{2} &=& \frac{\sqrt{3}m_{q}}{g_{5}} - b_{1},\\
b_{3} &=& \frac{g_{5}\sigma}{b_{2}\sqrt{3}}.
\end{eqnarray}
Given (\ref{equgauss}), the dilaton takes the form 
\begin{equation}
\frac{d\phi_{G}(z)}{dz} = \frac{b_{1}^{3} \kappa z e^{b_{3}z^{2}} + 8 b_{2}b_{3}^{2}z^{3}+ 3 b_{1}^{2}b_{2}\kappa z + 3 b_{1}b_{2}^{2}\kappa z e^{-b_{3}z^{2}} +b_{2}^{3}\kappa z e^{-2b_{3}z^{2}} }{2(b_{2} + b_{1}e^{b_{3}z^{2}} - 2 b_{2}b_{3}z^{2})},
\end{equation}
whose asymptotic behavior becomes
\begin{eqnarray}
\phi_{G}(z\rightarrow 0) &=& \frac{(b_{1}+b_{2})\kappa}{4}z^{2} + \frac{b_{2}b_{3}^{2}}{b_{1}+b_{2}}z^{4} + \ldots\\
&=& \frac{\sqrt{3}m_{q}\kappa}{4 g_{5}}z^{2} + \left(\frac{g_{5}^{2}\sigma^{2}}{3} - \frac{2 g_{5}^{3}\mu\sigma^{2}}{3\sqrt{3\kappa}m_{q}}\right)z^{4}+\ldots,\\
\phi_{G}(z\rightarrow \infty) &=& \frac{\kappa b_{1}^{2}}{4}z^{4} + \ldots\\
&=& \mu^{2}z^{2} + \ldots. \label{equgausslargephi}
\end{eqnarray}
We recover the quadratic dilaton in the large-$z$ limit. Using the equations of motion and the methods described in Chapter \ref{chzero}, we find the meson mass spectra for the scalar, pseudoscalar, vector, and axial mesons, listed in Table \ref{tblGauss}.

\begin{table}[h!]
\begin{center}
\begin{tabular}{|c||c|c|c|c|}
\hline
$n$ & $m_{f_{0}}$ (MeV) & $m_{\pi}$ (MeV) & $m_{\rho}$ (MeV) & $m_{a_{1}}$ (MeV)\\
\hline
\hline
1  &  660 &  141 &  450 & 1080 \\
\hline
2  & 1130 & 1402 & 1020 & 1500 \\
\hline
3  & 1440 & 1779 & 1380 & 1820 \\
\hline
4  & 1680 & 2037 & 1650 & 2060 \\
\hline
5  & 1890 & 2251 & 1870 & 2250\\
\hline
6  & 2080 & 2478 & 2060 & 2420 \\
\hline
7  & 2250 & 2721 & 2230 & 2570 \\
\hline
8  & 2410 & 2982 & 2400 & 2720 \\
\hline
\end{tabular}
\caption{Meson mass spectra produced by the Gaussian form (\ref{equgauss}).}
\label{tblGauss}
\end{center}
\end{table}

\chapter{Holographic Thermodynamics}

\section{Superpotential Method at Finite Temperature}
\label{appsuperpot}

An alternative solution technique can also be used to solve the five-dimensional Einstein's equations. It generalizes the superpotential method considered in \cite{Batell:2008zm} to the case of finite temperature \cite{Gursoy:2008za}. The method presents a simple way to convert a system of second-order differential equations into a system of first-order equations. First, we must take the metric \ref{equbhmetric} and define a new bulk coordinate $u$, where
\begin{equation}
a(z)dz = du.
\end{equation}
We express a new metric form,
\begin{equation}\label{equinitialmetric}
ds^2 = e^{2 A(u)}(e^{F(u)} d\tau^2 + d\vec{x}^2) + e^{-F(u)}du^2,
\end{equation}  
where $f=e^{F}$, $a=e^{A}$.
The Einstein equations become
\begin{eqnarray}
{\ddot A} + \frac{1}{6}{\dot\phi}^2 + \frac{1}{6}{\dot\chi}^2 &=& 0, \label{equEinfields} \\
12{\dot A}^{2} + 3 {\dot A}{\dot F} -\frac{1}{2}(\dot\phi^{2} + \dot\chi^{2}) + e^{-F}V &=& 0, 
\label{equEinpotential}\\
4 {\dot A} +{\dot F} +\frac{\ddot F}{{\dot F}} &=& 0, \label{equEinBH}
\end{eqnarray}
where ($\dot{}$) denotes derivatives with respect to $u$. A superpotential $\mathcal{W}$ can be introduced 
to replace the second-order equation (\ref{equEinfields}) by two first-order equations. 
The system of equations then becomes
\begin{eqnarray}
&&\qquad {\dot A} = -\frac{1}{6}\mathcal{W}(\phi,\chi), \qquad  {\dot\chi} = \frac{\partial\mathcal{W}}{\partial\chi}, \qquad
\dot\phi = \frac{\partial\mathcal{W}}{\partial\phi}, \label{superpoteqn}\\
&&\qquad\qquad\qquad\qquad\mathcal{W}(\phi,\chi) =\frac{3}{2}\left({\dot F} + \frac{\ddot F}{\dot F}\right),\\
&&-\frac{1}{2}\left[\left(\frac{\partial {\cal W}}{\partial\phi}\right)^2 + \left(\frac{\partial {\cal W}}{\partial\chi}\right)^2\right] +\frac{1}{3} {\cal W}(\phi,\chi)^2 -\frac{1}{2} {\cal W}(\phi,\chi) {\dot F} \nonumber\\
&&\qquad\qquad\qquad\qquad\qquad\qquad\qquad\qquad\qquad+ e^{-F} V(\phi,\chi) = 0.
 \label{equSPBH} 
\end{eqnarray}
Using this system, we can arrive at similar solutions for the functions $\phi$, $\chi$, $f$, and $V(\phi,\chi)$ as in Chapter \ref{chthermo}. In our research, we did not find that the superpotential method held an advantage over our method.

\section{The Arnowitt-Deser-Misner (ADM) Energy}
\label{appADM}
The ADM energy is a useful definition of energy for gravitational systems approaching an asymptotic, well-defined metric at the boundary \cite{Arnowitt:1961zz,Hawking:1995fd}. To verify the thermodynamic relation $E={\cal F}+T\,S$, we will compute the ADM energy for our
black-hole solution with respect to the thermal AdS solution. Considering a time slicing of the five-dimensional metric in ADM form,
\begin{equation}
     ds^2 = -N^2 dt^2 +\gamma_{ij}(dx^i-N^i dt)(dx^j-N^j dt),
\end{equation}
with $i,j=x_{1},x_{2},x_{3},z$. In our case, we have
\begin{eqnarray}
N=a(z)\sqrt{f(z)},\\
\gamma = a(z)^{2}\left(d\vec{x}^{2} + \frac{dz^{2}}{f(z)}\right).
\end{eqnarray}

The  expression for the ADM energy is given by \cite{Gursoy:2008za, Hawking:1995fd},
\begin{equation}
      E=-\frac{1}{8\pi G_5} \int d\Sigma_B N\left( \sqrt{-\gamma_{ind}} ^{(3)}K -  \sqrt{-\gamma_{0,ind}} ^{(3)}K_0 \right)~.
\end{equation}
The integral is performed over a three-dimensional surface at the bulk boundary $\Sigma_B$ embedded in the 4D constant time slice $\Sigma_t$, where $\gamma_{ind}$ is the induced three-dimensional metric. The superscript $^{(3)}$ refers to the dimensionality of the extrinsic curvature. Using three-dimensional equivalents of (\ref{defnintcurv}) and (\ref{defnormal}), we obtain
\begin{equation}
       E=\frac{2 N_c^2}{15\pi^2}\frac{V}{L^3} a^2(\delta) \sqrt{f(\delta)}\left( \sqrt{f(\delta)}a'(\delta) - a^2(\delta)
       \frac{a_0'(\delta)}{a_0^2(\delta)}\right),
\end{equation}
where we have matched the two solutions at the AdS boundary as specified in \ref{equmatch}. Taking the limit $\delta\rightarrow 0$ gives the 
final expression
\begin{equation}
       E=-\frac{N_c^2V}{15\pi^2} \left(f_{4}(z_h,\mathcal{G}) -\frac{2 \sqrt{6}}{3} {\cal G}(z_h)\right) = {\cal F} + T S,
\end{equation}
where we have used the fact that
\begin{equation}
S T = -\frac{4 N_{c}^{2} V}{45 \pi^{2}} f_{4}. 
\end{equation}

\section{Phenomenological Thermodynamics}\label{appPhenom}

As shown in Chapter \ref{chthermo}, obtaining an exact solution for the soft-wall model with a black hole is challenging. Thus, it is sometimes advantageous to follow the logic of \cite{Herzog:2006ra, Colangelo:2009ra}, where we are not concerned with dynamically generating the soft-wall geometry. In this case, we allow total freedom within the potential to satisfy all the necessary equations of motion. The advantage, of course, is removing an added complication; however, we sacrifice any thermodynamic relations useful in calculating the transition temperature, namely $\mathcal{F} = \int S dT$. 

To begin the phenomenological analysis of bhAdS, we start with the same action in the Einstein frame,
\begin{eqnarray}
S_E &=&-\frac{1}{16 \pi G_{5}} \int d^5x \sqrt{-g}\left(\mathcal{R}- \frac{1}{2}(\partial\phi)^2 - \frac{1}{2}(\partial\chi)^2- V(\phi, \chi)\right)\nonumber\\
&&\quad+\frac{1}{8 \pi G_5}\int d^4 x \sqrt{-\gamma} K,
\end{eqnarray}
using the same metric form,
\begin{equation}
ds^{2} = a(z)^{2}\left(-f(z)\,dt^{2} + d\vec{x}^{2} + \frac{dz^{2}}{f(z)}\right).
\end{equation}
We assume forms for the dilaton $\phi(z)$ and the metric warping factor $a(z)$,
\begin{eqnarray}
a(z) &=& \frac{R}{z}e^{\frac{-\phi(z)}{\sqrt{6}}},\label{eqappametric}\\
\phi(z) &=& \sqrt{\frac{8}{3}}\mu^{2}z^{2} - \mathcal{G}(z_h) z^{4},
\end{eqnarray}
 and then consider two of the equations of motion,
\begin{eqnarray}
12 \frac{a'(z)^{2}}{a(z)^{2}} - 6 \frac{a''(z)}{a(z)} &=& \phi'(z)^{2} + \chi'(z)^{2},\label{eqappfield}\\
f''(z) + 3 f'(z) \frac{a'(z)}{a(z)} &=& 0.\label{eqappmetricf}
\end{eqnarray}
Using (\ref{eqappfield}), we solve for the tachyon field $\chi(z)$,
\begin{eqnarray}
\chi(z) \approx z\sqrt{24\mu^2 - \frac{9\sqrt{6}\mathcal{G}(z_h)}{\mu^{2}}} + z^{3}\frac{G(z_h)\left(\frac{3G(z_h)}{2 \mu^{4}}-\sqrt{6}\right)}{\sqrt{24\mu^2 - \frac{9\sqrt{6}\mathcal{G}(z_h)}{\mu^{2}}}}, \label{equchiexpand}
\end{eqnarray}
where in (\ref{equchiexpand}), the integration constant from the $\chi$ equation has been chosen so that $\chi(z) \rightarrow \chi_0(z)$ as $z\rightarrow 0$ and (\ref{equchiexpand}) is the expansion about $z=0$. Imposing the black-hole conditions on (\ref{eqappmetricf}),
\begin{equation} \label{equfboundary}
f(0) = 1, \quad\quad\quad f(z_h) = 0,
\end{equation}
we find that 
\begin{equation} 
\label{equfgeneral}
f(z) = 1-\frac{\int_{0}^{z} dw\,{a(w)^{-3}}}{\int_{0}^{z_h} dw\,{a(w)^{-3}}} \equiv 
1 - \frac{\mathcal{P}(z)}{\mathcal{P}(z_h)}= 1-f_{4}(z_h) z^4 + \mathcal{O}(z^6).
\end{equation}
The integral expression in (\ref{equfgeneral}) can be evaluated using (\ref{eqappametric}) to give
\begin{eqnarray} 
\label{equSpecificf}
{\cal P}(z) &=&\frac{1}{2\sqrt{6} {\cal G} R^3}\Bigg\{ \sqrt{\frac{2\pi}{{\sqrt{6}\cal G}}}\mu^2e^{\frac{2 \mu^4}{\sqrt{6}\cal G}} \Bigg[{\rm Erf}\left(\sqrt{\frac{2}{{\sqrt{6}\cal G}}}\mu^2\right)  \nonumber\\ 
&&\quad\quad\quad - {\rm Erf}\left(\sqrt{\frac{2}{{\sqrt{6}\cal G}}}\mu^2 -\sqrt{\frac{3{\cal G}}{\sqrt{6}}} z^2\right)\Bigg]+\left(1-e^{3 \frac{\phi(z)}{\sqrt{6}}}\right)\Bigg\},
\end{eqnarray}
where $\rm Erf$ is the error function.
The other two independent equations of motion (\ref{equbkVphi}) and (\ref{equbkVchi}) are satisfied by specifying the needed potential, which could be temperature dependent.

\begin{figure}[h!]
\begin{center}
\includegraphics[scale=0.45]{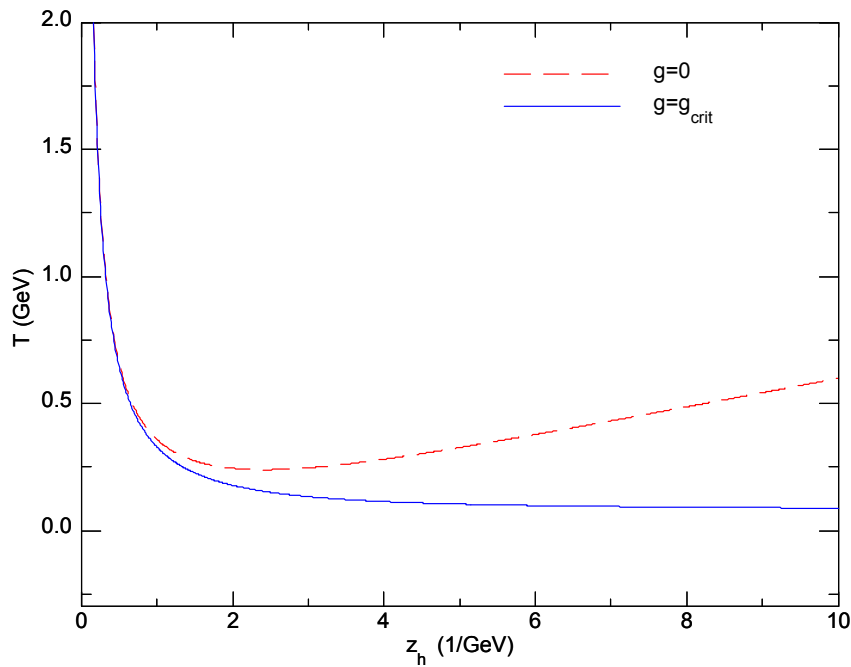}
\caption{The case of $g=0$ and $g=g_{{\rm crit}}$ are plotted. The large-$z_h$ behavior differs with the $g=0$ case exhibiting a second black-hole solution.}
\label{figAppT}
\end{center}
\end{figure}

After specifying the field behavior, $f(z)$, and $a(z)$,  we need to find an acceptable form for $\mathcal{G}$. Unfortunately, in the phenomenological picture, we have little guidance. Requiring $T(z_h)$ to go to the conformal limit in the small-$z_h$ region is the simplest constraint that we could impose. To satisfy the conformal limit, we find again that the leading order must take the form
\begin{equation}\label{eqappG}
\mathcal{G}(z_h) = g\frac{\mu^{2}}{z_{h}^{2}},
\end{equation}
as it did in Chapter \ref{chthermo}. Using (\ref{eqappG}), we find that the temperature,
\begin{equation}
T(z_h)  = \frac{1}{4\pi} \frac{\mathcal{P}'(z_h)}{\mathcal{P}(z_h)}
= \frac{1}{4\pi L^3} \frac{z_h^3 e^{3 \frac{\phi(z_h)}{\sqrt{6}}}}{{\cal P}(z_h)},
\end{equation} 
can be expanded in the small-$z_h$ limit,
\begin{equation}
T(z_h)=\frac{1}{\pi z_{h}} + \left(\frac{2\mu^{2}}{3 \pi}-\frac{\sqrt{6}g\mu^{2}}{4}\right)z_{h} + \ldots.
\end{equation}
The temperature has interesting behavior as $z_h\rightarrow\infty$. When $g=\sqrt{2/3}$, the temperature converges to a constant,
\begin{equation}
T(z_h\rightarrow\infty) = \frac{\mu}{\pi^{\frac{3}{2}}}.
\end{equation}
When $g$ takes any other value, the temperature decays exponentially, $T\sim e^{-z_{h}^{2}}$, so that the theory has no minimum temperature. Thus, we define a critical value for $g$,
\begin{equation}
g_{{\rm crit}} = \sqrt{\frac{2}{3}}.
\end{equation}
At $g_{{\rm crit}}$, we can express the temperature in its exact form,
\begin{equation}
\frac{\mu^{2}z_{h}}{\pi\left(e^{-\mu^{2}z_{h}^{2}}-1+\sqrt{\pi}\mu z_{h}{\rm Erf}(\mu z_h)\right)}.
\end{equation}
Figure \ref{figAppT} shows the plot of the temperature with and without a condensate function. We see that the behavior is similar until we reach large-$z_h$. The case when $g=0$ displays a linear growth, indicating a second, unstable black-hole solution 

\begin{figure}[h!]
\begin{center}
\includegraphics[scale=0.45]{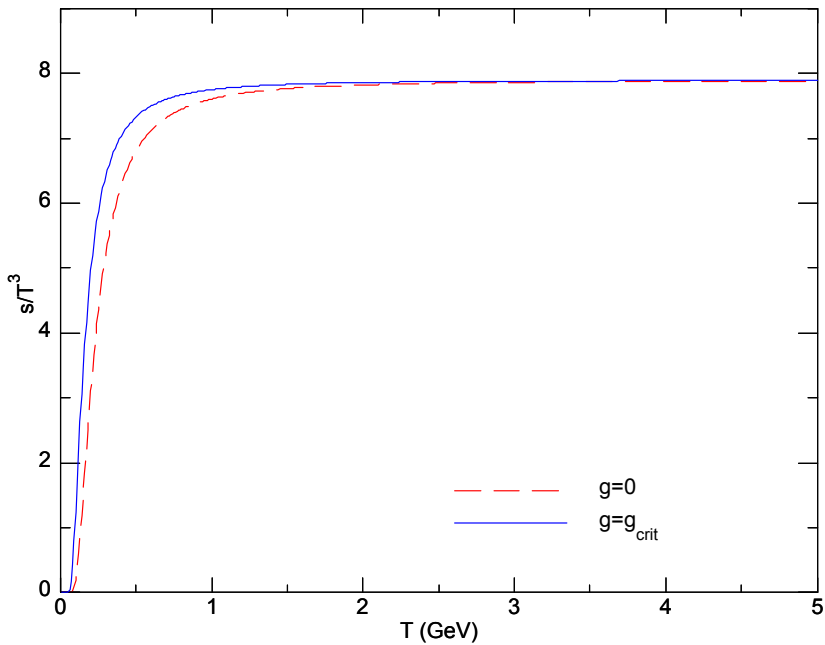}
\caption{A scaled entropy is plotted for two different values of $g$. Whether we consider the condensate function, there appears to be little difference in the behavior of the entropy.}
\label{figAppS}
\end{center}
\end{figure}

The entropy depends strictly on the form of $a(z)$,
\begin{equation}
s(z_h) = \frac{4 N_{c}^{2}V}{45\pi^{2}}\frac{e^{-\frac{3}{\sqrt{6}}\phi(z_{h})}}{z_{h}^{3}}.
\end{equation}
We expand this expression in terms of small-$z_h$ or large-$T$,
\begin{eqnarray}
s(z_h)&\approx& \frac{4 N_{c}^{2}V}{45\pi^{2}}\left(\frac{1}{z_{h}^{3}} - \frac{2\mu^{2}}{z_h}+ \frac{3\sqrt{2}}{2\sqrt{3}}\frac{g\mu^{2}}{z_h}\right),\\
s(T)&\approx& \frac{4 N_{c}^{2}V}{45\pi^{2}}\left( \pi^{3}T^{3} - 2\mu^{2}\pi T + \frac{3\sqrt{2}}{2\sqrt{3}}g\mu^{2}\pi T\right).
\end{eqnarray}
As expected, we obtain the conformal behavior in the high temperature limit, $s\sim T^{3}$. At $g_{{\rm crit}}$, the entropy becomes quite simple,
\begin{equation}
s(z_h) = \frac{4N_{c}^{2}V}{45\pi}\frac{e^{-\mu^{2}z_{h}^{2}}}{z_{h}^{3}}.
\end{equation}
Figure \ref{figAppS} shows two cases of the $s/T^{3}$ vs $T$. Besides a more rapid ascent to the conformal limit for $g=g_{{\rm crit}}$, each case shows little distinction. 

\begin{figure}[h!]
\begin{center}
\includegraphics[scale=0.45]{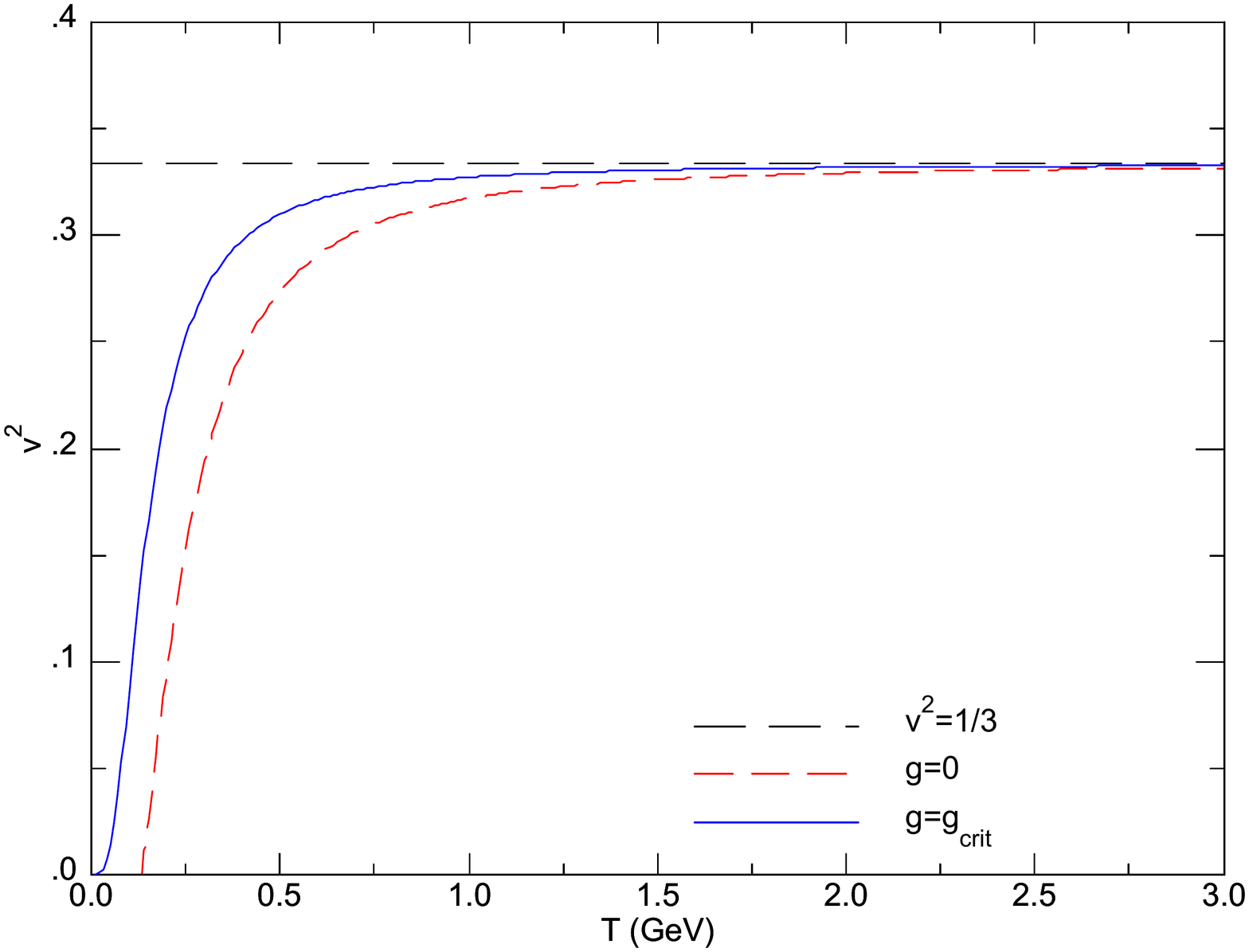}
\caption{The speed of sound is plotted for two different values of $g$. Each case approaches the conformal limit of $1/3$. When $g=g_{{\rm crit}}$, there appears to be a steeper ascent to this limit, however.}
\label{figAppvs2}
\end{center}
\end{figure}

We also find the speed of sound in the phenomenological model also goes to the conformal limit as the temperature increases,
\begin{eqnarray}
v_{s}(z_h)^{2} &\approx& \frac{1}{3} - \frac{8}{9}\mu^{2}z_{h}^{2} + \frac{5\sqrt{6}g}{18}\mu^{2}z_{h}^{2},\nonumber\\
v_{s}(T)^{2}&\approx& \frac{1}{3} - \frac{8\mu^{2}}{9\pi^{2}T^{2}} + \frac{5\sqrt{6}g\mu^{2}}{18\pi^{2}T^{2}}.
\end{eqnarray}
At $g_{{\rm crit}}$, we obtain an exact expression of $v_{s}^{2}$,
\begin{equation}
v_{s}^{2} = \frac{e^{\mu^{2}z_{h}^{2}}-1}{(3+2\mu^{2}z_{h}^{2})(\left(1+e^{\mu^{2}z_{h}^{2}}(\sqrt{\pi}\,\mu z_h \,{\rm Erf}(\mu z_h)-1)\right)}.
\end{equation}
Figure \ref{figAppvs2} shows the plot of the speed of sound versus temperature for $g=0$ and $g=g_{{\rm crit}}$. As with the entropy, we see similar behavior of $v_{s}^{2}$ in each case.

\begin{figure}[h!]
\begin{center}
\includegraphics[scale=0.45]{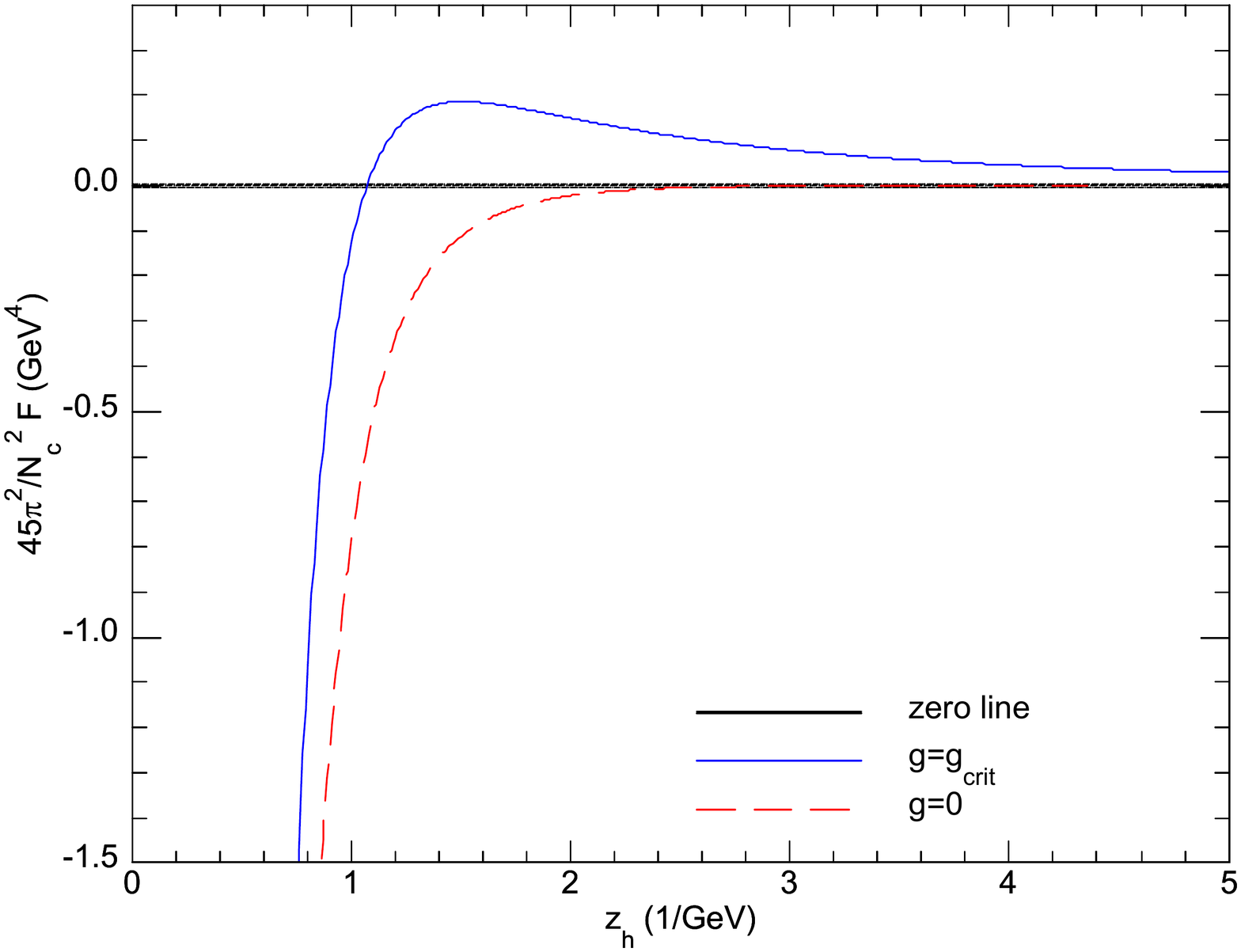}
\caption{The free energy is plotted for two different values of $g$. We find that only when the condensate function is nonzero ($g\ne 0$), do we recover a transition. For $g=g_{{\rm crit}}$, we find the transition occurs around $T_{c}\sim 300$ MeV. This value is high, but consistent with lattice calculations of $N_{f}=0$ QCD. }
\label{figAppF}
\end{center}
\end{figure}

We use the same general expression for the free energy, $\mathcal{F}$, (\ref{equfreeEmod2}), but using boundary matching, the cut-off in the thAdS solution becomes
\begin{equation}
\tilde{\delta} = \frac{ \sqrt{4\mu^{2}\delta^{2} - \sqrt{6}\mathcal{G}\delta^{4}} }{2\mu},
\end{equation}
where $\delta$ is the cut-off point in the bhAdS solution. As in Chapter \ref{chthermo}, we find that it reduces to the general expression,
\begin{equation}
\mathcal{F} = f_{4}(z_h) + 2\sqrt{6} \mathcal{G}(z_h).
\end{equation}
Using $g_{{\rm crit}}$, we find that a transition occurs at $z_{h}\approx 1.07$ GeV$^{-1}$, which corresponds to a transition temperature of $T_{c}\approx 308$ MeV. We plot two cases of the free energy in Figure \ref{figAppF}. The phenomenological model also indicates a condensation function is needed to obtain a transition. 

\end{document}